\documentclass[aps]{revtex4-2}
\usepackage{amsmath,amssymb,bbold}
\usepackage{color,graphicx}
\usepackage[export]{adjustbox}
\usepackage{geometry}
\newgeometry{tmargin=2cm, bmargin=3cm, lmargin=1cm, rmargin=1cm}
\newcommand{\be}{\begin{equation}}
\newcommand{\ee}{\end{equation}}
\begin{document}
 \title{Tamed Feynman-Kac diffusion processes: Killing-branching intertwine.}
 \author{Piotr Garbaczewski   and Mariusz \.{Z}aba}\thanks{P. G.  ORCID 0000-0003-4545-1107, M.Z. ORCID 0000-0001-5314-329X;  email addresses: pgar@uni.opole.pl, zaba@uni.opole.pl}
 \affiliation{Institute of Physics, University of Opole, 45-052 Opole, Poland}
 \date{\today }
 \begin{abstract}
 Relaxation to equilibrium of a drifted Brownian motion can be  quantified by a transition probability density function,  whose main   multiplicative (Doob-like weighted)  is an inferred Feynman-Kac kernel of the Schr\"{o}dinger semigroup operator.  The pertinent  kernel  captures  a complete    information about  the time   evolution and actually  controls  the  asymptotic equilibration.  The implicit  Feynman-Kac potential ${\cal{V}}(x)$, which is confining,  continuous  and bounded from below,  may take negative values.  If  positive, ${\cal{V}}(x)$  can  be considered  as the  killing rate  of the decaying  diffusion process. In the case of  relaxing diffusion,   killing effects need to be  tamed.  For unbounded random motion, the taming   unavoidably  appears in conjunction with  the existence of  the negativity subdomains of ${\cal{V}}(x)$ in $R$.  If  locally   ${\cal{V}}(x) < 0$, we  assign  a    probabilistic meaning to  the  sign inverted potential  $- {\cal{V}}(x)$, and interpret it  as  the branching (cloning, alternatively  trajectory bifurcation)  rate.  This points towards the   killed  diffusion  with branching,  as  a possible path-wise  background  for  the time evolution of the considered Feynman-Kac kernel.  The emergent tamed  Feynman-Kac diffusion scenario is  discussed in detail  for the exemplary quadratic  potential ${\cal{V}}(x)$.  The killing/branching algorithm, introduced in the present paper,   induces a statistical ensemble of random paths, whose  time evolution and asymptotics   prove  to be consistent with the  exact  analytical outcomes.
  This sets a rationale for a subsequent  computer-assisted path-wise  analysis of the validity  of   the  tamed F-K diffusion concept,  for   a number  of nonlinear  models    in one space dimension, where analytic results   are   scarce. Special  attention is paid to  Feynman-Kac  potential shapes  in the double  well form, where an  analytic   access to  eigenvalues and eigenfunctions of the related  Schr\"{o}dinger semigroup generator  is unavailable beyond the semiclassical (or perturbative) regime. Throughout the paper the dynamics refers to the real time $t \geq 0$.  Since  the   Newton-type equations of motion for admissible  classical trajectories have a  Euclidean form   (due to the sign inverted force term), we give a brief resume  of a couple of  their  explicit   solutions, without recourse to "imaginary"  (Euclidean) time intuitions, and the instanton lore of related quantum model systems.

   \end{abstract}
 \maketitle

\section{Introduction.}
\subsection{Basics.}

 Our departure point is   the conceptual framework set by the pseudo-Schr\"{o}dinger reformulation of the Fokker-Planck dynamics, \cite{risken,pavl,faris}, and the subsequent   exploitation of the Feynman (respectively Feynman-Kac) path integration route in the derivation of integral   kernels of  intimately   related  semigroup (motion) operators $\exp(tL^*)$ and $\exp(-tH)$, \cite{klauder,hunt} and \cite{glimm,olk},  c.f. also  \cite{mazzolo0}-\cite{stef}.    Here $L^*$  stands for the Fokker-Planck generator, while $H$  for the associated  Schr\"{o}dinger-type  Hamiltonian.

 We point out that the integral kernels of semigroup operators  in question are: (i) transition probability densities $p(y,s,x,t)=  ( \exp[L^*(t-s)] )(y,x),  0\leq s<t $ of the diffusion process,  and (ii) (Euclidean)  propagators $k(y,s,x,t) =  ( \exp[-H(t-s)] )(y,x)$ of the generalised Schr\"{o}dinger equation. In connection with (ii), we  employ   $H=-\nu \Delta + {\cal{V}}$, where $\nu =1/2$ is  predominantly in use  in the present paper.   A  continuous bounded from below   potential  function  ${\cal{V}}(x)$   may take negative values on subdomains of $R$,  \cite{gar,pre24,huillet}.

 Let us consider a  Markovian  diffusion process   $X(t)$, associated with the stochastic differential equation of the Langevin-type, (here interpreted in terms of infinitesimal  time increments) $dX(t) = b(X(t)) dt + \sqrt{2\nu } dW(t)$.  We presume the forward  drift  to be time independent and  conservative,  $b(x)= - \nabla \phi (x)$,    $\nu $  stands for   a  diffusion constant ($2\nu  $ is  interpreted as the variance parameter), and  $W(t)$  is the normalised   Wiener noise   in $R$, defined by   expectation values $\left< W\right> =0$ and $\left< W(s)W(t)\right> = \delta(s-t)$.

From now on we rescale the diffusion coefficient to the value $\nu =1/2$, to conform with the notation of \cite{glimm,olk,zaba1,gar,pre24,zaba,stef}.  Accordingly, if an initial probability density   function  $\rho_0(x)$ is given,   its  time evolution
$\rho_0(x)= \rho (x,0) \rightarrow  \rho (x,t)=  [\exp(tL^*)\rho_0](x)$  follows  the  Fokker-Planck equation:
\be
dX(t) = b(X(t)) dt +  dW(t)  \,   \Longrightarrow  \,     \partial _t \rho = {\frac{1}2} \Delta \rho - \nabla (b \rho ) = L^* \rho ,
\ee
where the  operator $L^*=  (1/2) \Delta  - \nabla (b \, \cdot )$, in view of $\nabla  (b \rho ) = (b  \nabla ) \rho +
\rho (\nabla b)$, can be rewritten as follows:
\be
L^* ={\frac{1}2} \Delta - b \nabla -  (\nabla b) =
{\frac{1}2}(\nabla  - b)^2  - {\cal{V}}.
\ee
The emergent potential function ${\cal{V}}(x)$  reads:
\begin{equation}
{\cal{V}}(x) =   \frac{1}{2} \left(b^2  + \nabla b\right) = \frac{1}{2} \left[(\nabla \phi )^2  -  \Delta \phi \right].
\end{equation}

Given $\rho (x,t)$ solving Eq. (1), let us  introduce an osmotic  velocity  field  $u(x,t) =  \nabla \ln \rho ^{1/2} (x,t)$  and  the current velocity field  $v(x,t)=b (x) - u(x,t)$, with $b= - \nabla \phi  $, where $\phi =\phi (x)$ does not depend on time.  We can  rewrite the F-P equation as the  continuity equation $\partial _t \rho = - \nabla j$, where  $j= v\cdot \rho $  has   a standard interpretation of a probability current.

We assume that the diffusion process asymptotically relaxes to the stationary (invariant)  strictly positive  pdf,   $\rho (x,t) \to \rho _*(x)$ as  $t\rightarrow \infty $.  In the  stationary  regime we have $j\rightarrow j_*=0$ and thence   $v\rightarrow v_*=0$. Since   $b $  is time-independent, the drift field  potential  $\phi (x)$   (presumed to be confining)   becomes correlated with $\rho _*$:  $b=u_* = \nabla  \ln \rho _*^{1/2} =  - \nabla \phi $.

Accordingly,  a stationary solution  of the Fokker-Planck equation actually appears in the form  $ \rho _*(x) = (1/Z) \exp[ - U(x)]$, where  $Z = \int_R \exp[-U(x)] dx$, with    $U(x) = 2\phi (x) $. (This stems from  the customary  form of the  Gibbs-Boltzmann weight $(1/Z) \exp(- \phi (x) /\nu )$,  once we  set $\nu = 1/2$, \cite{zaba}.)

\subsection{Pseudo-Schr\"{o}dinger route.}

 The  Fokker-Planck  time evolution (1) can be quantified by means of  a transition probability density function  $p(y,s,x,t)$, $0\leq s<t\leq  T$, ($T\rightarrow \infty $), so that   $\rho (x,t) = \int p(y,s,x,t) \rho (y,s) dy$.  We presume $p(y,s,x,t)$ to be a (possibly fundamental, be aware that this needs quite stringent growth restrictions imposed upon the drift function) solution of the F-P equation, with respect to
   variables $x$ and $t$, i.e.  $ \partial_t  p(y,s,x,t)  = L^*_{x}p(y,s,x,t)$.

 Following a standard procedure \cite{risken,pavl}, given a stationary probability  density $\rho _*(x)$,   one can  transform the Fokker-Planck  dynamics   into an associated Hermitian (Schr\"{o}dinger-type) dynamical  problem in $L^2(R)$,  by means of a  factorisation:
\begin{equation}
\rho (x,t) = \Psi (x,t) \rho _*^{1/2}(x).
\end{equation}
Indeed, the Fokker-Planck evolution  of  $\rho (x,t)$   implies the validity  of  the   generalized diffusion (Schr\"{o}dinger-type)   equation
 \begin{equation}
 \partial _t\Psi=  {\frac{1}2} \Delta \Psi - {\cal{V}} \Psi  = - H \Psi ,
\end{equation}
for  $\Psi (x,t)=  [e^{(-tH)}\Psi ](x)$, with $\Psi (x.0)= \rho (x,0)/ \rho^{1/2}_*(x)$. \\

 Asymptotic features of the semigroup dynamics  driven by $\exp(-tH)$,  critically depend  on  spectral properties of the Schr\"{o}dinger-type operator $H$. The knowledge of lowest eigenfunctions and eigenvalues is of utmost importance. See e.g. our discussion of the quadratic case in Section II.

  For nonlinear models of Sections III and IV, the  available analytical data are extremely limited. To overcome these shortcomings of the theory,  we shall   employ an efficient  computer-assisted   Strang splitting method (thoroughly tested in Refs.  \cite{jmp14,jmp15}) to get an  access to the  spectral data of $H$. Quite  accurate approximations of lowest eigenfunctions and eigenvalues of Schr\"{o}dinger-type operators considered in Sections III and IV, have been achieved  this way.  \\

   The relaxation asymptotics   $\rho (x,t) \rightarrow \rho _*(x)$ as $t\rightarrow \infty $,  needs to be  paralleled by  $\Psi (x,t)  \rightarrow  \rho _*^{1/2}(x)$, hence $\Psi (x,t)$  itself exhibits  the  relaxation  behavior (its path-wise implementation is actually the  main focus of  the present paper).

In view of $\partial _t \rho _*^{1/2}=0$,     $\rho _*^{1/2}$ is a  strictly positive  eigenfunction  of  the Schr\"{o}dinger-type operator  $H= -(1/2) \Delta + {\cal{V}}$, corresponding to the eigenvalue zero, $H\rho_*^{1/2}=0$.    This implies that the    potential function  ${\cal{V}}(x)$ necessarily   derives in the form:
\begin{equation}
{\cal{V}}(x) = {\frac{1}2} \,  {\frac{\Delta \rho _*^{1/2}}{\rho _*^{1/2}}},
\end{equation}
which  actually  is another (equivalent) form of  Eq. (3)  (that in view of   $b=u_* = \nabla  \ln \rho _*^{1/2} =  - \nabla \phi $).

The    potential function ${\cal{V}}(x)$ of Eqs. (3) and (6),  bounded from below and continuous (this  is  secured by the properties of $\rho _*(x)$ and thence $\phi (x)$),  takes negative values on  bounded subsets of $R$.  This  local {\it negativity property}  of ${\cal{V}}(x)$   will be of relevance in our further discussion. We shall relate it to the concept of trajectory cloning (branching/bifurcation) events, for  sample paths of a killed diffusion process,   visiting  the negativity area  of ${\cal{V}}(x)$,  cf. \cite{pre24}.\\

  In the dual    picture provided by the  Feynman-Kac  (Schr\"{o}dinger)  semigroup,   a priori chosen   $\rho _*(x)$   determines  the  Feynman - Kac potential    of  Eq. (6), and   thence the  "potential landscape"  set  by the spatial profile of ${\cal{V}}$.
 The  Feynman-Kac  relaxation process refers to the time rate at which  the bottom  eigenfunction       $\rho _*^{1/2}$    of   $H$, Eq. (6), is    approached in the course  of the time  evolution by  $\Psi (x,t) \rightarrow   \rho _*^{1/2}(x)$.
 The    spectral property  $H \rho _*^{1/2} = 0$ is here  of vital importance, \cite{klauder,faris,gar,stef}.\\

{\bf Technical comment 1:}\\
 The identification of Eq. (3) with Eq. (6) is possible only in the stationary regime, which is maintained by suitable drift fields. It is worthwhile to invoke  more general framework of Ref. \cite{zaba1,physa}, in which drifts, still represented by gradient fields, may be time-dependent. The Fokker-Planck equation (we use $\nu =1/2$)   $(1/2)\Delta \rho  - \nabla (b \rho )$, with $b(x,t)= - \nabla \phi(x,t)$   takes the familiar form of the continuity equation   $\partial _t\rho = - \nabla (v \rho )$ upon a substitution $v= b- {\frac{1}2}[(\nabla \rho )/\rho ]$.  The current velocity obeys an equation (we follow the notation of \cite{physa,qpot,recoil})
 $$
 \partial_tv + (v\nabla )v  = \nabla (\Omega - Q),
  $$
 where  $\Omega(x,t) = - \partial_t \phi + {\frac{1}2} (b^2 + \nabla b)= - \partial_t\phi + {\frac{1}2} [(\nabla \phi )^2 - \Delta \phi ]$,
and (remembering that $u={\frac{1}2}\nabla \ln \rho $) we have   $Q(x,t)= {\frac{1}2}(u^2 + \nabla u) = {\frac{1}2} {\frac{\Delta \rho ^{1/2}}{\rho ^{`1/2}}}$.
We realize that in the stationary regime,  $Q(x)= \Omega(x)= {\cal{V}}(x)$.
\noindent
In case of the free Brownian motion, the drift contributions are absent, and we have $\partial_tv + (v\nabla )v  = - \nabla Q$.
\\

{\bf Remark 1:} We may  proceed  {\it  in reverse}, \cite{klauder,zaba,vilela}, and  choose  any bounded from below continuous potential   $V(x)$  as the Feynman-Kac exponent entry, cf. \cite{glimm,faris,klauder}. To follow the previous routine, we typically need to replace the original potential by a "potential with  subtraction" (e.g. "shifted potential")  $ {\cal{V}}(x)$ , \cite{faris,gar,zaba,pre24,stef}, so that the Schr\"{o}dinger-type operator with  a   subtraction, admits zero as the lowest eigenvalue.
 To this end   we must  know  the bottom eigenvalue of the original Hamiltonian   $H_0= -(1/2)\Delta + V$ , (with $V$ not necessarily positive).  In case of nonnegative confining potentials, we need to  know  a couple of lowest  positive eigenvalues.  Then, given $V(x)$, having deduced the  bottom  eigenvalue  $\epsilon $   of $H_0$, together with the corresponding eigenfunction $\psi (x) \sim \rho _*^{1/2}(x)$,    we  can   modify    the identity (6) to  encompass the  "shifted potential":  ${\cal{V}}(x)= V(x)- \epsilon = [  \Delta \rho _*^{1/2}]/ 2 \rho_*^{1/2}$.  The resultant   Schr\"{o}dinger type operator  $H= H_0 - \epsilon = -(1/2) \Delta + {\cal{V}}(x)]$ admits zero as the bottom eigenvalue:  $H \rho_*^{1/2}= 0$.\\
 We shall  explore this "shift"   routine  in below,  while discussing  superharmonic and double-well (where   the negative bottom eigenvalue is to be   handled) candidates for F-K potentials, cf. Sections III.B,   IV.A and IV.B of the paper. The above  reasoning can be readily  adopted to the  harmonic case, see e.g.  Fig. 1.\\

{\bf Remark 2:}
In the Langevin modeling of the   Brownian motion,  we prefer a rescaling of  the diffusion constant to the form $\nu =1/2$, while quite often $\nu =1 $ happens to be in use.
For clarity of discussion, and  to facilitate  a  comparison with arguments  of Refs.  \cite{gar,stef}, we recall  that prior to the ultimate rescaling  of $\nu $ to a convenient form, an  asymptotic pdf reads   $\rho _*(x) = (1/Z) \exp[-\phi (x)/\nu]$, while the drift derives from $\rho _*(x)$ according to  $b(x)=- \nabla \phi (x) = 2\nu \nabla \ln \rho _* ^{1/2}(x)$.\\
In view of   $ H\,  \rho _*^{1/2} = 0$,  the admissible   functional form of the  potential function  ${\cal{V}}(x)$   derives  as a function  of $\rho _*^{1/2}(x)$. We have  ${\cal{V}}(x) = \nu  [\Delta \rho _*^{1/2}]/[\rho _*^{1/2}] =     (1/2) [b^2/2\nu  + \nabla b]$.  Here  $b(x)= 2\nu \nabla \ln \rho _*^{1/2}(x)= -\nabla \phi(x)$.\\
For a trivial verification, consider  the harmonic case  $\phi (x) = x^2/2 \rightarrow b(x)= - x$, and   set $\nu =1/2$. Then  ${\cal{V}}(x) = (1/2) (x^2-1)$, while $\rho _*(x)= \pi ^{-1/2} \exp (-x^2)$, see e.g. section II in below.\\

\subsection{Path integration.}

Fokker-Planck  transition probability density functions  and probability densities,   for diffusions with (non)conservative drifts, are  known to be    amenable to   Feynman's  path integration  routines, \cite{hunt,zaba1}. In case of  conservative drifts, this  can be achieved  by means of a   multiplicative (Doob-like)  conditioning of the related (strictly  positive)   Feynman-Kac kernels,  \cite{zaba,gar,mazzolo0,hunt,olk,zaba1,glimm},  provided the existence of   stationary  pdfs is granted.

The   path integral context  for  drifted diffusion processes   has been  revived in Refs. \cite{hunt,monthus,zaba1},
 through the formula   "for the propagator associated with the Langevin equation" (1) (e.g.  the  integral kernel of the operator $\exp(tL^*)$  with $L^*$ given by Eqs. (2) and (3)):
\be
p(y,0,x,t)=  \exp(L^*t)(y,x)=  \int_{x(\tau =0)=y}^{x(\tau =t)=x}
   {\cal{D}}x(\tau ) \,  \exp \left[ - \int_0^t  d\tau {\cal{L}}(x(\tau ), \dot{x}(\tau )) \right],
\ee
where the $\tau $-dynamics stems  from  the  {\it Euclidean}  (this is a folk term)   Lagrangian ${\cal{L}}$:
\be
{\cal{L}}(x(\tau ), \dot{x}(\tau )) = {\frac{1}2} \left[ \dot{x}(\tau ) - b(x(\tau ))\right]^2  + {\frac{1}2}  \nabla b(x(\tau ))=
{\frac{1}2}\dot {x}^2(\tau )  - \dot{x}(\tau ) b(x(\tau )) + {\cal{V}}(x(\tau )),
\ee
with ${\cal{V}}(x)$ given by  Eq. (5).
We stress   that the "normal"   (e.g. non-Euclidean)  classical Lagrangian would have the form $L = T - V$ with $T=  \dot{x}^2 /2$   and $V(\dot{x},x,t)= {\cal{V}}   - \dot{x} b$.

Let us consider the action functional  (e.g.  the  minus exponent) in Eq. (7). In view of  $b = - \nabla \phi = \nabla  \ln \rho _*^{1/2}$, we readily infer that the term $\dot{x}(\tau )\,  b(x(\tau ))$ in the Lagrangian (8) contributes:
\be
\int_0^t \dot{x} [- \nabla \phi (x(\tau ))]  d\tau =
  - \int_0^t {\frac{d}{d\tau }} \phi (x(\tau )) d\tau  = \phi (x(0)) - \phi (x(t)).
\ee

Therefore, the related probability density function  (path integral kernel of $\exp(tL^*)$) can be rewritten in  the form:
\be
p(y,0,x,t)=  e^{\phi (y) - \phi (x)}\,  k(y,0,x,t)
\ee
where the new function $k(y,0,x,t)$ is no longer a transition probability density (does not integrate to one) but an integral kernel of another  motion operator (actually $\exp(-tH)$, c.f. Eqs. (5), (6)):
 \be
 k(y,0,x,t) =   \int_{x(\tau =0)=y}^{x(\tau =t)=x}
   {\cal{D}} x(\tau ) \,  \exp \left[ - \int_0^t  d\tau {\cal{L}}_{st}(x(\tau ), \dot{x}(\tau )) \right],
\ee
where
\be
{\cal{L}}_{st} (x(\tau ), \dot{x}(\tau )) =
{\frac{1}2}\dot{x}^2(\tau )  +  {\cal{V}}(x(\tau ))
\ee
and ${\cal{V}}$ is given by  Eq.(3).

On the  operator  level, the passage from the  transition kernel $p$ of  (7) to $k$ of (11-13), amounts to the similarity transformation,  \cite{pavl,monthus,zaba,zaba1}:
$H = e^{\phi } L^* e^{-\phi }=  - {\frac{1}2} \Delta  + {\cal{V}}$.
This  outcome can be readily verified by resorting to the operator identity  $e^{\phi } {\nabla } e^{- \phi } =  c{\nabla } - ({\nabla }\phi )$.

The   formula  (11)  can be redefined    as   a   (Feynma-Kac) weighted integral over sample paths  of the  Wiener process (colloquially, the free  Brownian motion), with the  conditional   Wiener path measure  $\mu _{(y,0,x,t)}(\omega)$    being involved, \cite{olk,glimm,faris,klauder}:
\be
  k(y,0,x,t) =  [\exp(- t H)](y,x)  =
       \int   \exp[-\int_0^t {\cal{V}}(\omega(\tau  )) d\tau ]\,   d\mu _{(y,0,x,t)}(\omega).
        \ee
 Here  paths  $\omega $ originate from $y$ at time $t=0$ and their  final  destination  $x$  is to be reached  at time $t>0$).
In contrast to the kernel function $k(y,0,x,t)$, transition pdfs  $p(y,0,x,t)$ are not symmetric functions of $x$ and $y$.

One may here  try to  imagine  a pictorial view  of the    Brownian motion in potential energy  landscapes, as   set by  Feynman-Kac potential  ${\cal{V}}(x)$ spatial profiles. The Wiener path measure in Eq. (13) refers to paths of the free (undisturbed) Brownian motion, and it is the exponential factor which represents,  \cite{klauder},  "the distortion of the distribution of free-particle paths, introduced by the potential".   The detailed mechanism  of this "distortion" is not  uniquely specified. \\

 Eq. (13) admits potentials which are continuous and bounded from below,  \cite{glimm}, while ${\cal{V}}(x) \geq 0$ is required for a standard  killed  diffusion  interpretation  of the Feynman-Kac formula and suitable  "distortion" features of the potential   (amplified by looking at  ensemble of surviving  random paths).

 As explained  in Refs \cite{klauder,ito},  the quantity ${\cal{V}}(x)$ may be assigned a meaning of  the  killing  time rate  at a point $x$. Accordingly, a diffusing path  approaching  $x$ at time $t$ has some chance to vanish with the given killing rate. Actually  we admit that a  random mover (whatever this term may mean), reaching    the point  $x$ at time $\tau $  along the sample path,  may  get  killed at $x$ in the time interval $\delta \tau $ (infinitesimally $d\tau $) with the probability ${\cal{V}}(x)\delta \tau $.  The killed path is removed from the ongoing  ensemble of  admissible  (still surviving)   Wiener paths.

 That modifies the statistics of paths-in-existence to the extent, that at the terminal  time $t$, the Feynman-Kac average (13) over Wiener paths is taken exclusively with respect to  these  paths, which have  survived the full period $[0, t ]$ to complete  their travel from $y$ to $x$.

The factor $ \exp [-\int_0^t {\cal{V}}(\omega (\tau ))d \tau ]$ in Eq. (13) is a probability that a random mover  actually  completes its   (continuous)  path from $(y, 0)$ to $(x, t)$, cf. Ref. [4]. However, for nonnegative potentials, in the asymptotic limit $t\to \infty$, no surviving paths are left. In passing, we mention that this paths extinction  is  reflected  in the continual exponentially modulated  decay of so-called  quasi-stationary distributions, \cite{stein,collet}.

It is clear that a  classical   redefinition of the Feynman-Kac kernel with ${\cal{V}}(x) \geq 0$,   in terms of the   killed  diffusion process, \cite{klauder,ito},  does not  comply with the relaxation scenario (4), (5).  To  tame  the  killing and  secure the relaxation  for the spatially unbounded  random motion, the inferred  potential (3), (6)  needs to have  negativity domains,  while being bounded from below.

   We shall henceforth  pursue   an interpretation of the emergent tamed Feynman-Kac process, in terms  of  the  path-wise  killing-branching scenario, with mutually exclusive killing or branching  spatial areas  (subdomains of the F-K potential), which is  superimposed on the  otherwise free Brownian (Wiener) evolution, \cite{gar,pre24}, see also \cite{huillet,kesten,ber,zoia,mazzolo1}.  \\

\section{Handling the linear drift  $b(x)= -x$.}

\subsection{Tamed Feynman-Kac diffusion vs the Ornstein-Uhlenbeck process.}

Eq. (1) with a linear drift  $b(x) = -x$, refers to the Ornstein-Uhlenbeck process (in its configuration space version) where  $\phi (x)= x^2/2$, $\nabla b = -1$, and ${\cal{V}}(x)=  {\frac{1}2}  (x^2 -1)$. The path integral action in Eq. (8) takes the form ${\cal{L}}(x(\tau ), \dot{x}(\tau )) = {\frac{1}2} \left[ \dot{x}(\tau ) + x(\tau ))\right]^2  - {\frac{1}2}$, while  ${\cal{L}}_{st} (x(\tau ), \dot{x}(\tau )) =
{\frac{1}2}[\dot{x}^2(\tau )  +  x^2(\tau ) -1]$.

The  additive   renormalization  ${\frac{1}2} x^2 \,  \rightarrow \,  {\frac{1}2} (x^2-1)$ of the  potential  (considered before  in the literature \cite{glimm}),   introduces  the notion of the "shifted potential"("potential with subtraction"),  \cite{faris,gar,pre24}. This    secures the validity of the relaxation scenario  for both $p$ and $k$ in Eqs. (11), (13).

The above  (looking innocent) subtraction refers to the presumed existence of the spectral solution of the Schr\"{o}dinger Hamiltonian $H= - (1/2) \Delta + {\cal{V}}$, assigning the value  zero  to the isolated  lowest (bottom) eigenvalue, \cite{glimm,zaba,stef,gar,pre24}.

For concreteness, let us recall that given a   spectral solution for the harmonic oscillator problem with $H_0= (1/2)(- \Delta + x^2)$,  the  integral kernel of $\exp(-tH_0)$ acquires  the  time-homogeneous form
\be
k_0(y,x,t)= \exp[-  H_0 t](y,x) = \sum_j \exp(- \epsilon _j t) \, \psi _j(y) \psi _j(x).
\ee
where $\epsilon _j= j + {\frac{1}2}$,  $j=0,1,...$  and $\psi _j$ are respective eigenfunctions (being  real-valued in the harmonic oscillator problem).   Here $\epsilon_0=1/2$ and   $\psi _0(x) = \pi ^{-1/4} \exp(-x^2/2)$  and $x^2/2 = \phi (x)$  in the notation of Section I.A. ( $\psi _0^2(x)= \rho _*(x)$, which is a stationary solution of the F-P equation (2), with the drift $b(x) = -x$.)\\
 \begin{figure}[h]
\begin{center}
\centering
\includegraphics [width=0.4\columnwidth, valign=c] {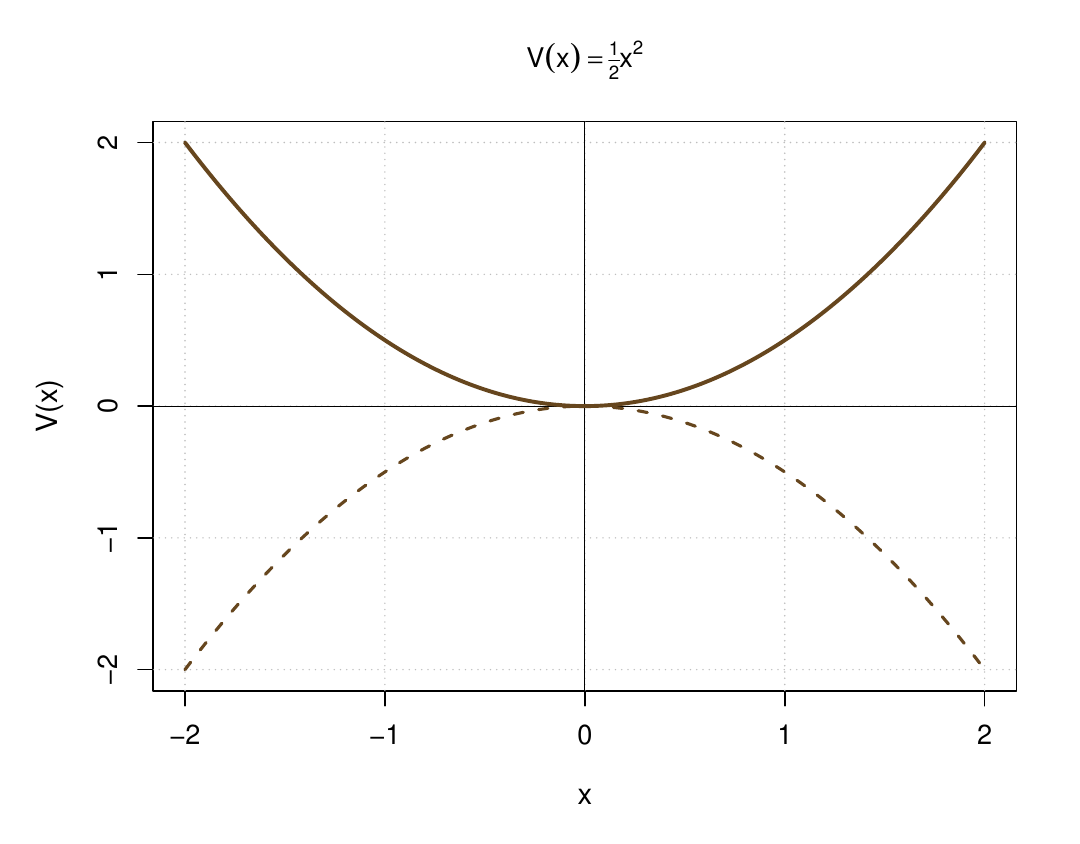}
\includegraphics [width=0.4\columnwidth, valign=c] {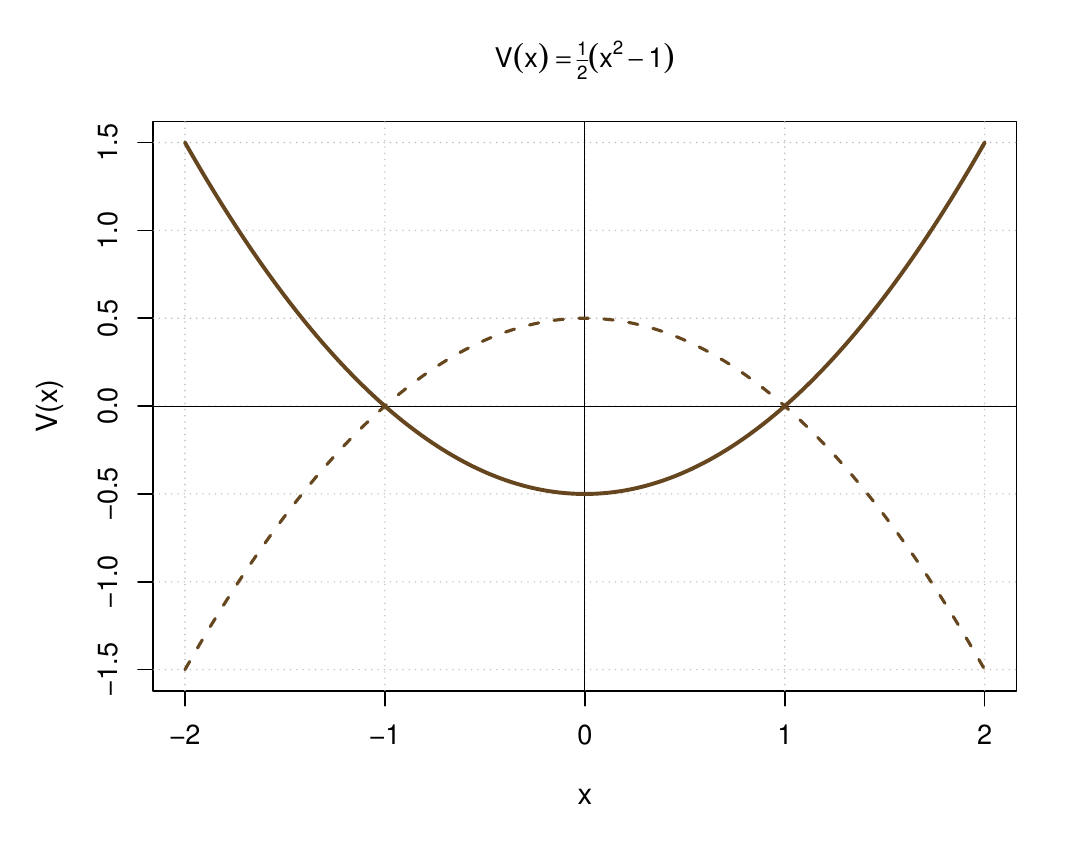}
\caption{Left panel: Harmonic oscillator potential $V(x)= x^2/2$ and its inverted (often termed Euclidean) partner. Right panel: "Harmonic potential with subtraction" and its inverted partner.
  Here, $(-1,1)$ is the negativity subdomain for  ${\cal{V}}(x) = V(x) - {\frac{1}2}= (1/2)(x^2 -1)$.  Remember that  $\epsilon_0= 1/2$ is the bottom eigenvalue of $H_0$. }
\end{center}
\end{figure}

 Since  $\epsilon _0 = 1/2$, multiplying $k_0(y,x,t)$  by $\exp(+t/2)$,  we pass to the integral kernel of the renormalized semigroup (where there  ultimately  appears a "potential with subtraction"), e.g.
\be
k(y,x,t) = k_0(y,x,t) \exp (+t/2)  = \left[\exp(- H t)\right](y,x) = \psi_0(y)\psi_0(x) + \sum_{j=1}^{\infty }  \exp[- (\epsilon _j - \epsilon _0) t] \, \psi _j(y) \psi _j(x).
\ee
Here,  $H= (H_0 - \epsilon _0)$ and   $(\epsilon_j - \epsilon_0)>0 $  for all $j>0$.  Clearly,   ${\frac{1}2}(x^2 -1) = {\cal{V}}(x)$,   as defined by  Eqs. (3), (6) with he choice of  $b(x)= -x$.  Compare e.g. Eq. (14).

Actually, we have in hands a complete analytical expression for  the harmonic oscillator kernel $k_0(y,x,t)$, Eq. (14):
\be
 k_0(y,x,t)
= {\frac{1}{(2\pi  \sinh t)^{1/2}}}  \exp  \left[ - {\frac{(x^2+y^2)\cosh t  - 2xy}{2\sinh t }} \right]
\ee
$$
 =\exp(-t/2)\, (\pi [1-\exp(-2t)])^{-1/2} \exp \left[{\frac{1}2} (x^2-y^2) -
   {\frac{(x- e^{- t}y)^2}{(1- e^{-2 t})}}\right].
 $$
We indicate the conspicuous  presence of the damping  factor $\exp(- t/2)$, which is responsible for an exponential decay of $k_0(y,x,t)$ as $t\rightarrow \infty $.
Indeed,   we  have   $ k_0(x,t) \rightarrow   [\pi ^{-1/4} \psi _0(x)] \exp (- t/2)$.     In Eq. (15), the  pertinent exponential  damping  factor is  counterbalanced by  $\exp(+t/2)$.

Recalling the definition of the transition probability density,  Eq. (10), we readily recover the final outcome of the path integration, Eqs. (8) and (9),  in the familiar form:
\be
p(y,0,x,t)= e^{\phi (y) - \phi (x)}\,  k(y,0,x,t) = (\pi [1-\exp(-2t)])^{-1/2} \exp \left[-{\frac{(x- e^{- t}y)^2}{(1- e^{-2 t})}}\right]
\ee
appropriate for the  (time homogeneous)  Ornstein-Uhlenbeck process, with $b(x)=-x$  and an asymptotic pdf $\rho _*(x)= \pi ^{-1/2} \exp (-x^2)= [\psi_0(x)]^2$.

\subsection{Euclidean trajectory  input in Feynman's derivation of $k_0(y,x,t)$.}

A specific feature of the previous analysis is that the   functional form of the integral kernel $k_0(y,x,t)$   of the harmonic oscillator semigroup (without subtraction)  directly derives from the standard (non-Euclidean, text-book) solution of the spectral problem, by  employing the Mehler formula  \cite{glimm}:
\be
k_0(y,x,t) = [\exp(-tH_0)(y,x)=   {\frac{1}{\sqrt{\pi }}} \exp[- (x^2+y^2)/2] \, \sum _{n=0}^{\infty } {\frac{1}{2^n n!}} H_n(y) H_n(x)
\exp(-\epsilon _n\, t),
\ee
where $\epsilon _n= n +  {\frac{1}2}$,  $\psi _n(x) = [4^n (n!)^2 \pi ]^{-1/4} \exp(-x^2/2)\, H_n(x)$ is the $L^2(R)$
 normalized Hermite  (eigen)function, while $H_n(x)$ is the n-th Hermite polynomial $H_n(x) =(-1)^n (\exp x^2) \, {\frac{d^n}{dx^n}} \exp(-x^2) $.  The $n=0$  ground state function reads $\psi _0(x)= \pi ^{-1/4} \exp (-x^2/2)$.

 We point out that Eq. (16) provides two visually  different, but   equivalent  forms  of  the kernel (19).

 It  may look somewhat surprising that the evaluation of the pertinent kernel, via Feynman's path integration recipe \cite{feynman}(Chap.3.2)  critically relies   on  an input of (Euclidean) classical paths of the inverted  harmonic oscillator. (We never refer to the concept of the inverted quantum oscillator, \cite{barton,gar1}.)

 Indeed,   the Lagrangian of the form (12) (we skip the {\it st}  subscript in ${\cal{L}}_{st}$ to simplify notation),  implies the Euler-Lagrange equations in the Euclidean form:
 \be
{\frac{\partial {\cal{L}}}{\partial x}} - {\frac{d}{dt}} {\frac{\partial {\cal{L}}}{\partial \dot{x}}}=0 \Longrightarrow
{\frac{\partial {\cal{V}}} {\partial x}}  - {\frac{d}{dt}} \left( {\frac{\partial {\cal{T}}}{\partial {\dot{x}}}}  +  {\frac{\partial {\cal{V}}} {\partial {\dot{x}}}}\right) = 0.
\ee
Accordingly, we have
\be
\ddot{x}= {\frac{\partial {\cal{V}}}{\partial {x}}},
\ee
which has the  Euclidean  (inverted, positive) right-hand-side sign. In the harmonic oscillator context, the   $-1/2$  renormalization of $x^2/2$  is irrelevant for Euclidean path  analysis, but has a profound (taming) effect  on the otherwise killed diffusion, cf. Eq. (16).

Dynamical equation (20) points towards the  idea of   considering "paths which make the largest contribution" to the action integral  (e.g. pinned random trajectories,  concentrated about an  extremal  Euclidean path).  This  can be safely exploited  in the  path-wise  evaluation of the integral kernel  $k_0(y,z,t)$.

At this point we refer to Chap. 3.2 of \cite{feynman} for the  computation  method  (actually adopted for the "path integral formulation of the  density matrix").
We scale away all original  dimensional constants to remain in our framework.

An extremizing path for the action  $S = \int_0^t  [{\frac{1}2} (\dot{x}^2 + x^2)]d\tau $, actually   is a solution of  Eq.(20), appearing in  the form  $\ddot{x}= +x$. Since  ${\cal{E}}= {\frac{1}2} (\dot{x}^2 + x^2)$ is a constant of motion, for  ${\cal{E}}=h^2/2 $, assuming $h>0$,  the solution can be chosen in the form $x(t)= h \sinh t$, for ${\cal{E}}= - h^/2 $, as $x(t)= - h \cosh t$, while for ${\cal{E}}=0$ as $x(t)= x_0 \exp[\pm (t-t_0)]$, \cite{barton}.

The  general solution  of  $\ddot{x}= +x$, in the time interval $(0,t)$:
\be
x(\tau ) = A \cosh \tau + B \sinh  \tau
\ee
must interpolate between the initial value $y= x(0)$ and the terminal value $x= x(t)$.  We do not impose any additional  restrictions on $\dot{x}(\tau ) = A \sinh \tau + B \cosh \tau $.

The initial condition $y =x(0)$ resolves as  $A=y$ and leaves  $\dot{x}(0) =B$ as  yet not specified.

The terminal condition $x= x(t) = A \cosh t + B \sinh t = y \cosh t + B \sinh t$, allows to retrieve $B$, so that:
\be
\dot{x}(0) = B= {\frac{x-y \cosh t}{ \sinh t}}  \Longrightarrow \dot{x}(t) = y \sinh t + (x-y \cosh t) \coth t.
\ee

Since  $x(\tau)$  obeys in $(0,t)$ the equation of motion  $- \ddot{x} + x =0$, we have
\be
\int_0^t {\frac{1}2} (\dot{x}^2 + x^2) d\tau = {\frac{1}2} x \dot{x}|_0^t +
\int_0^t {\frac{x}2} (-\ddot{x} + x)d\tau = {\frac{1}2} x \dot{x}|_0^t.
\ee
Thus, the classical Euclidean  path contribution to $S$ reads:
\be
S= S(y,0,x,t) ={\frac{1}2}[ x(t) \dot{x}(t) - x(0) \dot{x}(0)] =  (x^2 + y^2){\frac{\cosh t}{2\sinh t}}  -  {\frac{x y}{\sinh t}},
\ee
to be compared   with the expression in the exponent in Eq. (16).

For quadratic Lagrangians, we may proceed directly with the missing term,  comprising contributions of  random paths pinned to the prescribed  endpoints of the  $[0,t]$ segment of the Euclidean  classical trajectory. This term is actually a normalization factor, defined by the van Vleck formula \cite{monthus}, so that
\be
k_0(y,0,x,t) ={\frac{1}{ \sqrt{2\pi }}}   \left(- {\frac{\partial^2 S(y,0,x,t)}{\partial x \partial y}}\right)^{-1/2}  \,  e^{ - S(y,0,x,t)}.
\ee
Clearly, with the action integral for $S$ given by Eq. (24), the van Vleck term acquires  the form $(2\pi \sinh t)^{-1/2}$, in agreement with Eq. (16).

\subsection{Tamed Feynman-Kac process:  ${\cal{V}}(x) <0$ in $(-1,1) \subset R$.}

The integral kernel $k(y,0,x,t)$, involving the potential ${\cal{V}}(x) = (1/2)(x^2 -1 )$ instead of the plain harmonic potential  $V(x)= x^2/2$,  is obtained from $k_0(y,0,x,t)$ by means of a   damping  factor: $k(y,0,x,t)=  exp(+t/2) k_0(y,0,x,t)$, see Eq. (15). The outcome of this trivially looking operation,  sets the stage for the {\it  tamed } Feynman-Kac diffusion process.  Its killing/branching   interpretation   has  been elaborated in Ref. \cite{pre24}.

Presently, we shall outline   somewhat  odd (at the first glance)  probabilistic   features of the semigroup dynamics induced by the renormalized, \cite{glimm}, harmonic  oscillator Hamiltonian:
\be
k(y,x,t) = \left[\exp[- {\frac{1}2}(-\Delta + x^2 -1) t]\right](y,x)=
(\pi [1-\exp(-2t)])^{-1/2} \exp \left[ {\frac{1}2} (x^2-y^2) -
   {\frac{(x- e^{- t}y)^2}{(1- e^{-2 t})}}\right].
\ee
Let us adjust the kernel function to the  path-wise  situation, where  all sample paths emanate from $y=0$ at time $t=0$. The function itself
\be
k(x,t)= k(0,0,x,t)= (\pi [1-\exp(-2t)])^{-1/2} \exp (- {\frac{x^2}2} \coth t)
\ee
is not a regular probability density (needs an appropriate normalization in $L(R)$ or $L^2(R)$).
 At asymptotic times $t\rightarrow \infty $, in view of Eq. (16)  we have
\be
k(x,t) \rightarrow K(x)= (\pi )^{-1/2} \exp(-x^2/2)  = \psi_0(0) \psi_0(x)= (\pi )^{-1/4} \psi_0(x).
\ee

 \begin{figure}[h]
\begin{center}
\centering
\includegraphics [width=0.3\columnwidth, valign=c] {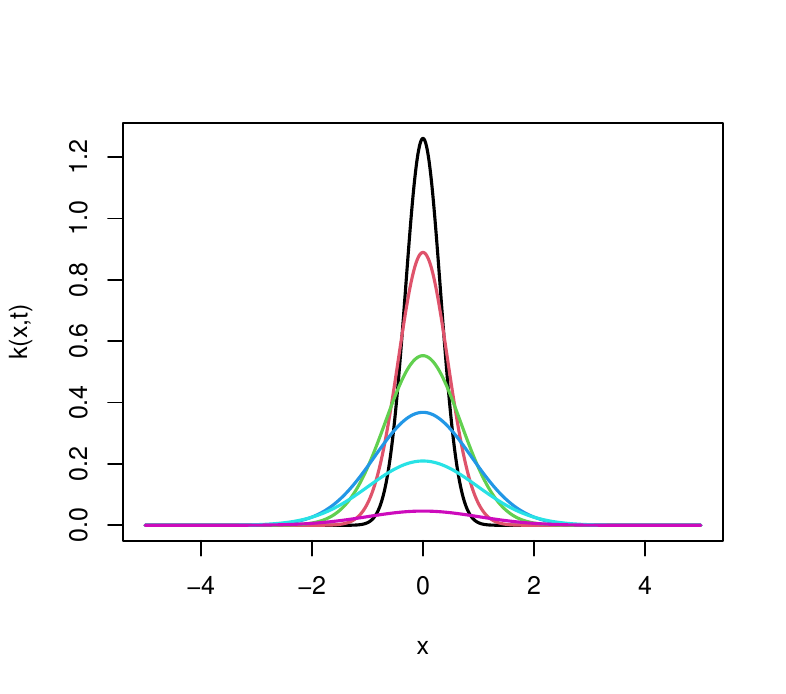}
\includegraphics [width=0.3\columnwidth, valign=c] {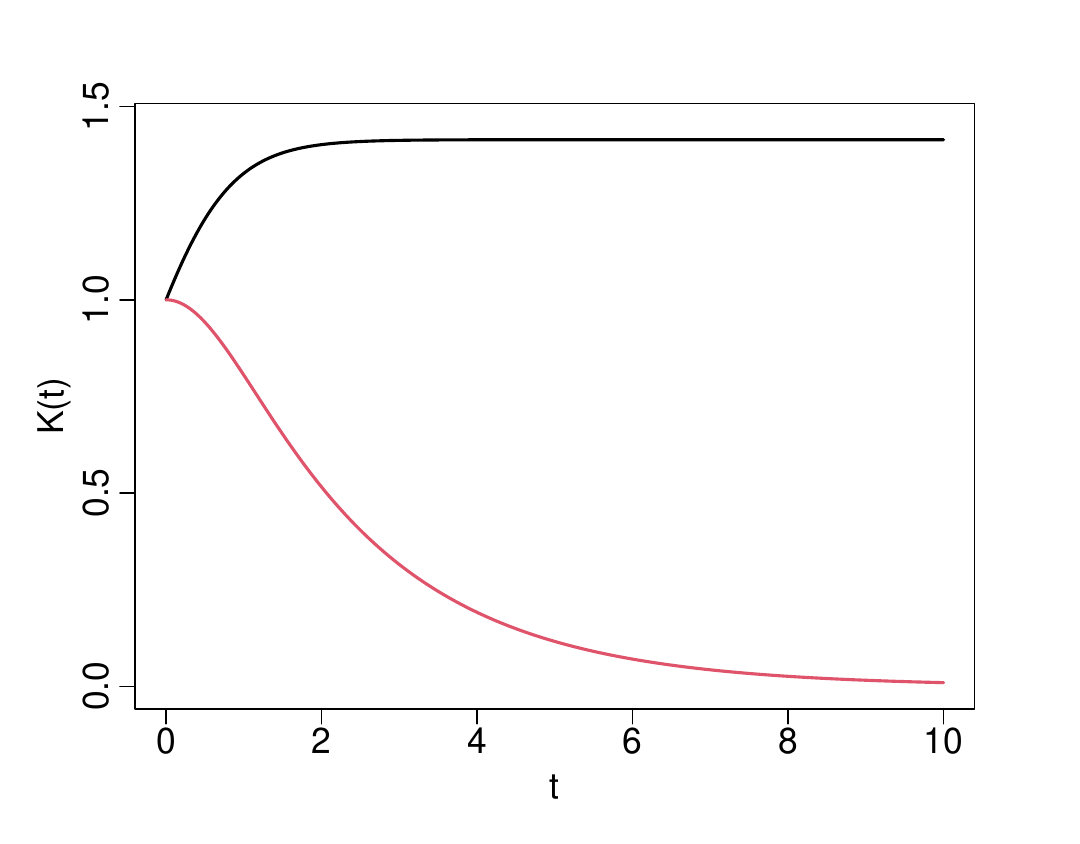}
\includegraphics [width=0.3\columnwidth, valign=c] {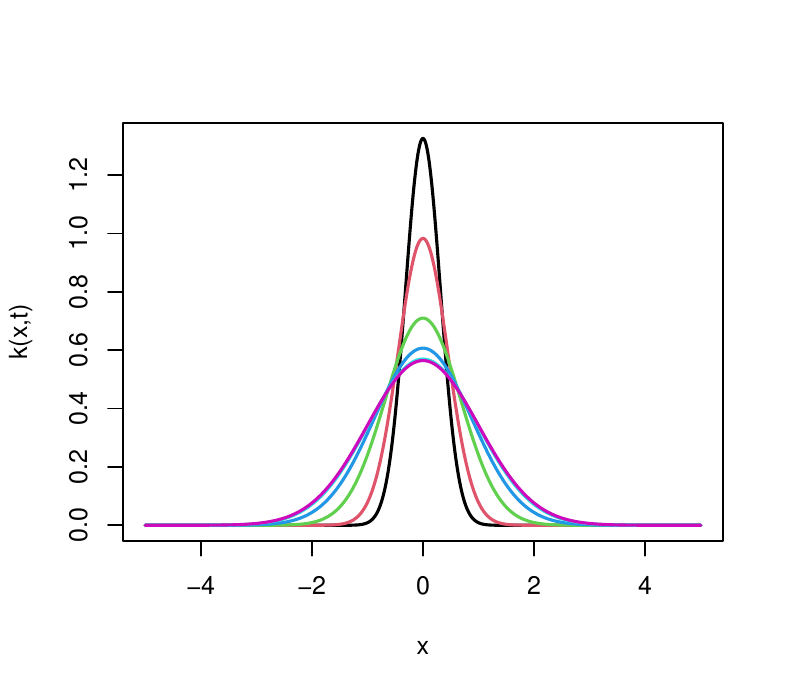}
\caption{Left panel: Exponential decay of $k_0(x,t)= \exp(-t/2)\, k(x,t)$, cf. Eq. (29);   Middle   panel:  A comparative display of $K_0(t)=\int k_0(x,t) dt \rightarrow 0 $ (decay, red) against $K(t)\rightarrow K= \int K(x) dx = \sqrt{2}$ (relaxation, black); Right panel: Relaxation of $k(x,t)$  to $K(x)$, maxima are to be followed from top to the bottom.}
\end{center}
\end{figure}

\subsection{The normalized Feynman-Kac  diffusion: Mean branching and killing  rates at equilibrium.}

Integrating $k(x,t)$ over all locations in $\mathbb{R}$, we get $K(t)= \int k(x,t) dx = (2/[1+ \exp(-2t)])^{1/2}$.  Hence $1\leq K(t) \leq \sqrt{2}$, and asymptotically, $K(t)$ reaches the upper bound $K= \sqrt{2} \sim 1.41421 > 1$.

In principle, we can employ $K(t)$ as the  $L(\mathbb{R})$ normalization  of $k(x,t)$,
and consistently introduce  a probability density
\be
\rho ^{norm}(x,t)= {\frac{k(x,t)}{K(t)}}  \Longrightarrow   \partial _t \rho^{norm} = [{\frac{1}2} \Delta  - ({\cal{V}} + {\cal{K}})]\rho^{norm} ,
\ee
with the positive  potential entry, reducing the branching-induced  surplus  of alive paths:
\be
{\cal{K}}(t) =  \partial _t\ln K(t) = + {\frac{1}{e^{2t} + 1}}.
\ee

The evolution rule for $\rho ^{norm}(x,t)$  derives  from the dynamics of $k(0,0,x,t) = [\exp(-tH)](0,x)$, where $H= (1/2)[- \Delta + (x^2-1)]$.  We have $\partial _t k = - H k$, so that Eq. (29)  follows.\\

  We emphasize, that the above  $L(\mathbb{R})$ normalisation  procedure    assigns a probability distribution  to the  original distribution of  surviving  paths of the tamed F-K diffusion. Its relevance can be tested numerically,by a direct counting of killing and bifurcation events, as depicted in terms of running averages in Figs. 5, 18, 20, 21-25.

The   usefulness of $\rho ^{norm}(x,t)$, can be seen in the demonstration that the mean killing and mean branching rates equal each other, at equilibrium, given in  exemplary  formulas  (35), (36), (52). \\

 {\bf  Technical comment 2:} \\
  Eq. (29) has the standard  form of the {\it generalised diffusion equation} (5),  except for the time-dependent contribution to the effective potential. It is useful to mention that the, imposed per force (plainly against the customary $L^2(\mathbb{R})$ routine),  $L(\mathbb{R})$ normalisation  $ 1 =  {A(t)}^{-1} \int \Psi(x,t) dx$ of  $\Psi (x,t)$ in Eq. (5) , would induce  the evolution equation for $\Psi ^{norm}(x,t)= \Psi (x,t)/A(t)$  of the form (29), with the    time-dependent potential entry   ${\cal{K}} \rightarrow {\cal{A}} = {\cal{A}} (t) =  \partial _t \ln A(t)$.  We hereby observe a  close  affinity with   evolution equations of the population dynamics,  \cite{huillet,jagers}. Moreover, the asymptotic $t\to \infty $  behavior of $\Psi ^{norm}(x,t)= \Psi (x,t)/A(t) \rightarrow  \rho _*^{1/2}$,   corresponds to  the so-called Yaglom limit  law   arising in the theory of killed diffusions, \cite{huillet,yaglom,stein},  which allows to recover (upon a "proper normalisation" !)  the lowest eigenfunction of $H$ as a probability density of surviving sample paths of the process in question. Compare e.g. Eq. (28). \\

Coming back to Eq. (29), we observe that   ${\cal{K}}>0$, while ${\cal{V}} <0$ for  $x\in (-1,1)$ and ${\cal{V}} >0$, if $x \notin [-1,1]$, which  implies  the   killing-branching intertwine, where  positive  and negative  contributions of the effective potential $({\cal{V}} + {\cal{K}})$  counterbalance each other.

Because of the $L(R)$ normalization of $k(x,t)/K(t)=  \rho^{norm} (x,t)$, we may  possibly name  the random motion  compatible with Eq. (30), the {\it  normalized  Feynman-Kac  diffusion}, which we abbreviate through  the superscript {\it norm} in $\rho^{norm}(x)$ .

In view of $K(t) \rightarrow   K= \sqrt{2}$, where $K= \int K(x) dx$ (cf. Eq. (29)), an asymptotic $t\rightarrow \infty $ limit  of $\rho ^{norm}(x,t)$  reads
 \be
 \rho ^{norm}(x,t)  \rightarrow   \rho ^{norm} _*(x)= K(x)/K =(2\pi )^{-1/2} \exp (-x^2/2) = {\cal{G}}(x).
 \ee

 We deal here  with a legitimate $L(\mathbb{R})$  probability distribution,  whose pdf is a normalized Gaussian, with mean $0$ and variance equal $1$.   The ground state function of the harmonic oscillator has the   $L^2(\mathbb{R})$ form $\psi _0(x)= \pi ^{-1/4} \exp (-x^2/2)$, while    $\rho_*^{norm}(x) =(4\pi )^{-1/4} \psi_0(x)$.  Actually $\rho_*^{norm}(x)= {\cal{G}}(x)$  is the $L(\mathbb{R})$-normalized expression for the function $A\exp(-x^2/2)$,  where $1/A= \int_{R} \exp(-x^2/2)dx = \sqrt{2\pi } $. \\

{\bf Remark 3:}  Since  $\rho ^{norm} _*(x)$ is a probability density function, it is instructive to    have  a path-wise motivated insight   into the   meaning of   the probability  $\rho ^{norm} _*(x) \Delta x= [K(x)/K] \Delta x$,  associated with events confined to  a small  spatial interval  $\Delta x$.   Namely, numerical  experiments outlined  in Section III of Ref. \cite{pre24} and the Appendix A, give support to the approximation of $K(t)$ by $N(t)/N(0)$, which is a relative number of alive trajectories at time $t$,  evaluated against the initial number $N(0)$. Denoting an asymptotic limit $ K= N/N(0) = \sqrt{2}$, with $N(0)= 10^5$,   and $N(t) \rightarrow N$, an approximate asymptotic number of alive trajectories  is $ N \sim  \sqrt{2 } \cdot 10^5\sim   141 421$. Let us introduce the notation  $K(x)= N(x)/N(0)$. We readily arrive at:
\be
\rho ^{norm} _*(x) = {\frac{N(x)}N}= {\cal{G}}(x)
\ee
where $N(x) \Delta x$ may be interpreted as a number of   alive  paths,  passing through an   interval $\Delta x$ about the  location $x$.  Accordingly  $\rho ^{norm} _*(x) \Delta x$ stands for a  {\it fraction} of  the asymptotic  population $N$  of alive paths, crossing  $\Delta x$ about $x$, {\it in the equilibration regime}. \\

 For the considered harmonic case, killing and branching (trajectory bifurcation)  options  are mutually exclusive, hence  it appears useful to  evaluate the mean branching rate  (the branching "speed")  $ - \left< {\cal{V}} \right>_{branching}$  at equilibrium,  by  integrating  contributions from  the negativity interval  $(-1,1) \subset \mathbb{R}$ of ${\cal{V}}(x) = (1/2)(x^2 -1)$,  (while taken  with a reversed sign).  The  $(-1,1)$-restricted   mean value $-{\cal{V}}$  of the branching time  rate, reads:
\be
 - \left< {\cal{V}} \right>_{branching} = \int_{-1}^{+1} \left[{\frac{1}2} (1-x^2)\right] {\cal{G}}(x)   dx.
\ee

Since $d{\cal{G}}/dx = -x {\cal{G}}(x) $, integrating by parts and making use of the identity   ${\cal{G}}(1) = {\cal{G}}(-1) $, we arrive at:
\be
  - \left< {\cal{V}} \right>_{branching} = \frac{1}{\sqrt{2\pi e }}  \sim  \sim 0,24197.
\ee
One  may readily  evaluate the mean value   $\left< {\cal{V}} \right>$  on  $\mathbb{R}$, with respect to the  normal  probability density  ${\cal{G}}(x)$, with mean $0$ and variance equal $1$.
The integration  outcome:
\be
\left< {\cal{V}} \right>= \int_{-\infty }^{+\infty } {\frac{1}2} (x^2 -1) {\cal{G}}(x)  dx = 0,
\ee
implies the mean killing  time  rate is  equal  the mean branching  time  rate
\be
\left< {\cal{V}} \right>_{killing} =  \int_{R \setminus [-1,1]} {\frac{1}2} (x^2 -1) {\cal{G}}(x)  dx =   \int_{[-1,1]} {\frac{1}2} (1 -x^2) {\cal{G}}(x)  dx  = - \left< {\cal{V}} \right>_{branching}.
\ee

In view of (33), (35), presuming  that  $K\sim N/N(0)$, \cite{pre24},   we have:
\be
N\cdot \left< {\cal{V}} \right>_{killing} = N\cdot  \int_{R \setminus [-1,1]} {\frac{1}2} (x^2 -1)\cdot N(x) dx  =    \frac{N}{K \sqrt{\pi e }}\sim \frac{N(0)}{\sqrt{\pi e }} .
\ee
  Since  $1/\sqrt{\pi e} \sim 0,34219$,  we get the mean branching/killing rate at equilibrium (per unit of time, which  equals $1$),  for the overall number $N$ of alive paths: $  N\cdot \left< {\cal{V}} \right>_{killing}  \sim   34 219$,  and  the same outcome for $- N\cdot \left< {\cal{V}} \right>_{branching}$.\\

{\bf Remark 4:} In Eqs. (33-37) we deal with mean values of time  rates, encompassing  branching and killing events   at equilibrium. To introduce probabilities of  such events,  in the short time interval $\delta t$, we should  consider  $p(\delta t)=  \delta t \cdot \left< {\cal{V}}\right>_{killing/branching}$.  Choosing  $\delta t = 10^{-4}$,  in view of  (34)  and (36),  we  get $p(\delta t) = \delta t \left< {\cal{V}} \right>_{killing}  =  \delta t  |\left< {\cal{V}} \right>_{branching}| \sim  0.24197 \cdot 10^{-4}$, i.e. the mean probability of branching (and killing) events in the time step $\delta t$, for each single sample trajectory.
 Since  the numerically  recorded number  of alive paths in the asymptotic regime is close to the theoretical prediction  $N  \sim 141421$, we can  consider $N \cdot p(\delta t)  \sim 3.422 $ as the expected  number   of trajectories  that are killed (and in parallel cloned via branching) at  the  dynamically maintained   equilibrium, in the time step $\delta t = 10^{-4}$ (this amounts to twice the number of $\sim 34 219$ counterbalancing events per $1$ second).  We point out that the  mean number  level  $3.422$ is depicted as a dashed  line in Fig. 5.  \\

\subsection{Killing vs branching  in the tamed F-K process: Briefly on computer-assisted
trajectory generation and  counting.}

Taking seriously  the interpretation of positive potential contributions  in the Feynman-Kac formula as killing rates of the diffusion process \cite{klauder,gar,pre24,ito,helms,helms1,nagasawa,zoia,mazzolo1}, we have faced  an issue of   the compensation of  killing. This, we   associate  with the branching (cloning) rates of sample paths of the random process, \cite{pre24} (see e.g.,    \cite{huillet,kesten,ber,nagasawa,jagers,zoia}),  which are  controlled by   values of the {\it sign inverted}  F-K  potential  ${\cal{V}}(x)$ in  its   negativity subdomain $(-1,1)\subset \mathbb{R}$. Killing and branching areas are mutually exclusive.
The    killing-branching intertwine  is   encoded   in the evolution equation (31),  with a  well defined asymptotic pdf $\rho _*^{norm}(x)$.  This dynamics refers to the  {\it  normalised} Feynman-Kac diffusion with killing and branching.

It is clear, that  for  positive   Feynman-Kac potentials,  the   untamed  killing  would  imply  a decay of the F-K kernel.  If potentials bounded from below  have negativity subdomains, branching (cloning)   may  not only compensate, but    overcompensate killing.   An additional killing term ${\cal{K}}$ in Eq. (30)   restores the balance  between killing and  branching effects in the  Feynman-Kac diffusion process in question, so that there is no   surplus (creation) or deficit (loss) of the  asymptotic  "probability mass", see e.g.,  \cite{ito,helms,helms1,kesten,ber,nagasawa}). In contrast to $k(x,t)$,   $\rho^{norm}(x,t)$ is a well defined probability density function on $\mathbb{R}$.\\

\begin{figure}[h]
\begin{center}
\centering
\includegraphics [width=0.45\columnwidth, valign=c] {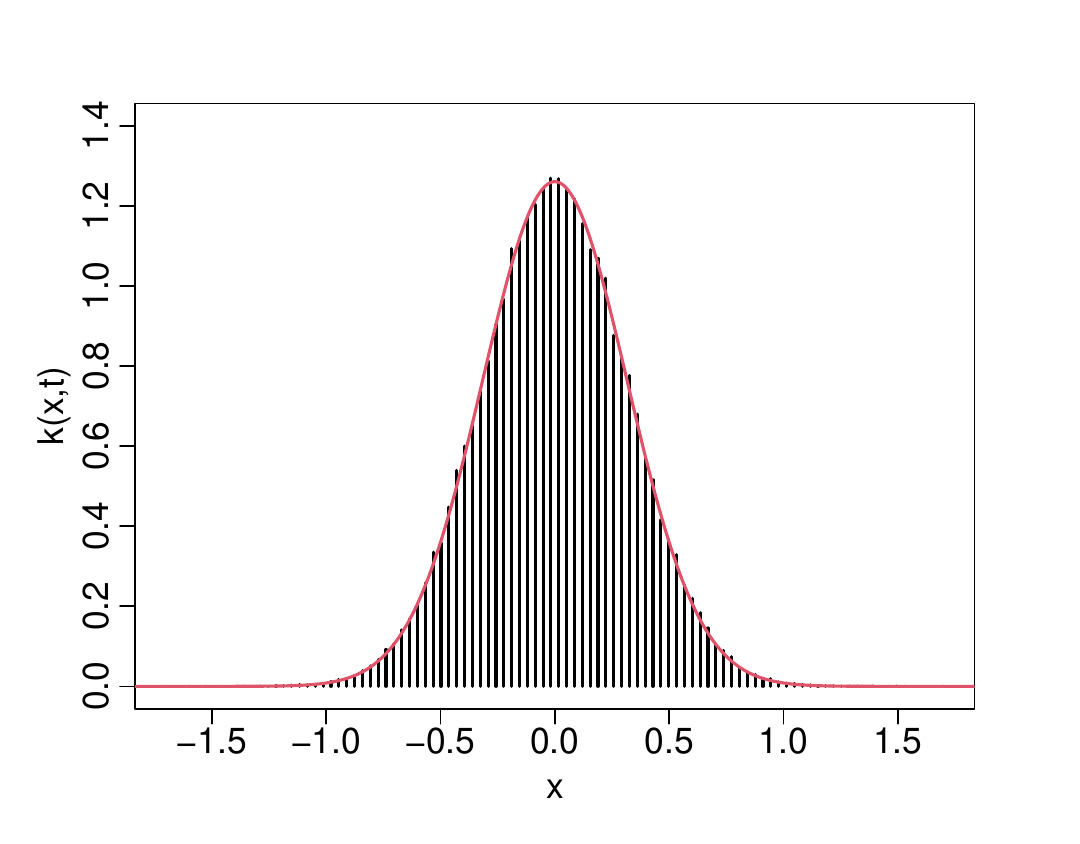}
\includegraphics [width=0.45\columnwidth, valign=c] {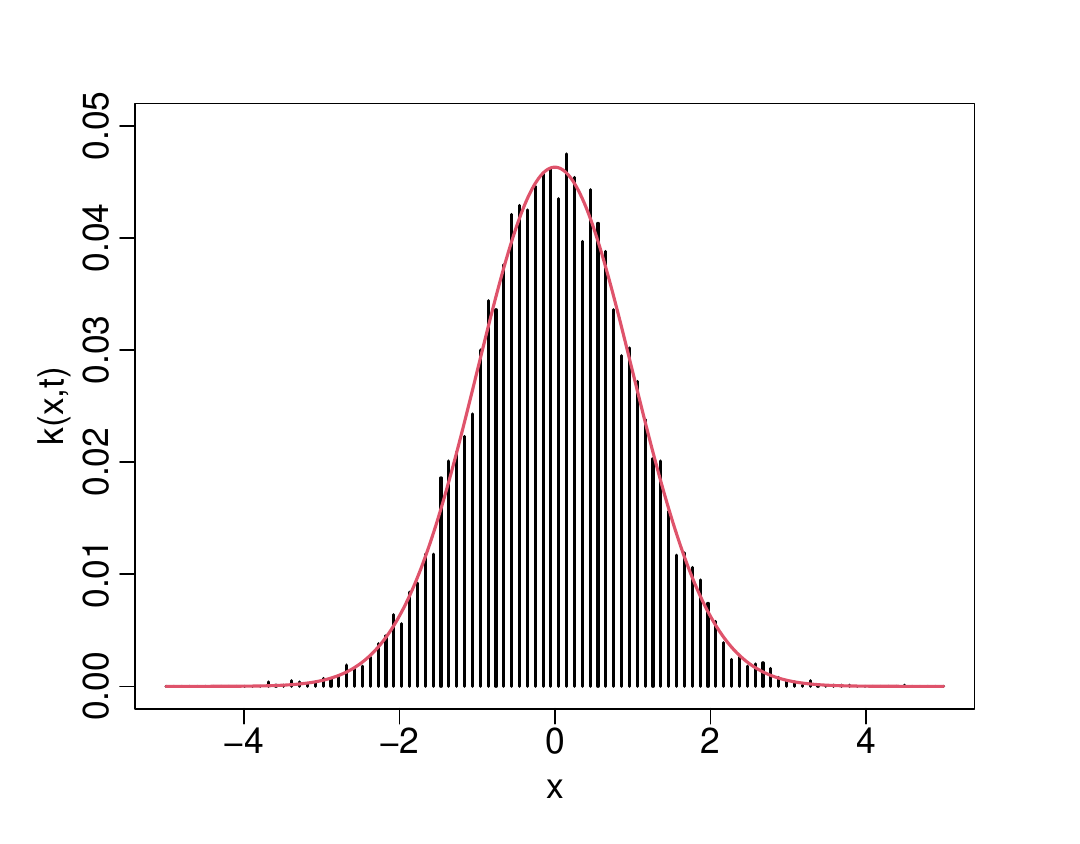}
\caption{Effects   of the killing rate  ${\cal{V}}(x)= x^2/2$ visualised for  the relative number  $N(t)/N(0)$  of    trajectories still alive at time $t$, against that of initially released.  Sample paths emanate from the point $x=0$ at time $t=0$, and  the number  of  consecutively released paths is $N(0)=10^5$.  Panels depict histograms, where   the height of each column depicts a relative number $h(x,t, \Delta x)= N(x,t,\Delta x)/[\Delta x \cdot N(0)]$ of counts  recorded  in the spatial widows $\Delta x$ about the column spatial coordinate  $x$, at a specified time instant (cf. Section III of Ref. \cite{pre24}). Here  $N(t)/N(0) \sim \sum [h(x,t,\Delta x)\cdot \Delta x $ along  the reference  spatial  interval $[x_{min},x_{max}]$.  Left panel: The reference interval is $[-1.5,1.5]$, $  \Delta x=0,03$, $t=0.1$; the overall  number of counted (alive) trajectories is  $N(0.1)= 99766$;\,  [The  accumulated data for  intermediate time  instants are:  $N(0.2)=98981$, $N(0.5)= 94085$, $N(1)=80462$, $N(2)=51472$]. \,  Right panel: The reference interval is $[-3,3]$,  $\Delta x= 0.1$, $t=5$, $N(5)= 11565$.   We realize that  $ K_0(t)= e^{-t/2} K(t)=  (\cosh t)^{-1/2} \approx N(t)/N(0)  \rightarrow 0$, \cite{pre24};   The  envelope (continuous curve) has an  exact  analytic form $k_0(x,t)= \exp(-t/2)\, k(x,t)$.  We indicate a  significant change of   scales along the coordinate axis  while passing from  Panel 1 to Panel 2. }
\end{center}
\end{figure}

\begin{figure}
\begin{center}
\centering
\includegraphics [width=0.45\columnwidth, valign=c]  {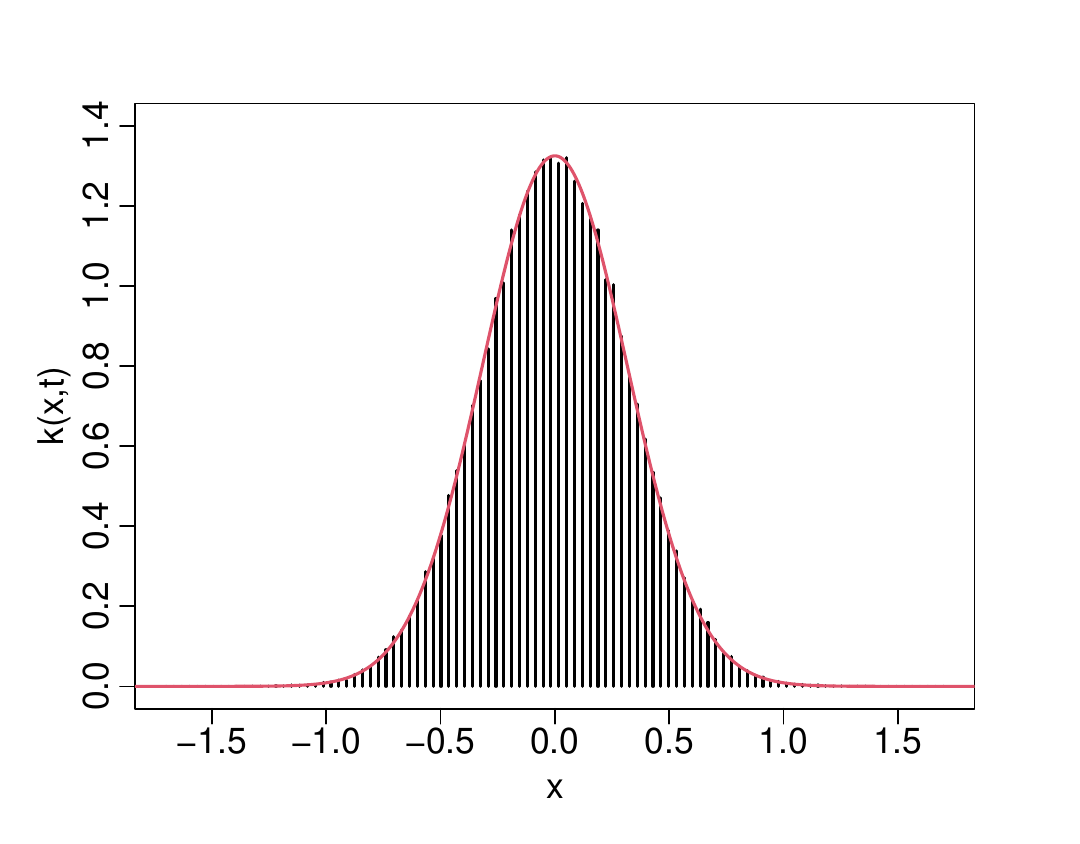}
\includegraphics [width=0.45\columnwidth, valign=c]  {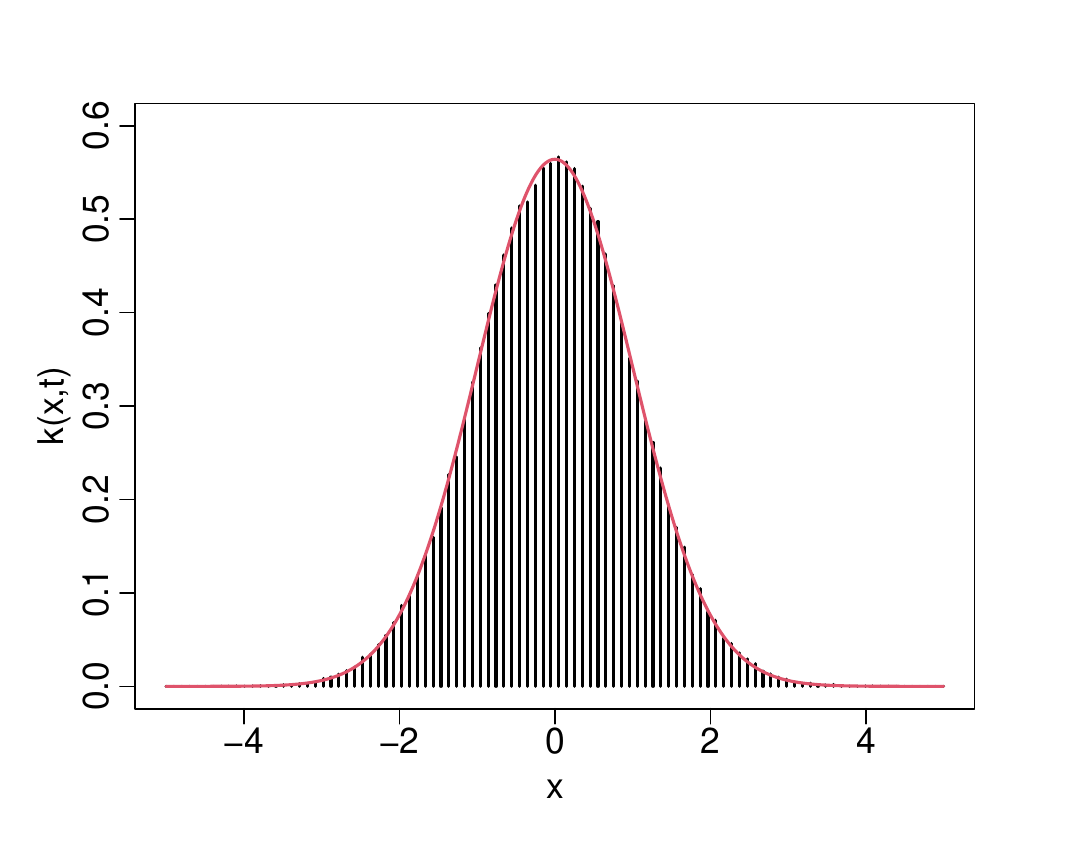}
\caption{Effects  (histograms) of the   killing/branching  rate  ${\cal{V}}(x)= (x^2 -1)/2$ in terms of  the  relative  number of   trajectories surviving till time $t$  (in the histogram, we depict the recorded relative  numbers $N(x,t,\Delta x)/N(0)$ of counts  in each consecutive  $\Delta x $ interval along the $x$-axis, in terms of the coulumn  height $h(x,t, \Delta x)= N(x,t,\Delta x)/[\Delta x \cdot N(0)]$  ). Left panel: the reference interval is  $[-1.5,1.5]$, $\Delta x=0,03$ , $t=0.1$. Right panel: The reference interval is $-3,3]$, $\Delta x = 0,1$,  $t=5$.  The  envelope (continuous curve) has an  exact  analytic form $k(x,t)$,  as  given by Eq. (28). Here  $K(t) \sim N(t)/N(0) \rightarrow  K= \sqrt{2}  \sim 1.41421 $, revealing an evident surplus  in the number of alive trajectories up to $\sim 141 421$,  if compared with   the initial number  $N(0) = 10^5$.   $K>1$  is a symptom of the overcompensation of killing by branching  ("probability mass" surplus).}
\end{center}
\end{figure}

The detailed description of the trajectory generation,  incorporating random  killing or branching (cloning)   events,  and  the counting  procedure of alive sample  paths at various time instants, can be found in Section III of Ref. \cite{pre24}. \\
 We emphasize that  killing and branching options are mutually exclusive in our procedure, the restriction seldom met in the standard  population dynamics literature, \cite{jagers,kesten,ber,zoia,mazzolo1}. There one may meet trajectories in which killing and branching occur at the same space-time point. As well,
one may consider the option of a multiple offspring (not
merely a bifurcation of a trajectory into two) at each branching
instant.\\

The numerically executed   path-wise counting outcomes give support to the killing/branching scenario advocated in Ref. \cite{pre24}. This happens  quite apart from the fact, that for confining potentials, the probability of a killing event in a small time interval $[t,t+\delta t] $ (tacitly presumed to be "almost infinitesimal", while typically  $\delta t \leq 10^{-4}$),   reads
\be
 p(t)= \min[1, {\cal{V}}(X(t)) \delta t ],
 \ee
  and cannot  selectively  account for large values of ${\cal{V}}(x)$.  This slightly  reduces  the sensitivity  of simulations  with respect to  the functional shape of ${\cal{V}}(x)$,   beyond  a finite domain in $\mathbb{R}$, whose size depends on the fine tuning  of $\delta t$, unless set  well below $10^{-3}$.

 In reference to negativity domains of the F-K potential, it is
   \be
   q(t)=\min[1,- {\cal{V}}(X(t)) \delta t ],
   \ee
   which  stands for a probability of the branching (cloning) event, in which the incoming sample  path bifurcates into two independent  outgoing  paths. Each copy moves according to the  primordial  Brownian (Wiener) rule, which is as well the motion rule between branching events along any path, unless being killed (i.e.  ultimately  removed from  the paths counting  statistics at later times).

\begin{figure}[h]
\begin{center}
\centering
\includegraphics [width=0.4 \columnwidth, valign=c] {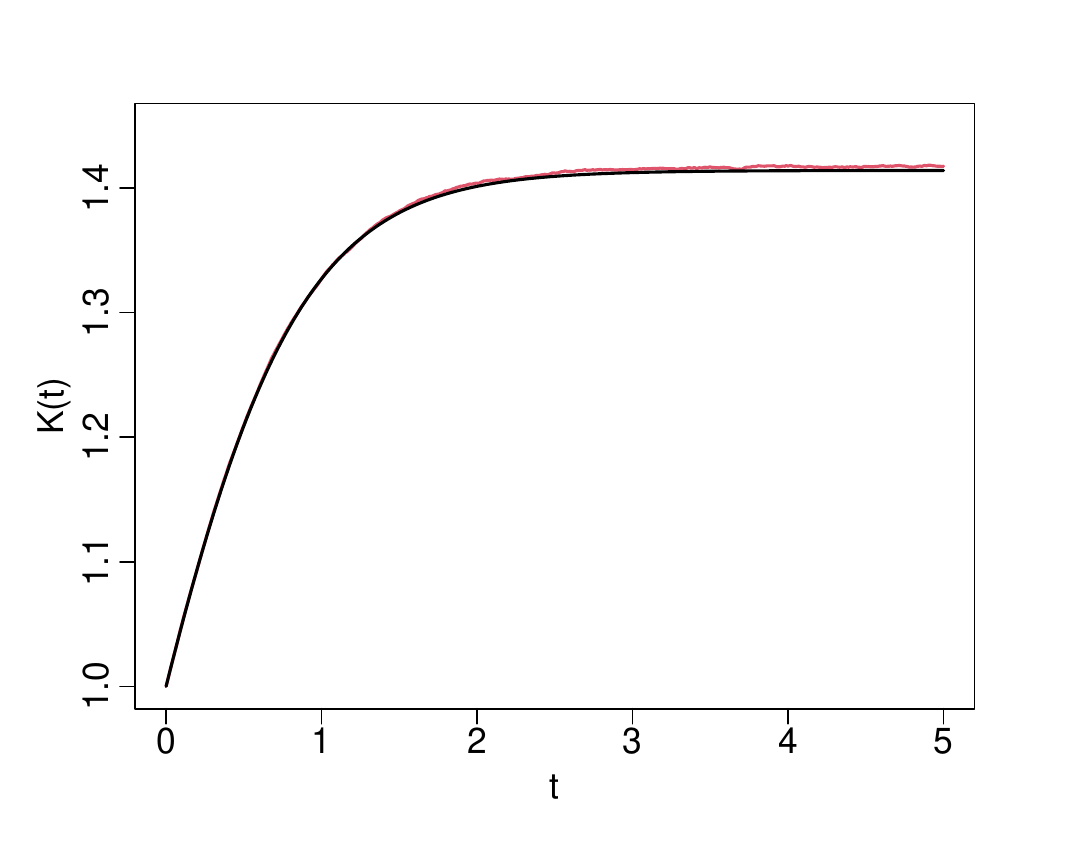}
\includegraphics [width=0.4 \columnwidth, valign=c] {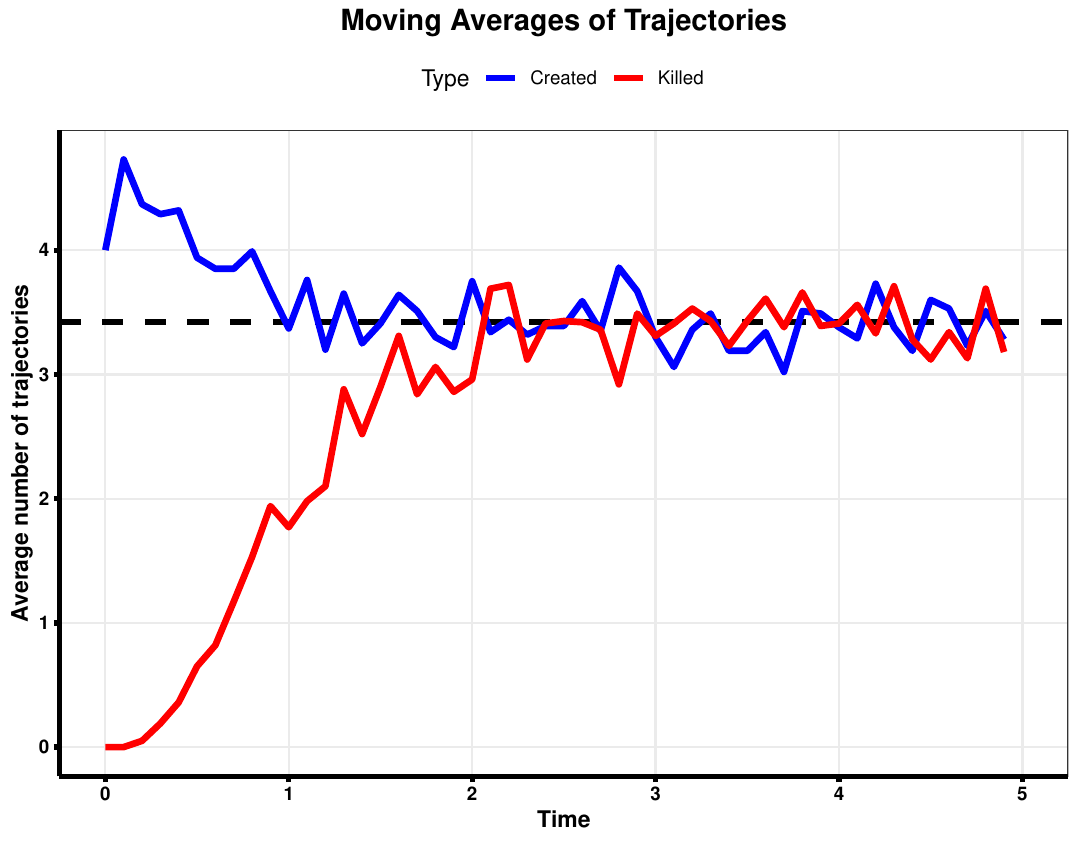}
\caption{The  path-wise approach to equilibrium. Left panel: A comparison of the analytic formula for  $K(t)=  e^{(+t/2)} (\cosh t )^{-1/2}$  (black curve), while set against the numerically retrieved  (trajectories counting) curve  $N(t)/N(0) \sim K(t) \rightarrow \sqrt{2}$   (red).  Right panel: Evolution in terms of   running averages (technical explanation is given below Fig. 6)  for  individually  counted  killing  (red)  and   branching  (blue) events, for $N(0)=10^5$ initially released trajectories. We note, that with the choice of $\delta t= 10^{-4}$, and the number of asymptotically alive trajectories stabilizing about  $N \sim 141 421$,  we recover  $N \delta t [- \left<{\cal{V}}\right>_{branching}] =  N \delta t /\sqrt{2 \pi e}\approx 3.422$. The dashed line  in the  figure    indicates that   $ 34 220$   actually is   the mean number of branching events (appearance of clones via   bifurcations of sample paths) in the time interval $\Delta t =1$.  At equilibrium, it is counterbalanced by roughly the same number of killing events, cf.  Eq. (37). }
\end{center}
\end{figure}

 To implement a computer-assisted  trajectory interpretation  of the killed diffusion process  with branching,  we  need to pass from the lore of  continuous nowhere differentiable  trajectories,  to their  space and time   discretized  approximations.  The simulation  procedure should   enable the trajectories counting, at any propagation time.

 Let $t\in[0,T]$, we set   $\delta t=T/n$ for a predefined value of $n\in\mathbb{N}$.  The notation $\delta t$ is  informal, but presupposes that  a  finite  time interval   $\delta t$ of interest  can be made arbitrarily small (with $n\gg 1$ we  bypass the usage of $dt$).
  The Brownian random  walk is defined according to $x(t+\delta t)= x(t)+\sqrt{\delta t} B$, where  $B$   is the random variable sampled from the normal distribution  $N(0,1)$, $x(0)=0$.

If the simulated random  trajectory  takes   the value  $x(t)=x$  for some  $t\in [0,T]$, its subsequent  behavior admits three  instances: killing, cloning, and  undisturbed  moving on, whose realization    in  each   simulation step  $[t, t+ \delta t)$ depends on  the  concrete value of the  potential  ${\cal{V}}((x(t))={\cal{V}}(x)$, where the sign of ${\cal{V}}(x)$ is of particular importance.

 We consider the following random  options for each sample path:\\

 \noindent
(I) {\it Killing}, ${\cal{V}}(x)> 0$.\\

\noindent
  (1) the trajectory is killed  at $x(t)=x$ with the probability $p(t)=\min(1, \delta t\cdot {\cal{V}}(x(t))$,  and  thence removed from the  trajectories population  at  time $t+\delta t$; \\
  (2) if the trajectory is not  killed, then it moves-on, by  following the evolution rule  $x(t+\delta t)=x(t)+\sqrt{\delta t} B $, (the trajectory  survival  probability at time $t$ is given by $(1-p(t))$).\\

\noindent
(II) {\it Branching} (trajectory bifurcation), ${\cal{V}}(x) <0$.\\

\noindent
(1) the  cloning (branching) event - the trajectory clones itself (produces an offspring)  at  $x(t)=x $  with the probability  $q(t)=\min(1, -\delta t\cdot {\cal{V}}(x(t)))$, subsequently  both the clone and  the parent  trajectory independently   move-on  from the branching point,   in accordance with the   rule  $x(t+\delta t)=x(t)+\sqrt{\delta t} B$, up to time $t+\delta t$. At $t+\delta t$  we thus need to handle two independent trajectories instead of  one;  \\
(2)  no offspring -   the trajectory follows the evolution (I.2).\\

We are interested in the statistics of  all "alive"  trajectories at  each (coarse-grained) time instant of time $t \in [0,T]$.\\

The   path-wise  analysis of the tamed   Feynman-Kac diffusion $k(x,t)$,   controlled  by the harmonic potential  (with and without subtraction)   of Section II, has revealed a striking fact, \cite{pre24}.  The curves $k(x,t)$, (28), together with $K(t)= \int k(x,t) dx$, may be  quite accurately mimicked  by  outcomes of  computer-assisted trajectory counting experiments. These refer to sample paths, which are  released from $y=0$ at $t=0$   and subsequently  subject to     (mutually exclusive)  branching and  killing  events in their time evolution, with  a distinctive number  of   remnants  (dominated by the  branching offspring),   at time $t$. \\

We indicate  that   $K(t) \sim N(t)/N(0) $,   stands for  a relative number of  alive (at $t$) random paths. Provided $N(0)$ is the overall number of trajectories started at $y=0$, while $N(t)$ is a total number of still alive trajectories at time $t$. In particular, with $N(0)=10^5$,  the limiting value of $K= \sqrt{2}  \sim 1.41421 $ has been set in correspondence with the  stabilized   approximate number  $\sim 141 421$ of counted, at sufficiently large time  $t$,  (still alive) trajectories.\\

{\bf Remark 5:} To construct  the histograms in Figs. 3 and 4,  we need  the relative measure of the trajectories  number increase/decrease. To this end,  we evaluate $h(\Delta x) = N(x,\Delta x )/ (10^5 \cdot   \Delta x )$, which  is a quantitative measure of a  fraction  of trajectories counted  in  a   small spatial  segment (say   $\Delta x \sim 10^{-2}$) about $x$,   while set  against their  overall initial  number  $10^5$, per the length of $\Delta x$.  The  number $h(x,\Delta x)$ corresponds to the height  of   the respective   vertical bar about its   $x$-location, actually depicted in the pertinent histograms. \\

{\bf Remark 6:} The moving averages display in Fig. 5 has been constructed as follows. We start from $N(0) = 10^5$ sample paths. The simulation time covers an interval  $[0,5]$. Choosing time steps of the size $\delta t =10^{-4}$, we arrive at $50 000$ control points in the simulation. The counting outcomes are highly irregular and thus beyond the optical resolution of the figure.  To heal this defect, we have slightly "regularised" the counting data. We could   count all  killing and branching events within  each single  $\delta t$ step.  To make the data  of  $50 000$ consecutive time steps visually   digestible, we introduce a "running  window", incorporating $100$  consecutive  time steps $\delta t$.  In  each window we  evaluate a mathematical average (divide by $100$) of the accumulated number of  branching  events, next calculating  the same average for killing events. After time $100 \delta t$, the running  window is shifted by one  $\delta t$ step, so that the next average pertains to time steps  $2$ to $101$, subsequently $3$ to $102$, $4$ to $103$  and so on. The number of the  running  window-averaged outcomes is still very large. In the  right panel of Fig. 6 we impose a "sieve" and   depict  running windows averages, with a time span $10^3 \delta t = 10^{-1}$ between each consecutively displayed point. Hence, instead of $50 000$ data points, we actually display only $50$. The equilibration  tendency of (averaged)  branching and killing events is clearly seen beginning from $t=2$. \\

{\bf Technical comment 3:}\\
In the  trajectory evolution  simulations the typical time step was predominantly $\delta t = 0,001$,  unless indicated otherwise. The  standard adopted by us  time interval $ t \in [0,5]$ has thus involved $5 000$ steps of the algorithm. (In simulations of eigenfunctions of
the Schr\"{o}dinger operator, by means of the Strang splitting method, time steps were much smaller, typically of the size $\delta t = 10^{-7}$.)

Sample path simulations started from $10^5$ repetitions of the algorithmic procedure, except for $N(0)= 10^6$, as denoted in Fig.25.
In trajectory simulations there is no finite  boundary truncation mechanism,  the  motion is a priori unbounded. Nonetheless, for potentials quickly escaping to infinity, a survival chance  for trajectories reaching locations  $x > |5|$ in their evolution, rapidly decays to zero. For all   practical purposes (fapp), at asymptotic times (time about $t=5$) the spatial interval $x\in [-5,+5]$   encompassed  almost all (fapp) surviving trajectories.

 An issue of the sampling noise has been analysed. In accordance with the law of large numbers this  impact  goes down like $1/\sqrt{N}$ , where $N$ is the number of paths. We have tested numerically  quite a number of trajectory populations, starting from $N(0) = 100 000$, for each specific choice of the F-K potential. The recorded minor deviations
 in the histograms shapes, are statistical-sample-dependent,  and have no relevance for our ultimate conclusions. Presented  in  Figs. 3, 9, 12, 18, 20, 21-25    numbers of surviving paths  may vary from one  statistical sample to another, but up  to the unavoidable statistical uncertainty (fuzziness) both numbers and full histograms   well fit  to analytically obtained enveloping curves   in the asymptotic regime.

\section{Nonlinear problems: Superharmonic potentials.}

By the  arguments of Sections I.A and I.B, given  $\phi (x)$, we have  in hands   the drift  field  $b(x)= - \nabla \phi (x)$  of   the F-P equation (2), and  an  associated Feynman-Kac potential ${\cal{V}}(x)$. The   eigenfunction   $\psi_0(x)$ of $H= -(1/2) \Delta + {\cal{V}}$ corresponding to the  bottom  eigenvalue zero, ($L^2(\mathbb{R})$  normalisation is tacitly presumed),  has the   functional form  $\psi_0(x) \sim \exp (-\phi )$. The  related asymptotic Fokker-Planck  pdf  reads $\rho _*(x)  = [\psi_0(x)]^2 $.\\

This line of reasoning amounts to inferring the Feynman-Kac potential, intimately   associated with the a priori given  equilibrating  Markovian diffusion process. We point out that the {\it  reverse} route has been analyzed  in Ref. \cite{klauder} (see also \cite{zaba}), with the presumed a priori choice of a candidate  Feynman-Kac  potential, which could have nothing in common with the previous ${\cal{V}}(x)$ notion.

 The induced   diffusion process needs to be   subsequently constructed, under fairly restrictive conditions on allowed drift fields. This was motivated by  the standard theory of stochastic differential equations,  and an issue of the existence of the probability density function $p(y,s,x,t)$, where  one attempts to grant the uniqueness and non-explosiveness of the diffusion process, c.f. \cite{klauder,olk}). Our approach is less restrictive, since  we establish links between F-K potentials and drift fields of the Fokker-Planck equation,  while staying   on the   dynamical  semigroup level.

\subsection{Handling nonlinear drifts:  $\phi (x) \sim x^{m}$,  $m=2n  \geq 2$.}

Let us consider a family of  potentials   of the form  $\phi (x) = {\frac{\alpha }2} x^m,\quad m=2n,\quad n\in\mathbb{N},\alpha >0$  for  the  Fokker-Planck  drift field $b(x) = - (m \alpha /2) \, x^{m-1}$.

 We know from the start the  asymptotic pdf of Eq. (2),   $\rho_{*}(x)\sim\exp\left(-\phi (x)/\nu \right)$, while  rememebering about our choice of $\nu =1/2$.
The $L(\mathbb{R})$  normalisation of $\rho_{*} $  follows via the  evaluation of (twice) the integral over $\mathbb{R}^+$:
  $2 \int_0^{\infty }  \exp(-\alpha x^m)  dx$.  This can accomplished by changing the integration variable  from $x$ to $y=\alpha x^m$, and exploiting the integral definition of the Euler Gamma function:
\be
\int_R \exp(-\alpha x^m) dx = {\frac{2 \alpha ^{-1/m}} m} \int_0^{\infty } e^{-y}  y^{{\frac{1}m} -1} dy = 2 \alpha ^{-1/m}  \Gamma(1+ {\frac{1}m}),
\ee
where $z\Gamma(z)= \Gamma (1+z)$ with  $z=1/m$.

For future convenience, we point out that  for even $k$ and $\alpha >0$, we can  readily evaluate the  improper  integral
\be
\int_0^{\infty }x^k e^{-\alpha x^m} dx = {\frac{1}{m \alpha^{(k+1)/m}}} \Gamma \left( {\frac{k+1}m} \right).
\ee

With the integral (40) in hands, we immediately recover the $\alpha$-family of $L(\mathbb{R})$ normalized stationary pdfs corresponding to  the drift potential  $\phi (\alpha ,x) ={\frac{\alpha }2}  x^m$:
\be
\rho_{*}(\alpha ,x)= {\frac{\alpha^{1/m}}{2\Gamma \left(1+{\frac{1}m}\right)}}
\exp\left(-\alpha x^m\right),
\ee
and accordingly,   the ground state function  $\psi_0(\alpha ,x)= \rho_{*}^{1/2}(\alpha ,x)$ of the emergent  Schr\"{o}dinger-type operator  $H=-\frac{1}{2}\Delta+\mathcal{V}(\alpha )$.
We have:
\be
\rho_{*}^{1/2}(\alpha ,x) ={\frac{\alpha^{1/2m}}{\sqrt{2\Gamma \left(1+{\frac{1}m}\right)}}}
\exp\left(-{\frac{1}2}\alpha  x^m\right),
\ee
where $\phi (\alpha,x)= {\frac{1}2}\alpha  x^m$  actually  defines the Fokker-Planck  drift $b(x) = - \nabla \phi (\alpha ,x)$. This  observation  allows to deduce
the $\alpha $-family of related  Feynman-Kac potentials, cf. (3) and (6):
\be
{\cal{V}}(\alpha ,x) = {\frac{1}2}\left( [\nabla \phi (\alpha,x)]^2 - \Delta  \phi (\alpha ,x)\right) =  {\frac{m\alpha }4} x^{m-2}\left({\frac{m\alpha }2} x^m + 1 -m \right).
\ee
This secures the vital spectral property $H\rho _*^{1/2}(x) = 0$.  The spectrum of $H$ is  discrete and non-negative,  begining from the bottom eigenvalue $0$. An (approximate) access to a couple of lowest  non-zero eigenvalues is possible by means of  numerical procedure. See e.g. Ref. \cite{stef}, and Ref. \cite{jmp14} where the tenets of the numerically  assisted procedure for the solution of the Schr\"{o}dinger eigenvalue problem, by means of the Strang splitting method, has been outlined.\\

In below, we shall mostly refer to the following specific choices of  $\alpha =2, 2/m,  2m$, which  identify superharmonic potentials  $\phi(x)= x^m, x^m/m, mx^m$ respectively, cf. \cite{zaba,stef}.\\

In passing, we mention  that the pertinent   superharmonic potentials with  $m=2n >2$,  induce   a family of   bistable-looking  functions of the form  ${\cal{V}}(x) = a(m) x^{2m -2} - b(m) x^{m-2};  a, b>0, m>2$.  The bottom  eigenfunctions of $H$, irrespective of any specific  choice  of $\phi (x) $, remain {\it  unimodal}  and have the functional form $\psi_0(x)   \sim \exp[ - \phi(x)]$, c.f. \cite{zaba,stef}.

At this point, we emphasize that the derived ${\cal{V}}_m(\alpha ,x)$,  Eq. (45)  is exactly the {\it tamed}  (not merely killing)  Feynman-Kac potential.

It takes the value $0$ at  points   $x=0$  and $\pm x_m$, where
\be
|x_m| = \left({\frac{2}\alpha }\right)^{1/m} \left({\frac{m-1}{m}}\right)^{1/m}.
\ee

 We  observe  the  negativity property   ${\cal{V}}(\alpha ,x) <0$  for $x\in  D_{neg} = (-x_m,0)\cup (0,x_m)$. The potential is positive   beyond  the interval $[-x_m, x_m]$  in $\mathbb{R}$.

 As discussed in Ref. \cite{pre24}, in the  negativity domain $D_{neg}$, we may implement the branching mechanism for  (Wiener/Brownian) random paths, with the branching rate $|{\cal{V}}(x)|$.  The probability of a branching event in a  small time interval $\delta t$,  is given by  $q(t)=\min[1,- {\cal{V}}(\alpha ,x(t)) \delta t ]$.  In view of somewhat "wild" properties of the F-K potential  (44)  for large values of $m$, the fine tuning of $\delta t$ is here   necessary to ensure  the  asymptotic equilibration  in numerical procedures.

 In    $\mathbb{R} \setminus  D_{neg}$ there is no branching at all  and we encounter rapidly  increasing  killing rates for $x>|x_m|$.

For all listed above  choices of $\alpha $,  we follow a  proviso that the spectrum of $H$ is  non-degenerate. Thence, we can  {\it  formally} consider a spectral representation of the Feynman-Kac kernel, in the form analogous to (15) (we take into account the fact that the lowest eigenvalue of  $H= -(1/2) \Delta + {\cal{V}}$ vanishes, i.e.   $\epsilon_0= 0$):
\be
k(y,0,x,t) =  [\exp(- t H)](y,x) = \psi_0(y) \psi_0(x) +   \sum_{j=1}^{\infty } \exp(- \epsilon _j t) \, \psi _j(y) \psi _j(x).
\ee

The expected asymptotic behaviour of the tamed  ($\epsilon_0=0$)   Feynman-Kac kernel  $k(0,0,x,t)= k(x,t)$  for $t\rightarrow  \infty $, can be deduced  without any knowledge of its detailed functional form  for finite time instants.  The presumed limit reads  $k(x,t) \rightarrow K(x)= \psi _0 (0)\cdot \psi _0 (x)$,  stemming from the $L^2(R)$ normalised entry $\psi_0(x) \sim \exp (- {\frac{\alpha }2} x^m)$.

\subsubsection{ $\phi(x)= x^m$.}

 \begin{figure}[h]
\begin{center}
\centering
\includegraphics [width=0.4\columnwidth, valign=c] {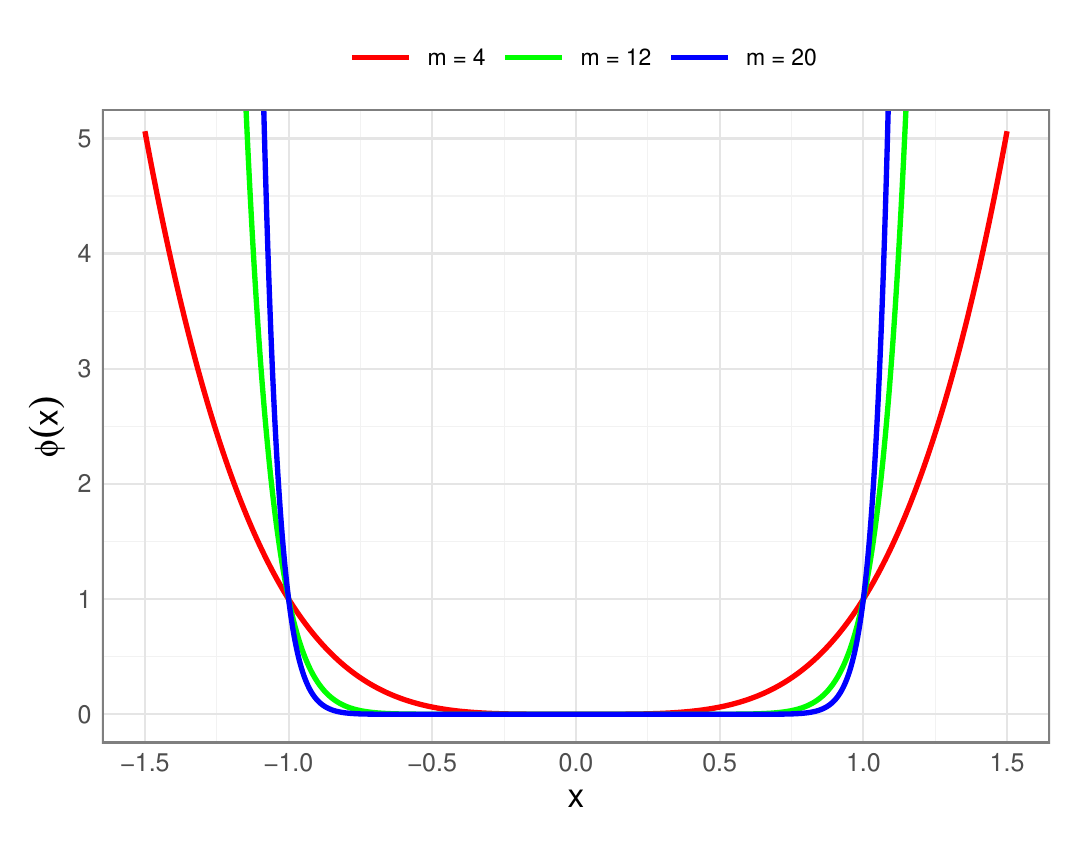}
\includegraphics [width=0.4\columnwidth, valign=c] {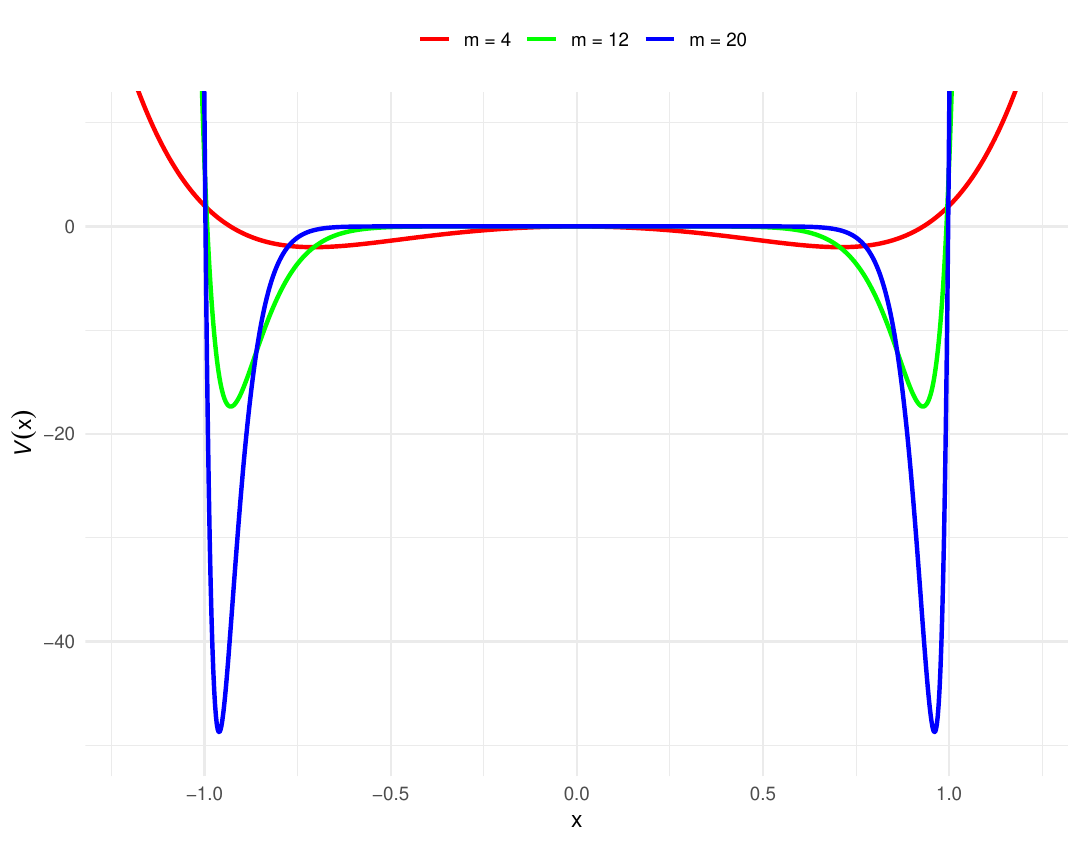}
\caption{Left panel: Superharmonic potential $\alpha =2 \rightarrow  \phi (x)= x^m$,  $m=2n$, for m=4, 12, 20;  Right panel:  Since $b(x)= - mx^{m-1}$, the  two well  Feynman-Kac potential ${\cal{V}}(x)$  derives directly from Eq. (3). To cope with very deep  (and excessively narrowing)  wells (c.f. \cite{zaba,stef}), we visualize the functional form of  ${\cal{V}}$  for lower values of $m$, namely $m= 4, 12, 20$, while  skipping (available)  $m=25, 50$.}
\end{center}
\end{figure}

 \begin{figure}[h]
\begin{center}
\centering
\includegraphics [width=0.3\columnwidth, valign=c] {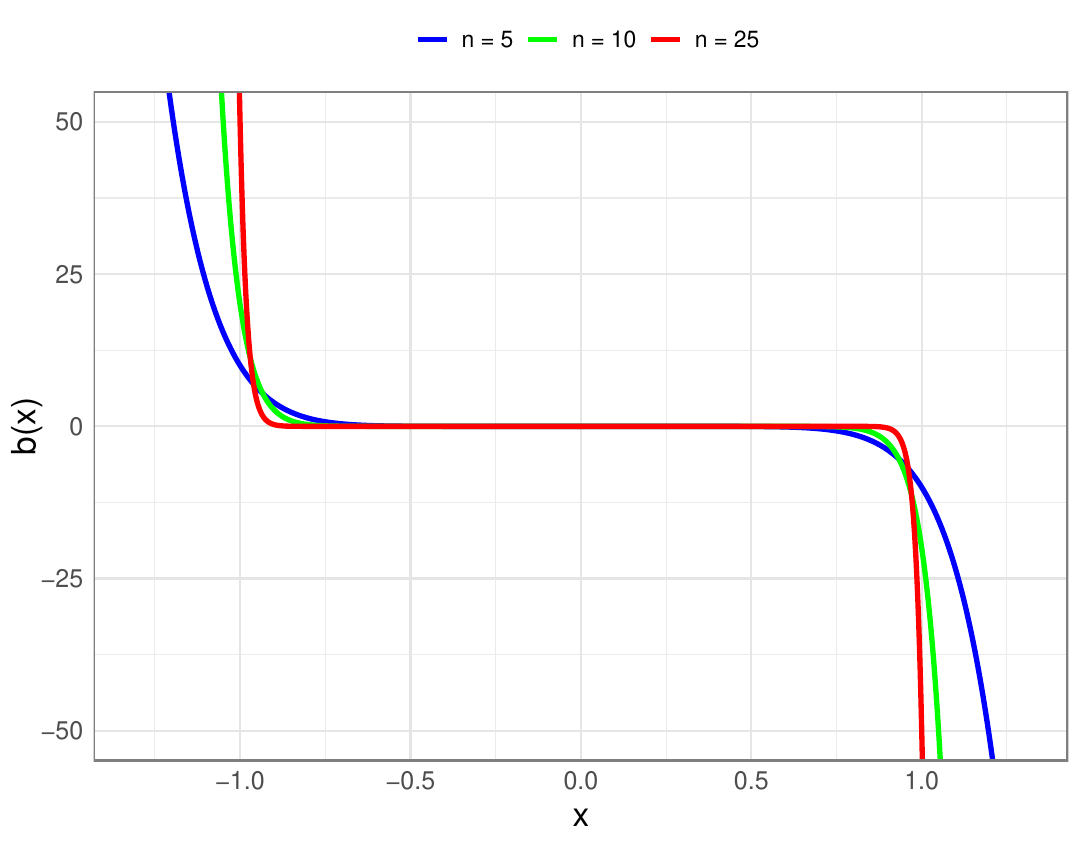}
\includegraphics [width=0.3\columnwidth, valign=c] {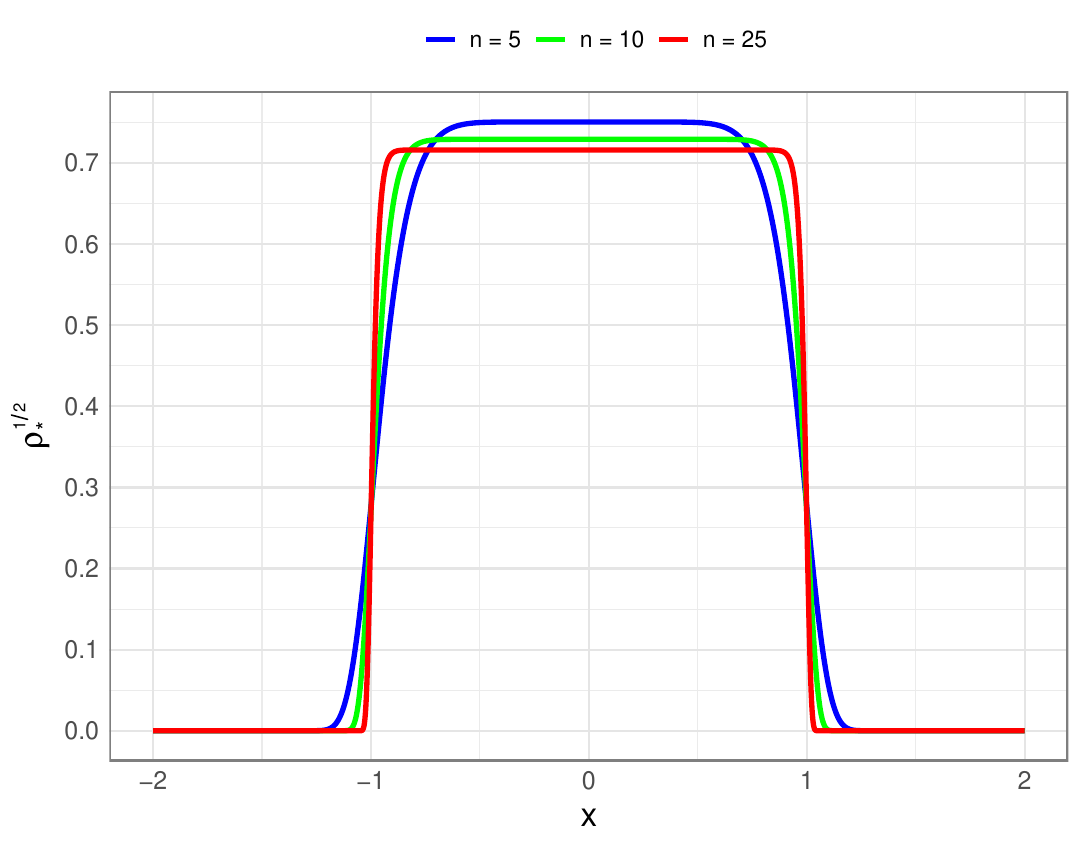}
\includegraphics [width=0.3\columnwidth, valign=c] {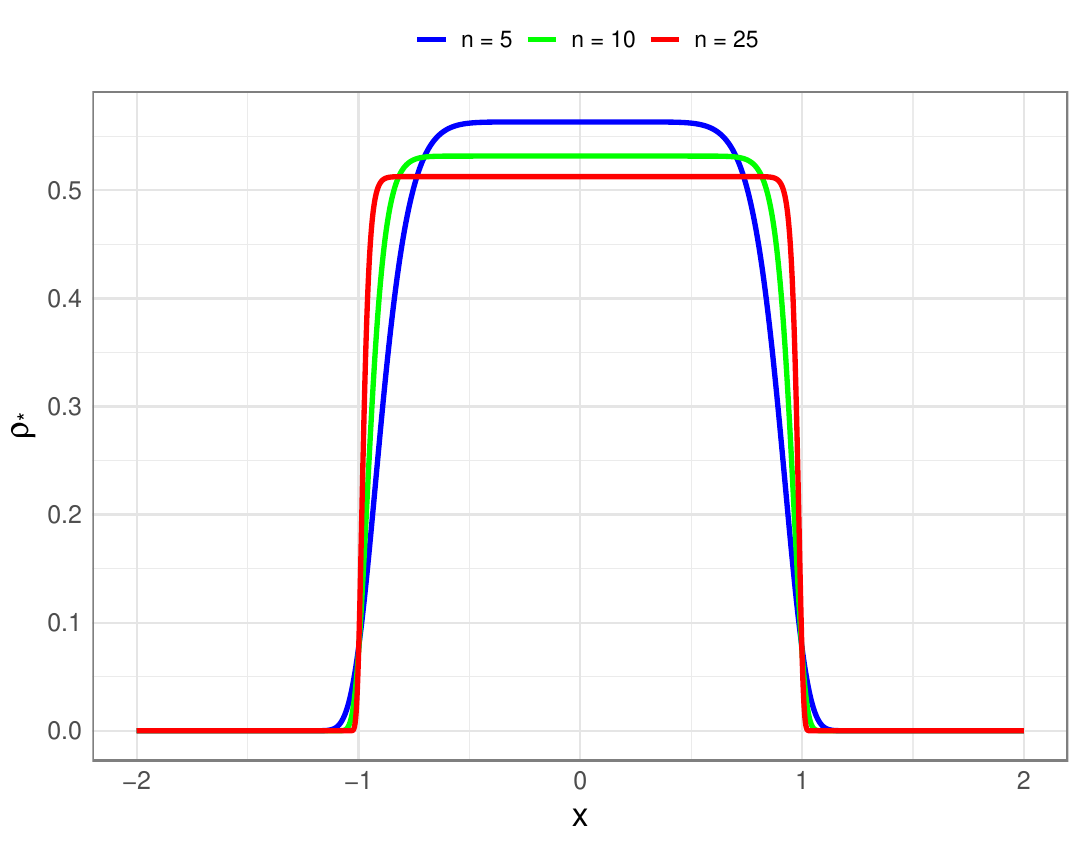}
\caption{$\alpha =2 \rightarrow \phi (x)= x^m$ for $m=2n$;  Left  panel: $b(x)=- m x^{m-1}$; Middle panel:  $\rho _*^{1/2}(x)= \psi_0(x) $;  Right panel: $\rho_*(x) = [\psi_0(x)]^2$.}
\end{center}
\end{figure}

Setting $\alpha =2$ in  Eq. (34),  we  arrive at:
\be
K(x) = \frac{1}{2^{1-\frac{1}{m}}\Gamma\left(1+\frac{1}{m}\right)}
\exp\left(-x^m\right) = \left[2^{1-\frac{1}{m}}\Gamma\left(1+\frac{1}{m}\right)\right]^{-1/2} \, \psi_0(x)
\ee
which differs from $\rho _*(x)$  (Eq.(43) with $\alpha =2$)  by the skipped factor $2$ in the exponent. Clearly, $K(x)$ itself is not a legitimate probability density, compare e.g. our discussion of Section II.C.\\

Since $\int_R  \exp(-x^m) dx =  2 \Gamma(1+1/m)$, we can integrate $K(x)$, with the  (asymptotic $K(t) \rightarrow K$) outcome:
\be
K= \int _R K(x) dx = 2^{1/m}  \in (1,\sqrt{2}).
\ee
Accordingly, we can introduce $\rho _*^{norm}(x)=  K(x)/ K $,  compare e.g. Eq.(32).

For $m=2$,  we   arrive at  the value  $K=\sqrt{2}$, which has appeared before  in connection with the harmonic oscillator potential $\phi(x)= x^2/2$, c.f. Section III.D.  A  consistency of Eqs. (29-36)   can be easily verified  for $m=2$,  with alternative choices of parameters  $\alpha = 1/2, 1, 2$,  while employing   $\Gamma(1/2) = \sqrt{\pi }$.

 Recalling our discussion of section  III.D, where $K(t)= N(t)/N(0)$  has been interpreted as the relative number of  alive trajectories  at time $t$, we can consider $K_m$  for each $m=2n $  as the relative number of  asymptotically  surviving trajectories. Like in  the harmonic case, $m=2, \alpha = 1/2$,  where  $K= \sqrt{2}>1$, the  considered  superharmonic  case, $m=2n >2, \alpha = 1$,   shows  equilibration features with $K_m \in (1, \sqrt{2}]$, hence  an   {\it overcompensation} of killing by branching,  as well. \\

We may repeat  steps (31-37) in the nonlinear setting as well. The normalized  pdf reads:
\be
\rho_*^{norm}(x) = K(x)/K = [2\Gamma(1+ {\frac{1}m})]^{-1} \exp(-x^m),
\ee
 cf. Eqs. (47-48).  We recall that $K(x)= \psi _0(0) \psi_0(x)$, where $\psi_0(x) = \rho _*^{1/2}(x)$ is tacitly  presumed.

  \begin{figure}[h]
\begin{center}
\centering
\includegraphics[width=60mm,height=60mm,valign=c] {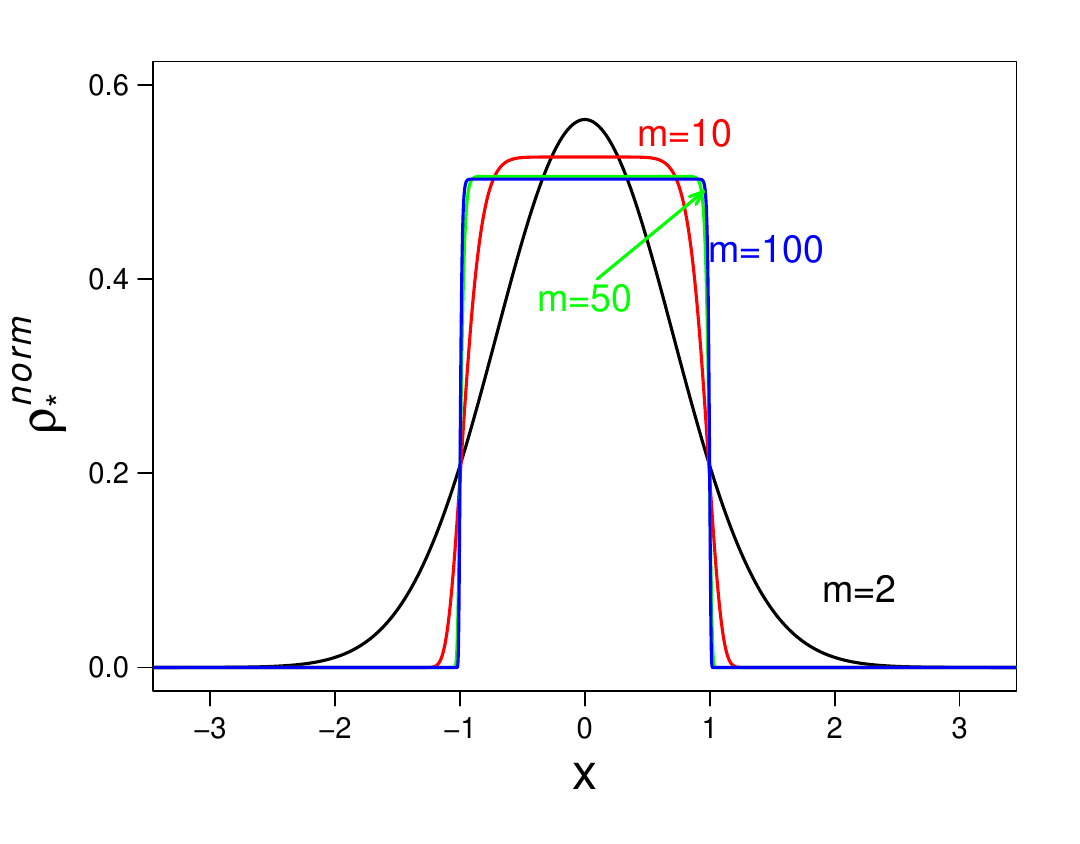}
\caption{ We depict  $\rho_*^{norm}(x)$ of Eq. (50). Actually it is the explicit $L(\mathbb{R})$-normalized expression for the function $A\exp(-x^m)$, $m=2n$, where $1/A= \int_{R} \exp(-x^m) dx $. The presumed  trajectory-wise  meaning of this probability density is analogous to that described in Remark 3, around Eq. (33): $\rho _*^{norm}(x)\Delta x = [K(x)/K]\Delta x $, with $K= N/N(0)$, $K(x)= N(x)/N(0)$, where  $N$ stands for  the (approximate) number of asymptotically alive paths.}
\end{center}
\end{figure}

 With  ${\cal{V}}(x)$ of  Eq.(44) in hands, we can evaluate the mean value $\left< {\cal{V}}\right>$ on $\mathbb{R}$:
\be
\left< {\cal{V}}\right>  = 2\int_0^{\infty} {\cal{V}}(x) \cdot \rho_*^{norm}(x) dx = {\frac{m^2}{2 \Gamma(1+{\frac{1}m})}} \int_0^{\infty }  \left [x^{2m-2} - {\frac{m-1}m} x^{m-2}\right]\cdot e^{-x^m} dx.
\ee
To complete the integrations, we  point out that directly from the integral identity  (41), valid for any $\alpha >0$  and $k$ even,  after  setting  $\alpha =1$,  there follows
$  \int_0^{\infty } x^k \exp (- x^m) dx =  {\frac{1}m} \Gamma \left( {\frac{k+1}m}\right)$.
Ultimately, we  arrive at:
\be
\left< {\cal{V}}\right> = {\frac{m^2}{2 \Gamma({\frac{1}m})}} \left[\Gamma(2-{\frac{1}m}) - {\frac{m-1}m} \Gamma (1-{\frac{1}m})  \right]  =0,
\ee
where   the identity  $\Gamma (2 -  {\frac{1}m}) = (1 -{\frac{1}m}) \Gamma (1- {\frac{1}m})$ has been employed.

Like in the harmonic case, Eq. (35),  in the superharmonic case    the  mean killing  time  rate is  equal to the mean branching  time  rate.  Indeed, since branching and killing  domains do not intersect, we can readily isolate respective  contributions to $\left< {\cal{V}}\right>$, which however refer to  (lower)  incomplete Gamma  functions. This, because the branching domain boundaries are set by  $ |x_m| =  \left({\frac{m-1}{m}}\right)^{1/m} <1$, (see  Eq. (45) for other $\alpha $ options).\\
   We have
\be
\left< {\cal{V}}\right>_{killing}  = - \left< {\cal{V}}\right>_{branching} =
 - 2\int_0^{|x_m|} {\cal{V}}(x) \rho_*^{norm} dx,
\ee
where, the  lowering   to $|x_m|$ the upper integration boundary in (50),  leads to integrals of the form :
\be
\gamma(\alpha, \beta ) = \int_0^B t^{\alpha -1} e^{-t} dt \Longrightarrow \int_0^B x^k e^{-x^m} dx = {\frac{1}m} \gamma \left({\frac{k+1}m}, B^m\right).
\ee
Plugging $B=|x_m|= {\frac{m-1}{m}}= 1-{\frac{1}m}$ in (54), allows us to rewrite (53) as follows:
\be
\left<{\cal{V}}\right>_{killing} =
 - {\frac{1}{2\Gamma(1+ {\frac{1}m})}} \left[m\gamma(2-{\frac{1}m},1-{\frac{1}m}) - (m-1) \gamma(1-{\frac{1}m},1-{\frac{1}m})\right].
\ee
Employing the identity:
\be
\gamma(s+1,u) = s\gamma(s,u) - u^s e^{-u}
\ee
we  cancel out the  explicit  $\gamma $ terms, so  arriving at:
\be
\left<{\cal{V}}\right>_{killing} ={\frac{m-1}{2\Gamma (1+{\frac{1}m}) \exp ({\frac{m-1}m})}} \left({\frac{m}{m-1}}\right)^{\frac{1}m}   \sim  {\frac{m}{2e}}.
\ee

{\bf Remark 7:}   With tabulated values of $\Gamma (1+1/m)$, we can in principle evaluate (56) without any approximations, but for large values of  $m$ the approximation $\sim  {\frac{m}{2e}}$ is reliable.  We mention that for $m=2$,    we get  the exact outcome $\left<{\cal{V}}\right>_{killing} = 2/\sqrt{2\pi e}$. This may be compared with the harmonic
result  $1/\sqrt{2\pi e}\sim 0.24197$, Eq. (34), for the potential ${\cal{V}}(x)= {\frac{1}2} (x^2-1)$.

\subsubsection{How deep are the local minimum wells ? Improving the resolution of
$q(t)=\min[1,- {\cal{V}}(X(t)) \delta t ]$ about the minima.}

The definition (44) of ${\cal{V}}(\alpha ,x)= a x^{2m-2} - bx^{m-2}$, with  $a= m^2\alpha ^2/8$,  $b=m(m-1)\alpha /4$,    $\{\alpha = 2, 2/m, 2m\}$, and  $\phi (\alpha,x)= {\frac{\alpha }{2}} x^m$,  allows us  to deduce the location of  deep  minima (symmetric, relative to $x=0$)   of the pertinent Feynman-Kac potential  (evaluate  ${\nabla \cal{V}}(\alpha ,x)=0$):
\be
|x_{min}| =\left[{\frac{b}{2a}}{\frac{m-2}{m-1}}\right]^{1/m} =   \left[{\frac{m-2}{m \alpha }}\right]^{1/m}.
\ee
Recalling that $\phi (x)= {\frac{\alpha }2} x^m$, we realize that $\alpha =2 \rightarrow |x_{min}|=[(m-2)/2m]^{1/m}$, $\alpha = 2/m \rightarrow |x_{min}|=[(m-2)/2]^{1/m}$, and $\alpha = 2m \rightarrow |x_{min}|= [(m-2)/2m^2]^{1/m}$.\\

The local minimum of ${\cal{V}}(\alpha ,x)$ at $|x_{min}|$, for all admissible choices of $\alpha $ reads:
\be
{\cal{V}}(\alpha ,|x_{min}|) = -{\frac{m(m-2)}8} \left({\frac{m \alpha }{m-2}}\right)^{2/m}.
\ee
The {\it maximal values}  of the time interval $\delta t_{max} = 1/{\cal{V}}_m$, set (respective, decreasing with the growth of $m$) upper  bounds for admissible time increments  $\delta t$, which   secure a probabilistic interpretation of  both  $p\sim {\cal{V}} \delta t$ and $q \sim -{\cal{V}} \delta t$, cf. Eqs (38)  and  (39).\\

{\bf Remark 8:} For concreteness,  let us  choose  $\alpha =2$, and  evaluate  ${\cal{V}}(2 ,|x_{min}|)$  for a couple of values of $m=20, 50, 70,100$.  We also  add the  corresponding values of $|x_m|$, Eq. (45):\\

\noindent
(i) $m=20$,   ${\cal{V}}_{20} \sim - 52,79  \rightarrow \delta t_{max} \sim0,1894$; ( $|x_{min}| \sim 0,961$, $|x_m|\sim 0,997$),  $\left<{\cal{V}}\right>_{k} \sim 3,679$, \\
(ii) $m=50$, ${\cal{V}}_{50} \sim - 411,9176   \rightarrow \delta t_{max} \sim 0,0024 $; ($|x_{min}| \sim 0,985$, $|x_m|\sim 0,9996$), $\left<{\cal{V}}\right>_{k} \sim 9,196$  \\
(iii) $m=70$, ${\cal{V}}_{70}  \sim  - 601,046 \rightarrow  \delta t_{max} \sim 0,00166 $; ($|x_{min}| \sim 0,9899$,  $|x_m|\sim 0,9998$), $\left<{\cal{V}}\right>_{k} \sim 12,876$ \\
(iv) $m=100$, ${\cal{V}}_{100}  \sim - 1242,6 \rightarrow \delta t_{max} \sim 0,0008$; ($|x_{min}| \sim 0,993$, $|x_{m}| \sim 0,9999$), $\left<{\cal{V}}\right>_{k} \sim 18,394$ .\\

In the above, the notation $\left<{\cal{V}}\right>_{k}$ refers to the mean killing/branching rate formula (52).\\

\begin{figure}[h]
\begin{center}
\centering
\includegraphics[width=0.35\columnwidth, valign=c] {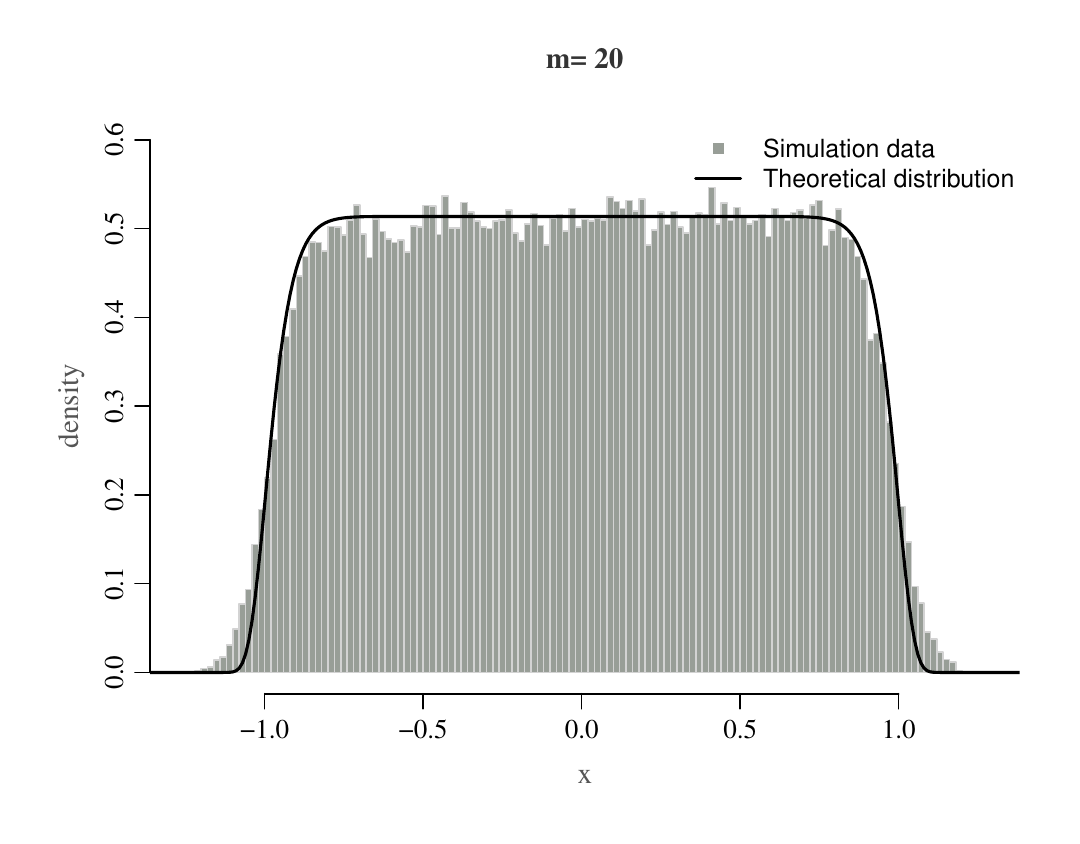}
\includegraphics[width=0.35\columnwidth, valign=c] {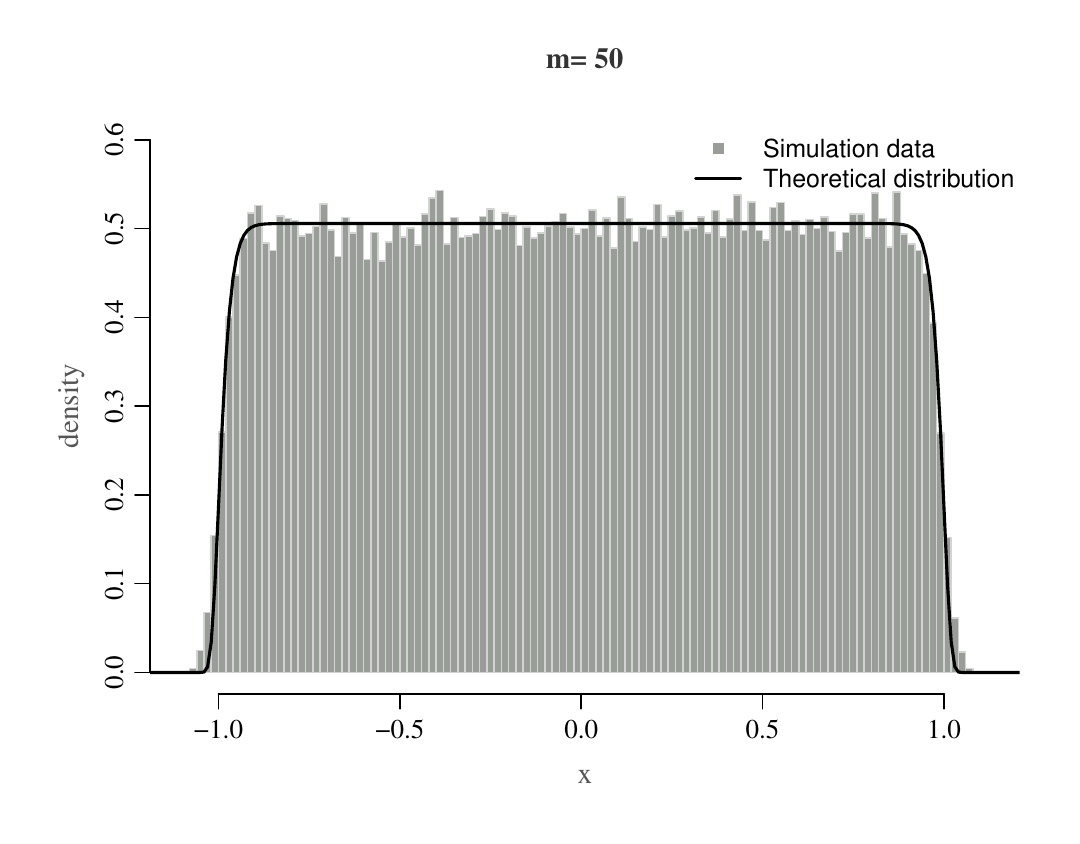}
\includegraphics[width=0.35\columnwidth, valign=c] {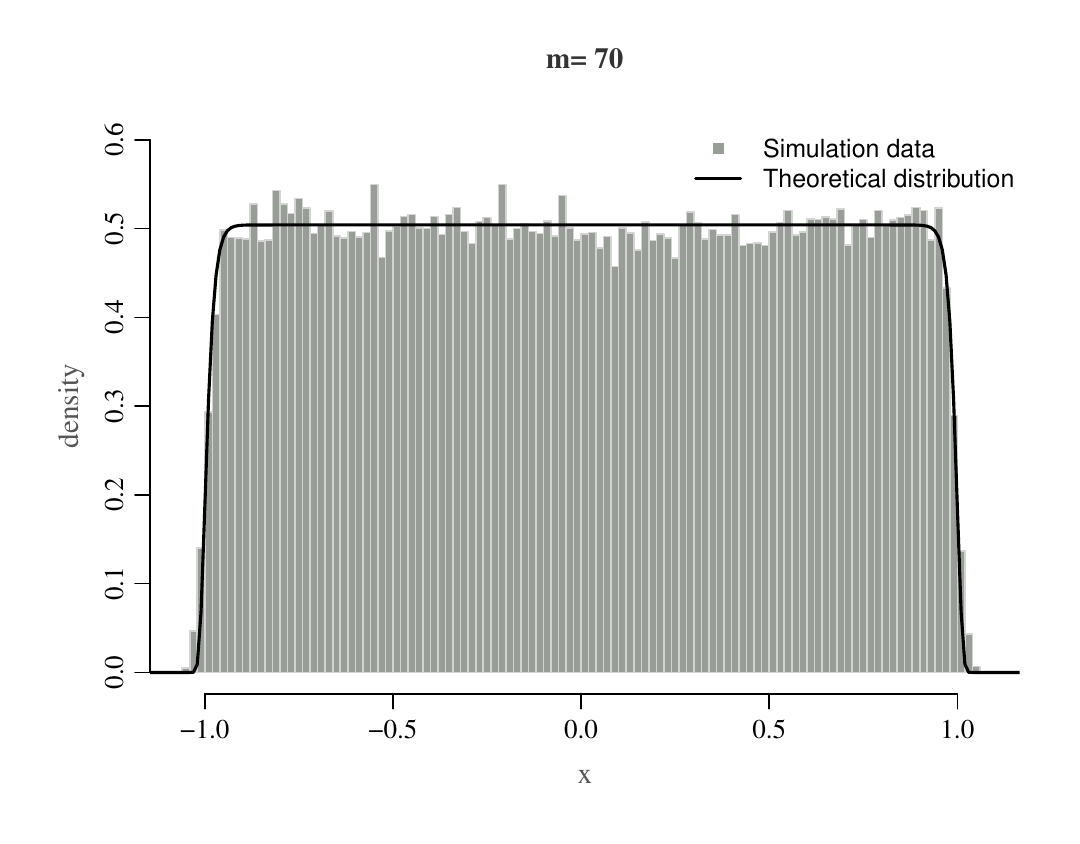}
\includegraphics[width=0.35\columnwidth, valign=c] {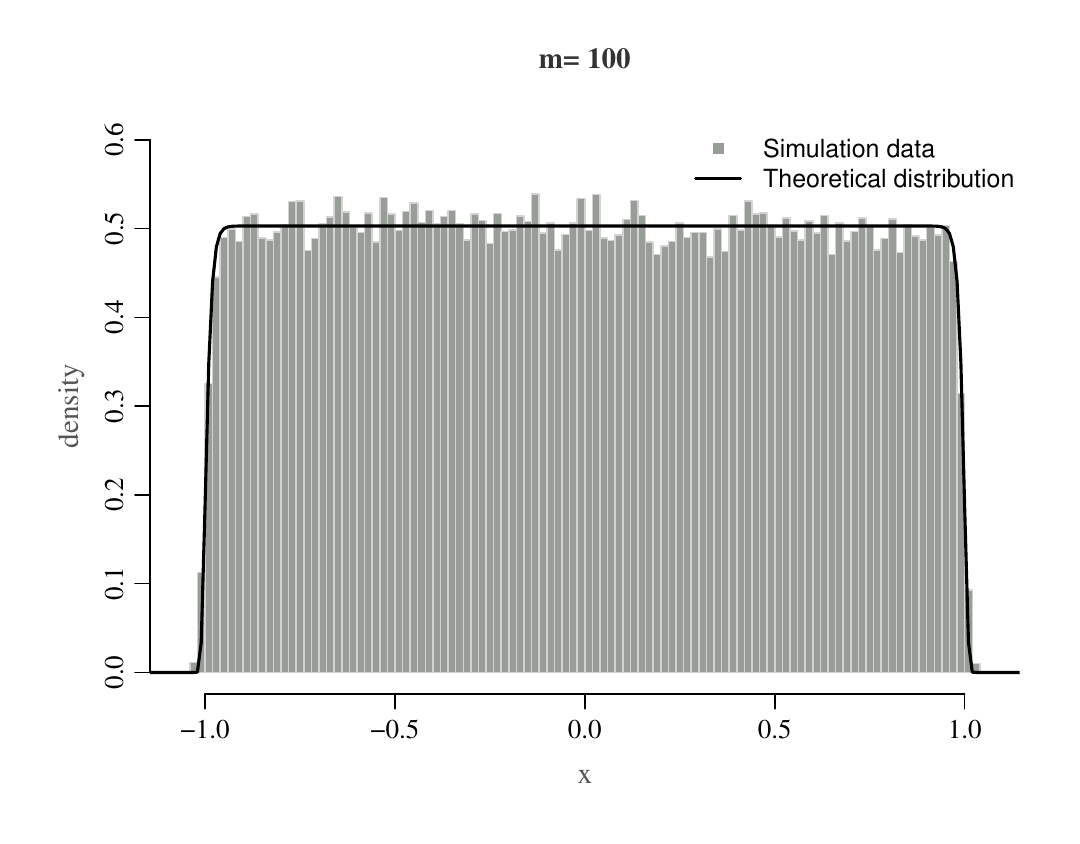}
\caption{$\alpha = 2$,$\phi(x) = x^m$, ${\cal{V}}(x)=  (m/2)x^{m-2}(mx^m +1 -m )$.  The branching/killing scenario  simulations of surviving paths  in the asymptotic regime $N(5)$. Simulation time increments $\delta t(m) \sim  8/ \pi m^2$  have been optimised to keep the final number of trajectories $N$ below $1,05 \cdot N(0)$ (we recall the interpretation of $K=2^{1/m} \sim   N/N(0)$.  The envelope for  each value of $m$  is  $\rho_*^{norm}(x)$, as given by Eq. (50).}
\end{center}
\end{figure}

{\bf Remark 9:}  In Refs. \cite{zaba} (Section 4.2) and \cite{stef}  (Section 4.3), the spectral closeness of large $m=2n$, $H=- (1/2)\Delta + {\cal{V}}(x)$  and $-(1/2)\Delta $ with Neumann boundary data  (implementing the the two-sided reflection) has been analyzed. The flat pdf equal $1/2$ is a signature of  reflecting Brownian motion. The transition kernel for the latter case cannot be obtained via (13), and needs the usage of  singular path integrals, with perturbations in terms of derivatives of  Dirac delta functions, with strength going to minus infinity, \cite{grosche,grosche1}.

\subsection{${\cal{V}}(x) = x^m - \epsilon_1$.}

In this subsection, we borrow some preliminary results from the Appendix B of Ref. \cite{zaba}, and develop them in the Feynman-Kac diffusion context, favoured in the present paper.

In conformity with the Technical Comment of Section I.B,  assume to have given $H_0= -(1/2)\Delta + V(x)$, with $V(x)= x^m$,  together with  its  bottom eigenvalue $\epsilon _1$  and the ground state function  $\psi (x) \sim \rho _*^{1/2}(x)$.  The reconstruction of the Fokker-Planck dynamics from its Feynman-Kac kernel entry, needs the "subtraction" trick, so that $V(x) \rightarrow {\cal{V}}(x) = V(x) - \epsilon_1$. This assigns the eigenvalue zero of $H = -(1/2)\Delta + {\cal{V}}(x)$ to the eigenfunction $\rho_*^{1/2}(x)$.

\begin{figure}[h]
\begin{center}
\centering
\includegraphics[width=84mm,height=70mm]{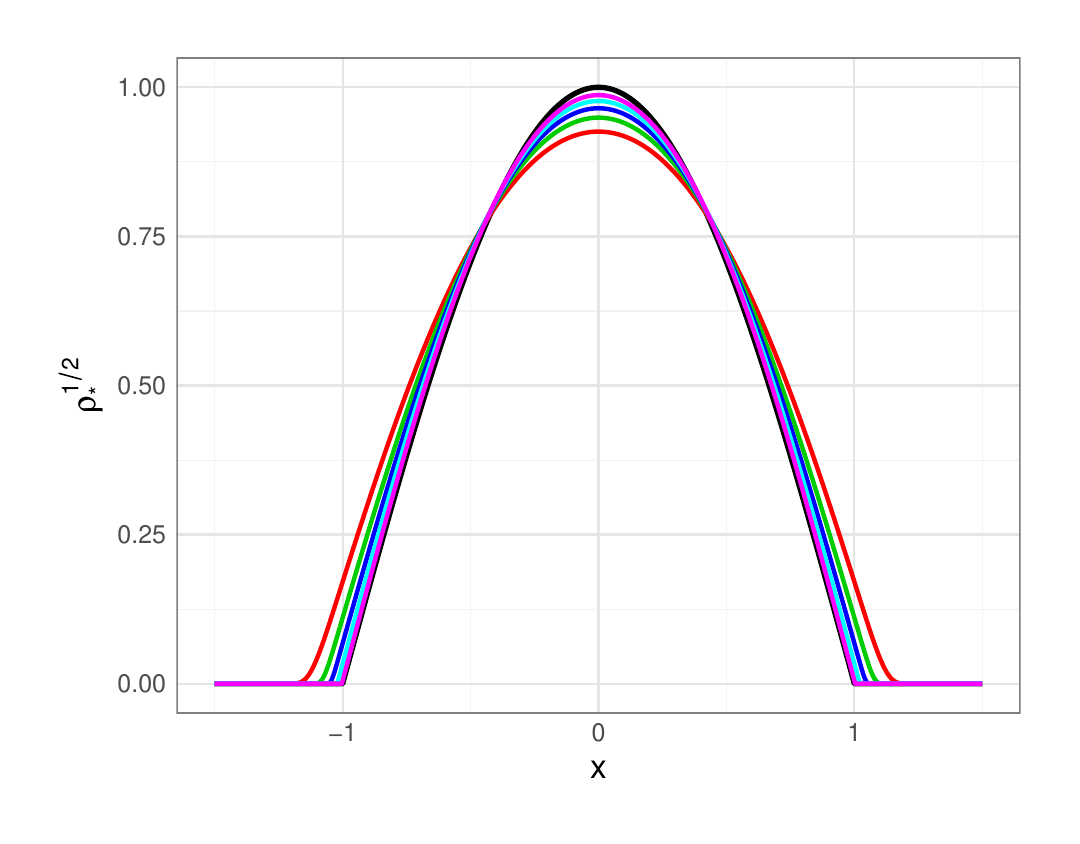}
\includegraphics[width=72mm,height=70mm]{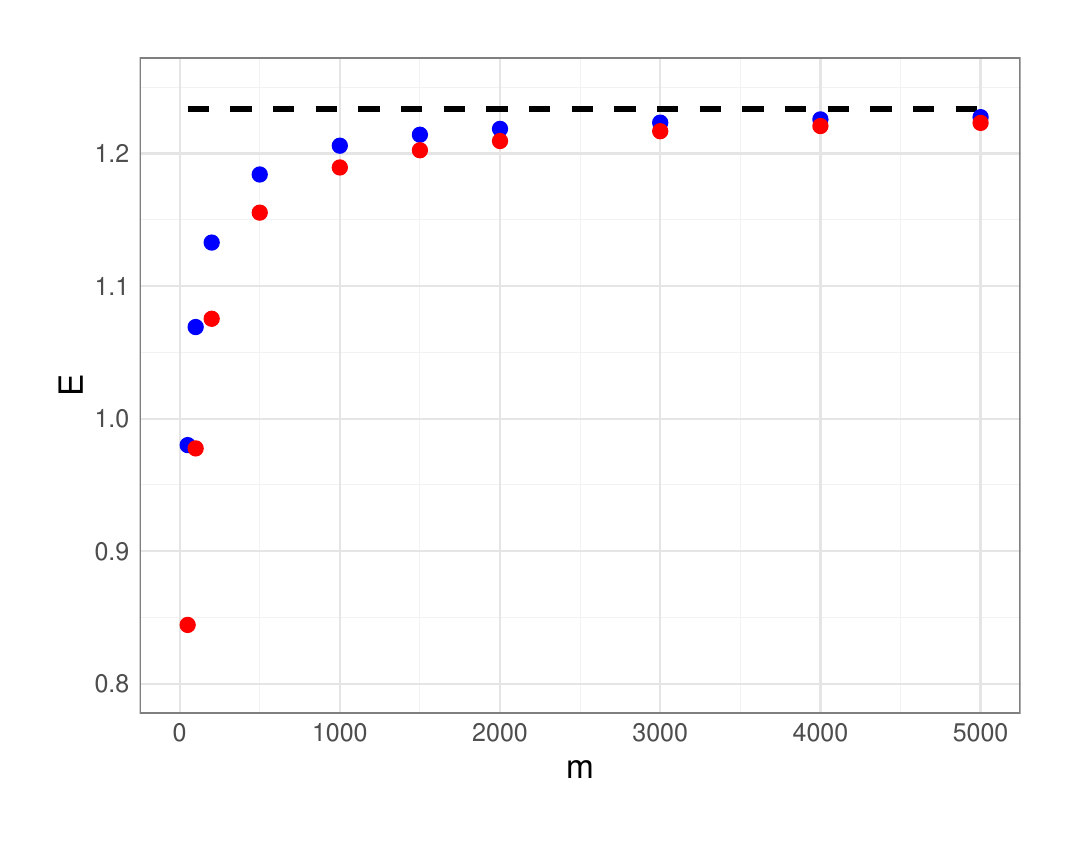}
\caption{Left panel: Ground state eigenfunctions  $\rho _*^{1/2}(x)$  of $\hat{H}= - {\frac{1}2} \Delta + x^m$, $m=50, 100, 200, 500, 5000$; one may assign $m$ to a concrete
 curve by following the maxima in the  increasing order, the  top maximum (black) curve
 corresponds to the  asymptotic $\rho _*^{1/2}(x) = \cos(\pi x/2)$. Right panel: The $m$-dependence of the ground state  eigenvalue $E=\epsilon_1(m)$  is depicted both for the potential  $x^m$ (blue)   and  $x^m/m$ (red). The convergence to the asymptotic  infinite  Dirichlet  well value   $\epsilon_1=\pi ^2/8 \simeq  1.2337$ ($\nu = 1/2$) is undisputable.  The reported eigenvalues  read (the case of $V(x)=x^m$, while  listed  in the growing $m$-order, the corresponding $m$'s are easily retrievable from the figure):  $\epsilon_1(m) = 0.980021,  1.06912, 1.13285, 1.18422, 1.20595, 1.21421, 1.21865, 1.22335, 1.2258, 1.22748$. }
\end{center}
\end{figure}

 \begin{figure}[h]
\begin{center}
\centering
\includegraphics [width=0.35\columnwidth, valign=c] {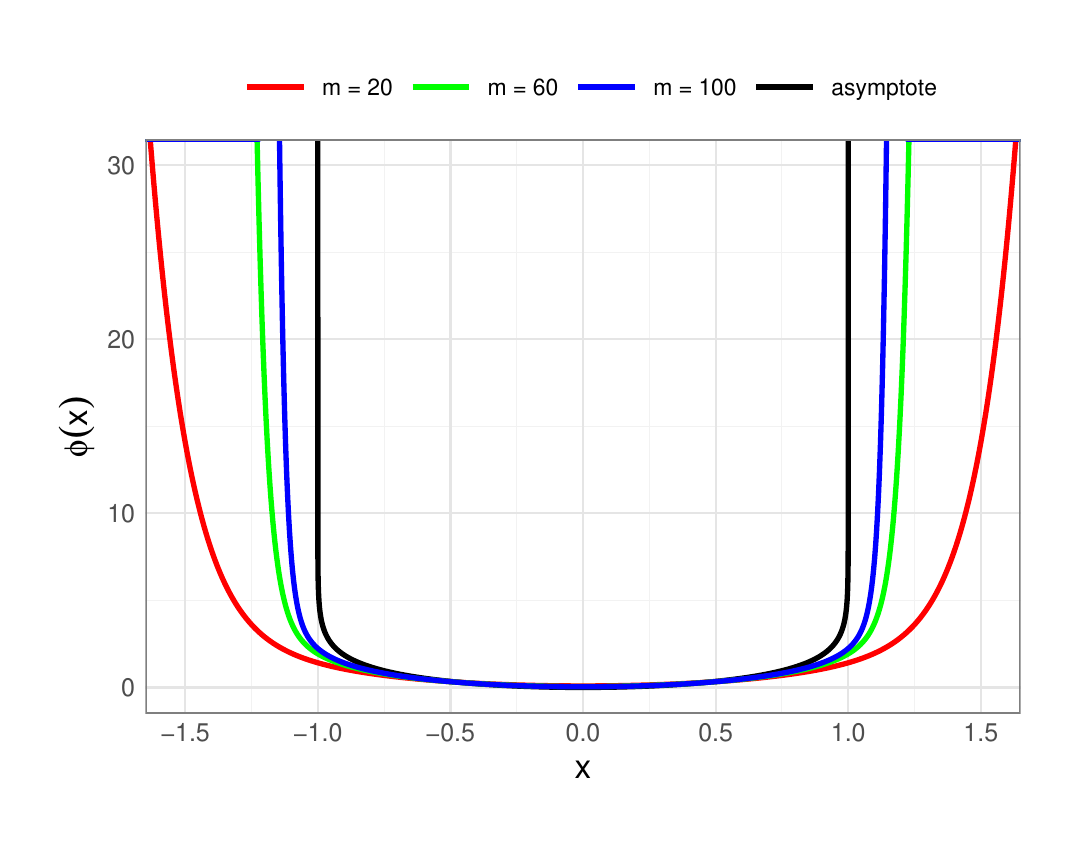}
\includegraphics [width=0.3\columnwidth, valign=c] {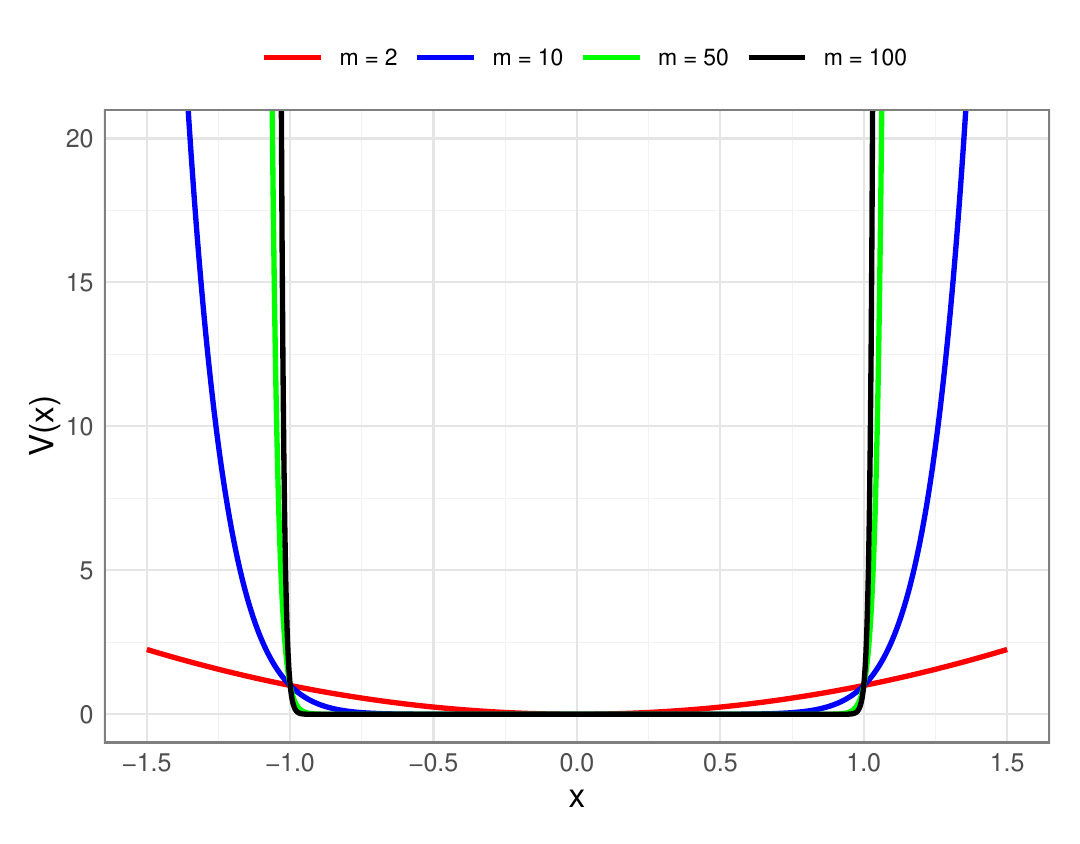}
\caption{${\cal{V}}(x)= V(x) - \epsilon_1$, where $V(x)=x^m$,  for $ m=20, 60, 100$;  Left panel:  Numerically retrieved $\phi(x)$ from which $b(x)= - \nabla \phi (x)$ follows; Right panel: $V(x)= x^m$.}
\end{center}
\end{figure}

\begin{figure}[h]
\begin{center}
\centering
\includegraphics[width=0.3\columnwidth, valign=c] {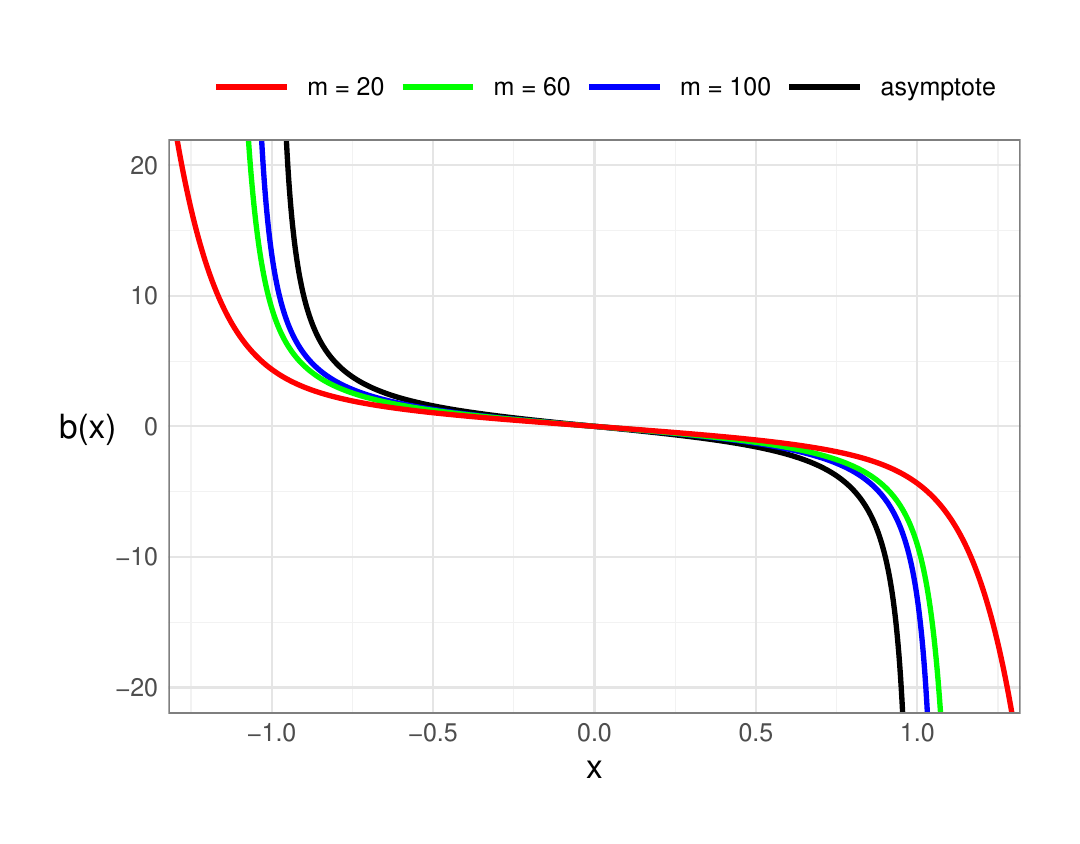}
\includegraphics[width=0.3\columnwidth, valign=c] {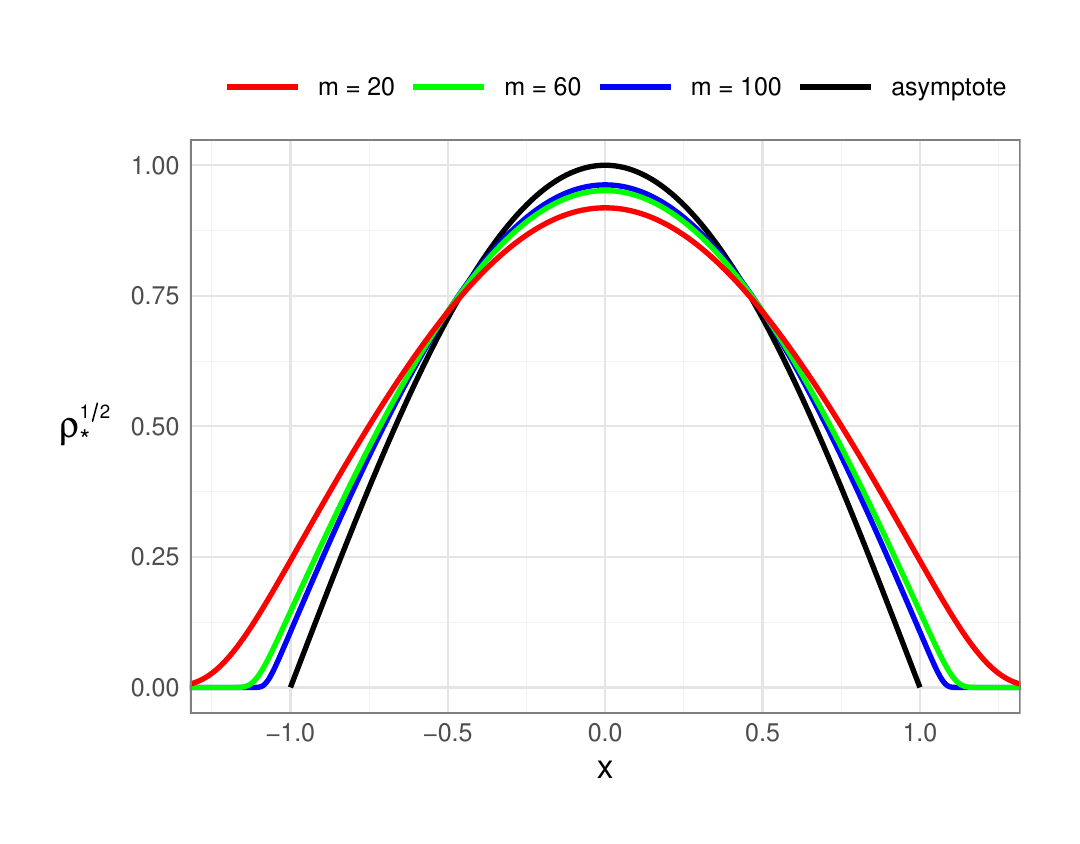}
\includegraphics[width=0.3\columnwidth, valign=c] {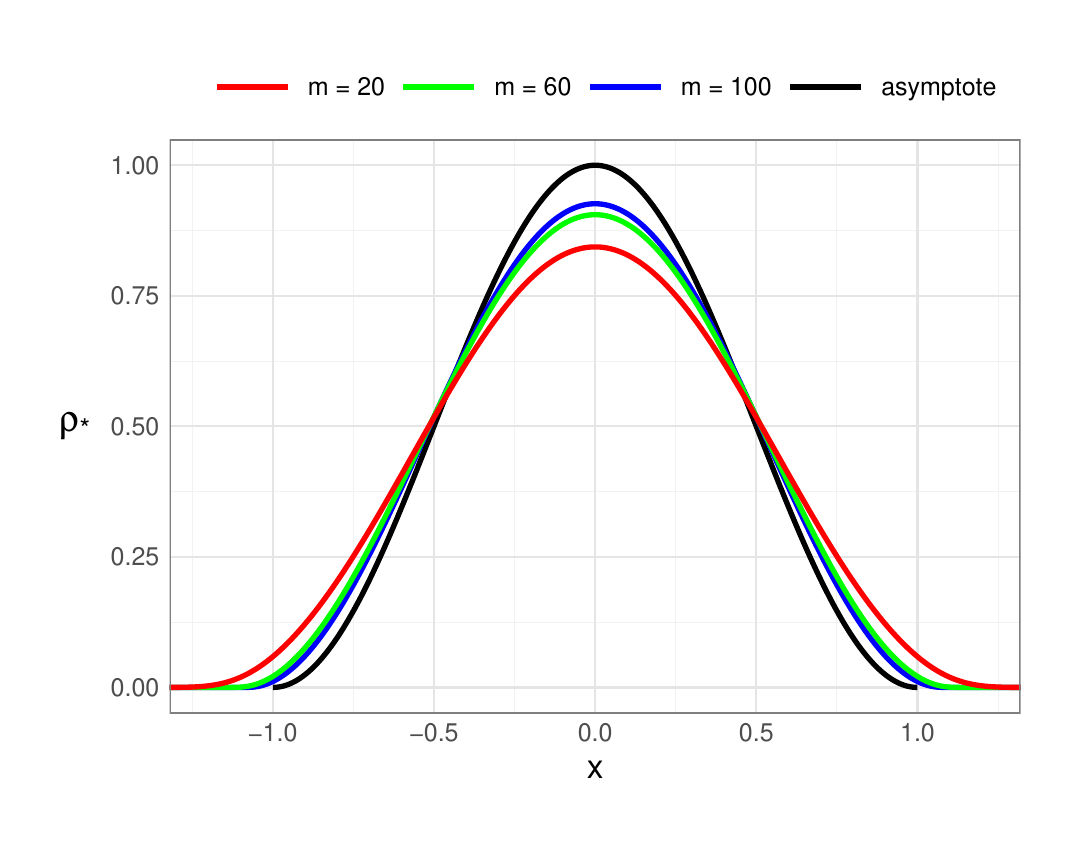}
\caption{ ${\cal{V}}(x)= x^m - \epsilon_1$  for $m=2n$;  Left  panel: Numerically retrieved  $b(x)= \nabla \ln \rho_*^{1/2}$, black asymptote is
$b(x)= -(\pi /2)\tan (\pi x/2)$; Middle panel:  $\rho _*^{1/2}(x)= \psi_1(x) $, black asymptote is $\cos (\pi x/2)$;  Right panel: $\rho_*(x) = [\psi_1(x)]^2$, black asymptote is $\cos^2(\pi x/2)$.}
\end{center}
\end{figure}

\begin{figure}[h]
\begin{center}
\centering
\includegraphics [width=0.33\columnwidth, valign=c] {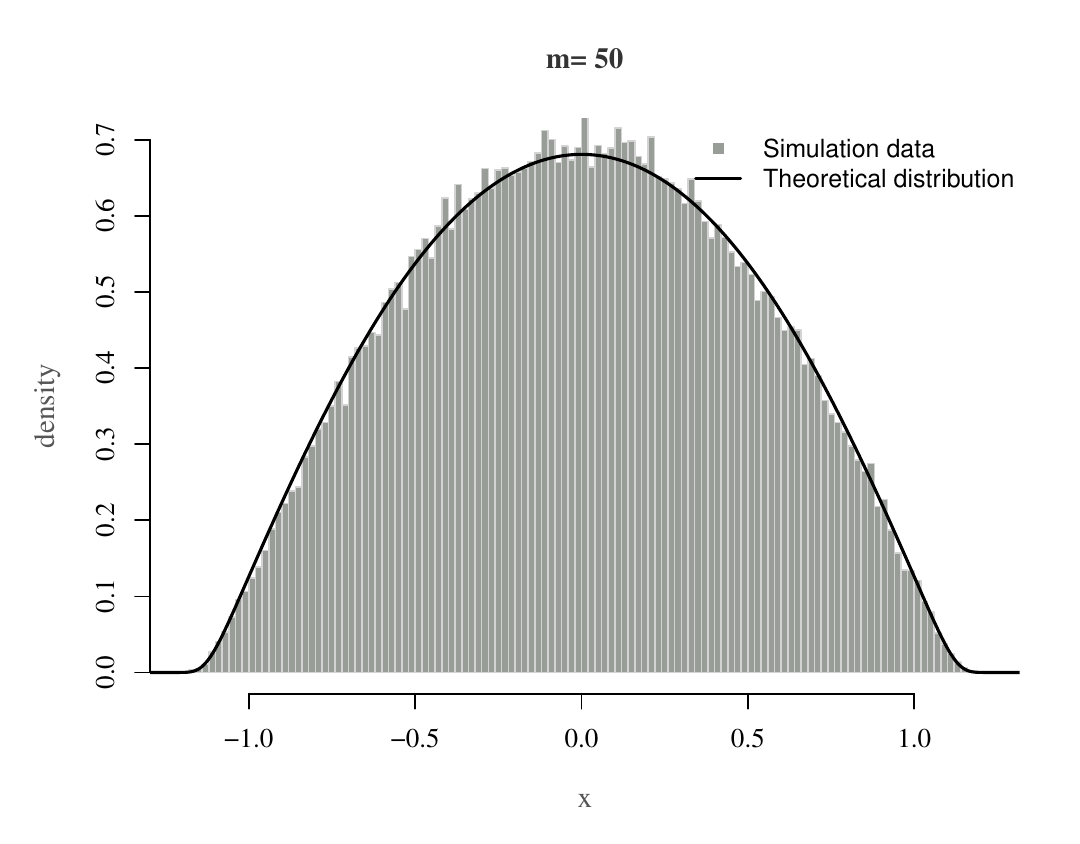}
\includegraphics [width=0.33\columnwidth, valign=c] {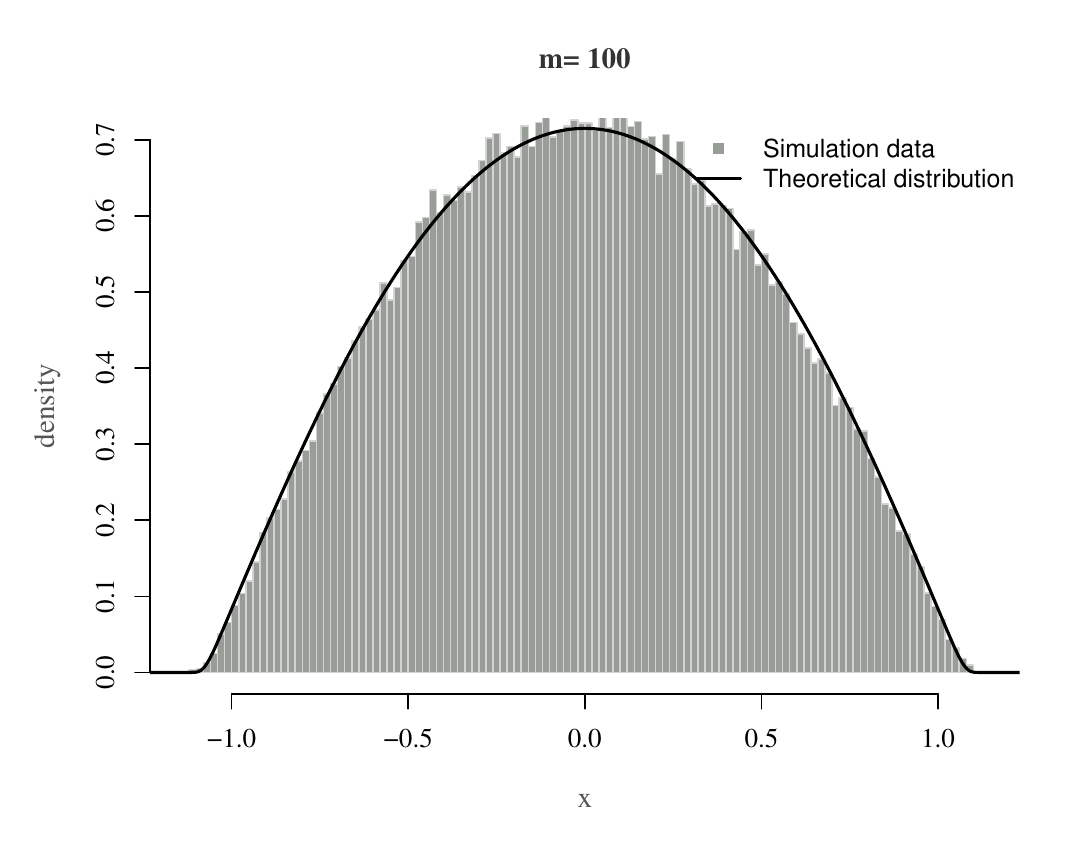}
\includegraphics [width=0.33\columnwidth, valign=c] {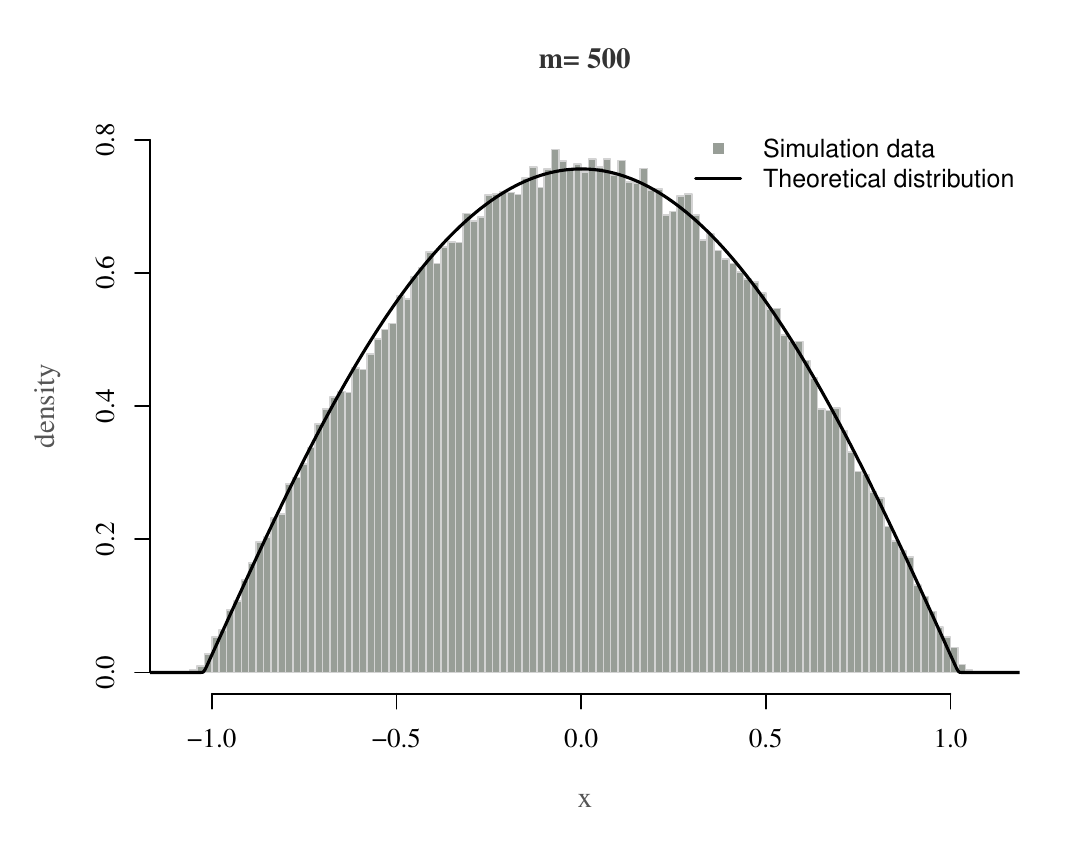}
\includegraphics [width=0.33\columnwidth, valign=c] {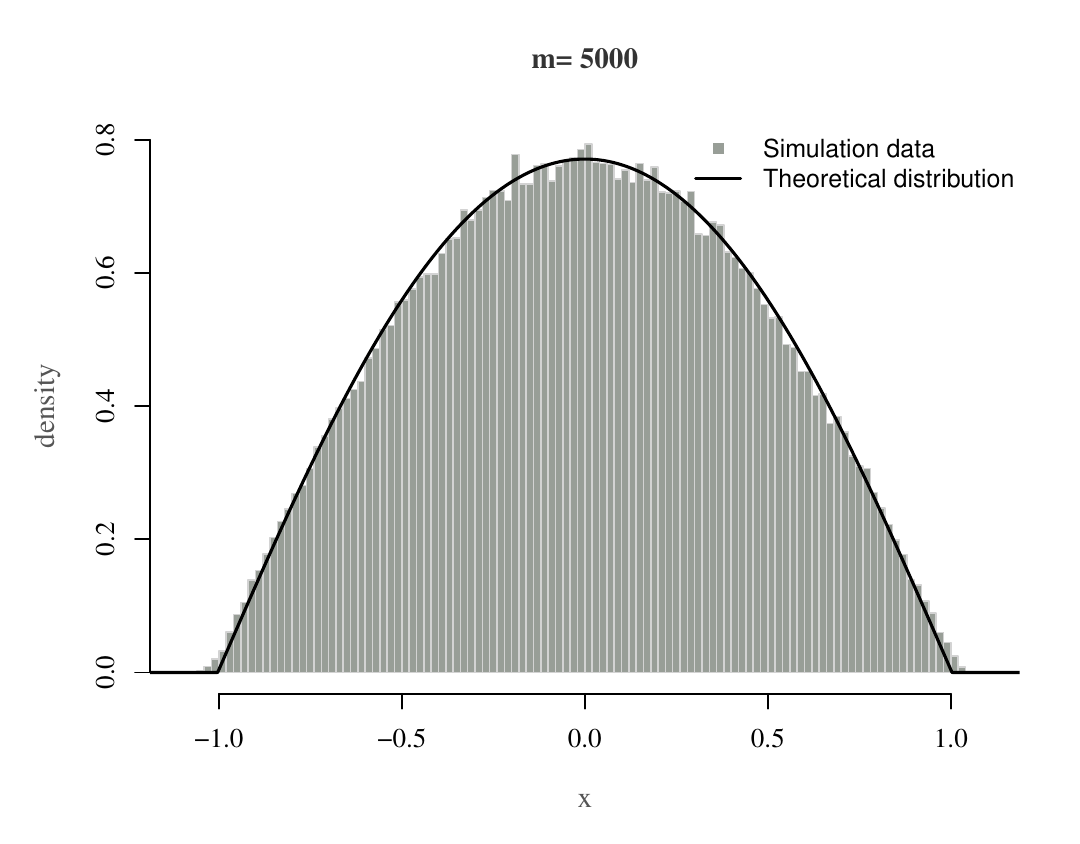}
\caption{${\cal{V}}(x) = x^m - \epsilon _1$. Equilibration in the Feynman-Kac diffusion. The  killing/branching path-wise simulation algorithm at work. The continuous curve represents $\rho_*^{norm}(x)= K(x)/K$ for  each $m=50, 100, 500, 5000$ case.  We recall that $K(x) =\psi_1(0) \psi_1(x)$.}
\end{center}
\end{figure}

The  previously employed  potential  $\phi (x)= x^m$ now appears in another role, while  transformed to  `  shifted (renormalized) Feynman-Kac  potential  form   ${\cal{V}}(x) = x^m - \epsilon_1$, necessary to  produce the bottom eigenvalue zero for  the Hamiltonian  $H= - {\frac{1}2}\Delta + {\cal{V}}$.  The spectrum of $H_0 = - {\frac{1}2}\Delta + x^m$ is discrete and positive. The  apparent problem is that there are no handy  analytic formulas for its eigenfunctions and eigenvalues.

 Therefore, we need  to rely on a  computer  assisted  route toward the (approximate) solution of the spectral problem for $H_0$. It is based on the Strang splitting method, employed by us before  in another context,  and explained in some detail in Refs. \cite{jmp14,jmp15}.

In Fig. 10 we reproduce final computer-assisted results,  with a focus on the $m$-dependence of the  strictly positive  ground state eigenvalue and the related eigenfunction. The ground state functions are depicted for $m = 50, 100, 200, 500, 5000$, two latter curves are indistinguishable from the asymptotic $cos(\pi x/2)$, being the ground state of the familiar infinite well spectral problem, cf. \cite{pre24}.
The ground state eigenvalues (comparatively for the cases  $x^m$ and $x^m/m$) were numerically
evaluated for m = 50, 100, 200, 1000, 1500, 2000, 3000, 4000, 5000. The convergence to
$\pi^2/8 \sim  1.2337$  is graphically confirmed.

The case of $x^m/m$ has been covered in a parallel computation, but made comparatively
explicit in Fig. 10, only on the level of eigenvalues. As far as the eigenfunctions are
concerned, their qualitative behavior for the potential $x^m/m$ is the same as that reported for  $x^m$.

The Strang method allows to deduce eigenfunctions, and in particular the ground state function $\psi_1(x)$ for each $H_0$, and in particular for  $H_0 - \epsilon_1$, as the $L^2(\mathbb{R})$ function. After normalization, its square $\psi_1 ^2(x)$ determines the  $L(\mathbb{R})$  pdf  $\rho_*(x)$.
In Fig. 12 we use the notation $\rho_*^{1/2}(x)$ and  $\rho_*(x)$  respectively.\\

As far as the Feynman-Kac diffusion context is concerned, we tentatively  invoke the  eigenfunction expansion of the (presumed to be well defined) Feynman-Kac kernel $k(y,x,t)$  with $\psi_1(x)$ replacing $\psi_0(x)$. This  tells us that in the large time asymptotic $k(0,0,x,t) \rightarrow K(x)=   \psi_1(0) \psi_1(x)$. In Fig 12 we have the numerically retrieved form of $\psi_1(x)=\rho_*^{1/2}(x)$.  Its maximum at $x=0$, equals $\psi_1(0)$, and can be retrieved from available numerical data. Hence, we have in hands $K(x)$, whose (numerical) integration gives us the value of $K$.  Thus, $\rho_*^{norm}(x)= K(x)/K$, in principle can be numerically retrieved as well.

Results of the path-wise  approach  (computer-assisted, by means of the killing/branching algorithm),  towards  the reconstruction of  $\rho_*^{norm}(x)= K(x)/K$, where $K(x)= \psi(0) \psi(x)$ and relevant functions are depicted in Fig.  (10),   are presented in below, in Fig. 13.  \\

{\bf Remark 10:} The presented analysis demonstrates that for large even $m$, the spectral problem for $H_0 = - {\frac{1}2}\Delta + x^m$   well approximates the spectral problem for the Dirichlet restriction of the  $- (1/2) \Delta $ to  the interval  (equivalently, the infinite well), \cite{zaba,pre24}. One should be aware, that for the infinite well  problem, the  uncritical usage of the definition (13) of the  F-K  path integral, with the F-K potential  ${\cal{V}} = - \pi^2 /8$  of Ref. \cite{zaba,pre24}, does not  literally  work, unless the boundary data are properly implemented.   In fact, the path integral analysis of Refs. \cite{grosche0,grosche,grosche1}  leads to correct answers for the  infinite well  path  integral kernel, only if the  Dirac delta  perturbations (with strength going to infinity), are added  to induce the effect of boundaries. These singular perturbations  in principle can be handled.\\

\section{Quartic double-well potentials.}

Previously, see also \cite{zaba}, we have encountered two well  ({\it not} double-well) potentials of the general form ${\cal{V}}(x)= bx^{2m-2} - ax^{m-2}$, $a>0, b>0$,  associated with  superharmonic  Brownian motion drifts  $b(x)= - \nabla \phi(x)$, where  $\phi (x) \sim x^m$, with $m=2n>2$. The latter restriction, has excluded the quartic double-well case  from the forgoing analysis. \\
From now on, we shall abandon the previous restriction and focus our attention on the disregarded  $m=2$ case, i.e.  the  double-well potential
\be
V(x)= bx^4-ax^2
\ee
 with  positive   $a$ and $b$. \\
 We shall consider these  "true"  double-well potentials  (alternatively) in the roles of  $\phi (x)$ or ${\cal{V}}(x)$),  in  the deeply nonperturbative regime. Then,  a substantial impact of both quartic and quadratic terms makes the  spectral problem   for $H= -{\frac{1}2} \Delta + V(x)$  not amenable to the  standard perturbation calculus, \cite{holstein}-\cite{shuryak}, see also \cite{turbiner1,turbiner2,turbiner3}.

 \begin{figure}[h]
\begin{center}
\centering
\includegraphics [width=0.45\columnwidth, valign=c] {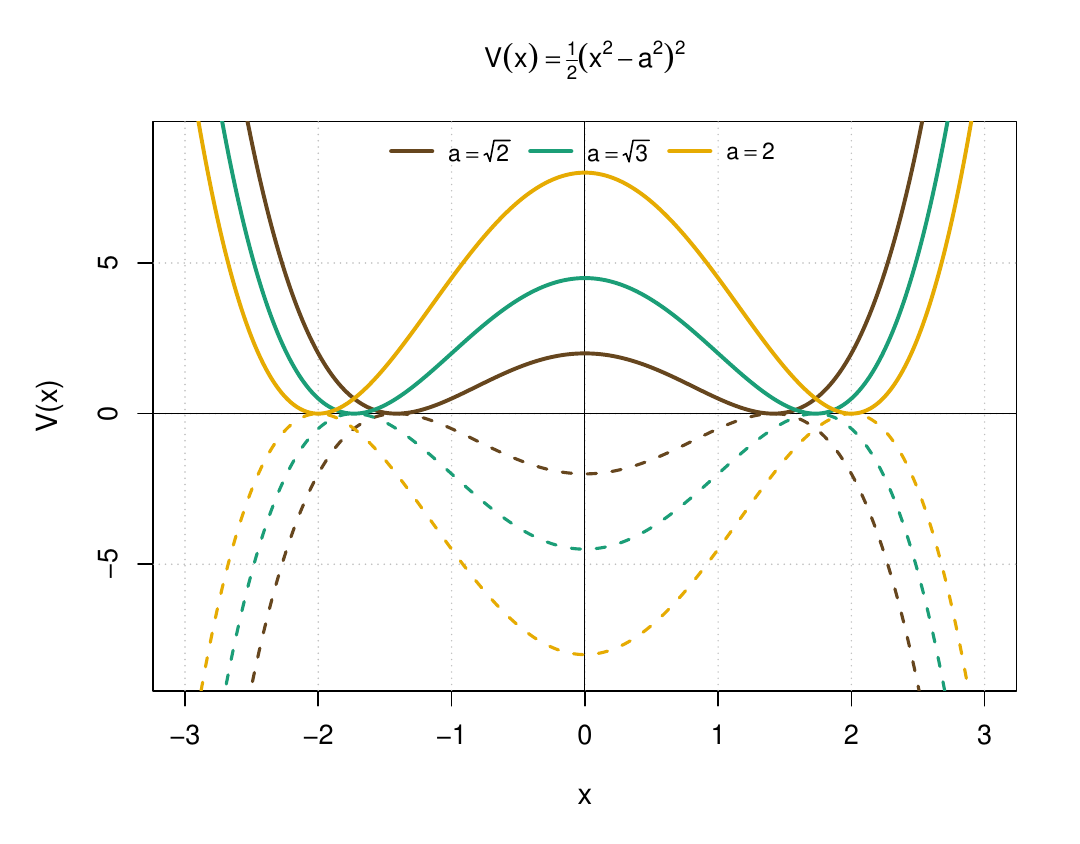}
\caption{The reference quartic potential  $V(x) ={\frac{1}2} \left(x^2 - a^2\right)^2$ with its inverted (casually termed Euclidean) partner},  for $a=\sqrt{2}, \sqrt{3}, 2$  (compare e.g. Fig. 1).  We point out the instanton approach of Ref. \cite{holstein}, as a tool  to analyze the removal of the bottom eigenvalue degeneracy (the tunneling issue). This is typically  accomplished by evaluating the Euclidean propagator (see e.g. Eqs (7-14))  connecting the local maxima $\pm a$ of the inverted double-well.
\end{center}
\end{figure}

For clarity of exposition, we mention that local maximum is located at 0, two negative-valued local minima (depth of the wells) equal  $ - a^2 /4b$ at $x_{1,2} =\pm \sqrt{a/2b}$, and the distance between the minima (e.g. an effective width of the central barrier) equals $2\sqrt{a/2b}$.
A nonnegative definite version of the double-well potential reads $V(x) + a^2/4b =
b(x^2-a/2b)^2$. For $b=1$, we recover  $(x^4-ax^2) +a^4/4 = \left(x^2- {\frac{a}2}\right)^2$.

 \begin{figure}[h]
\begin{center}
\centering
\includegraphics [width=0.45\columnwidth, valign=c] {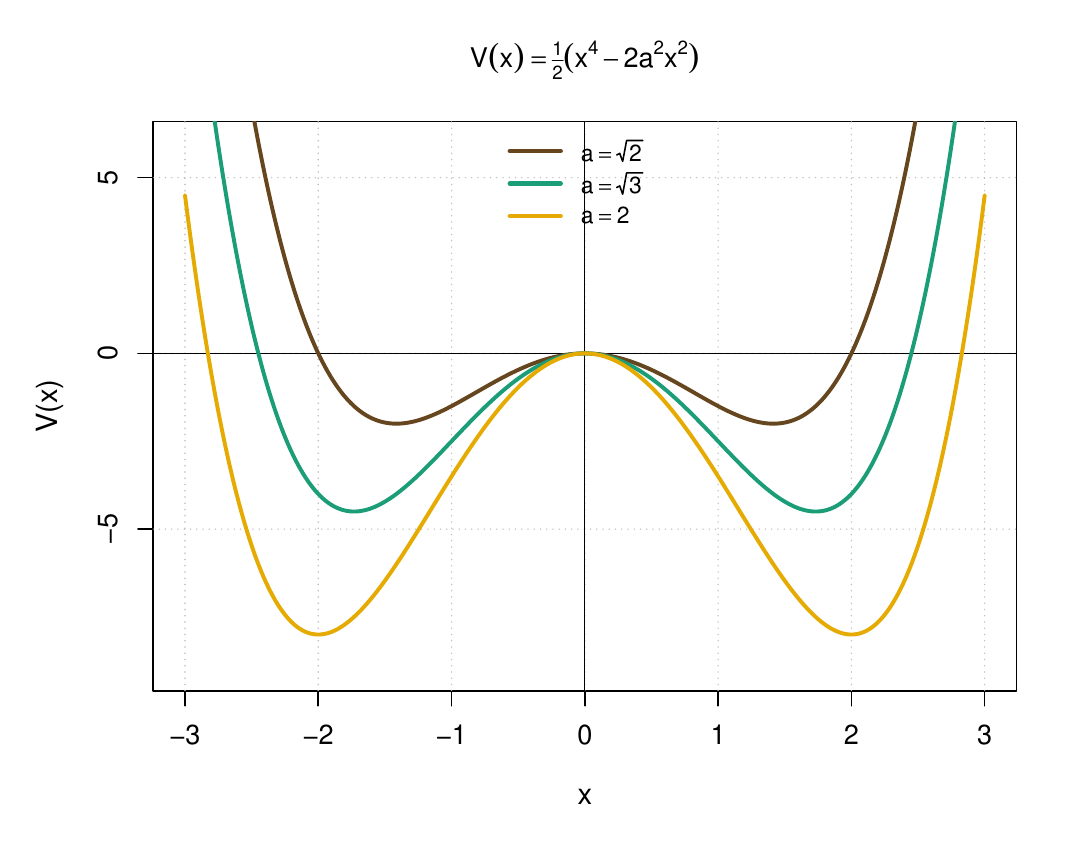}
\includegraphics [width=0.45\columnwidth, valign=c] {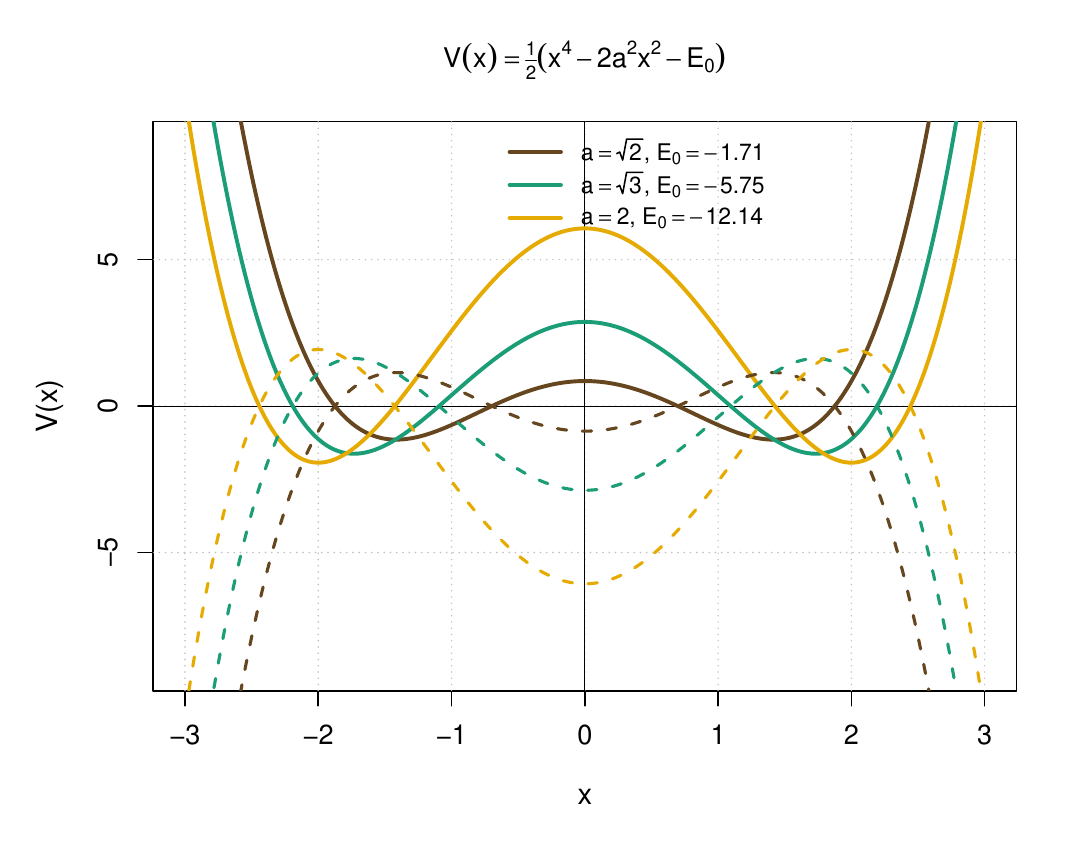}
\caption{Left panel: The reference quartic potential  $V(x)= {\frac{1}2}(x^4 - 2a^2x^2)$ for $a=\sqrt{2}, \sqrt{3}, 2$; Right panel: The potential "with subtraction", ${\cal{V}}(x)= V(x) - \epsilon_1$  together with its  inverted (Euclidean) version.  Bottom  eigenvalues $\epsilon_1 =(1/2)E_0$  of $H_0= -{\frac{1}2} \Delta + V(x))$ were computed by means of the Strang splitting method (cf. \cite{jmp14,jmp15}). All of them are negative.}
\end{center}
\end{figure}

Upon setting in Eq. (60) $b\rightarrow 1/2$ and $a\rightarrow 2a^2$ (securing the positivity of the pertinent  parameter), we pass to the  "canonical"  form of the double-well potential,
\be
V(x)= {\frac{1}2} [\left(x^2 - a^2\right)^2 - a^4]= {\frac{1}2}(x^4 - 2a^2x^2)
\ee
graphically reproduced in Figs 13 and 14,   for selected values of $a$.

\subsection{ $\phi(x)= {\frac{1}2}(x^4 - 2a^2 x^2)$.}

\begin{figure}[h]
\begin{center}
\centering
\includegraphics [width=0.45\columnwidth, valign=c] {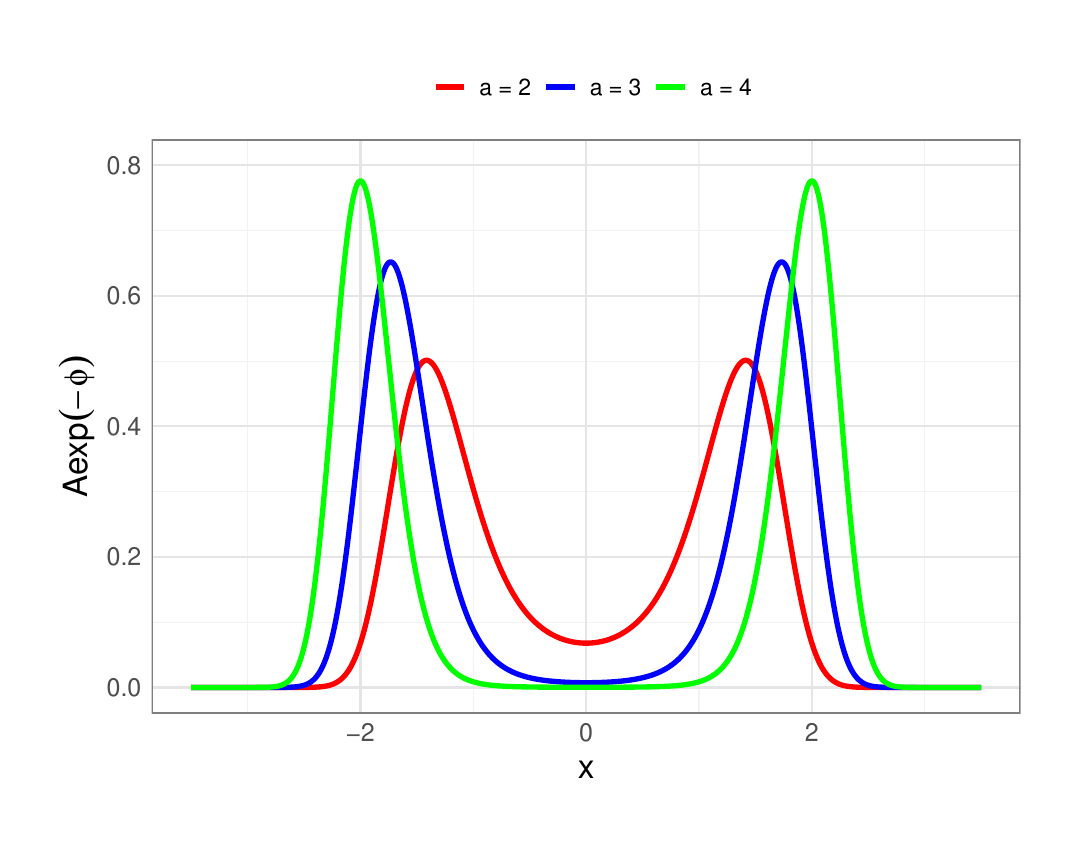}
\includegraphics [width=0.45\columnwidth, valign=c] {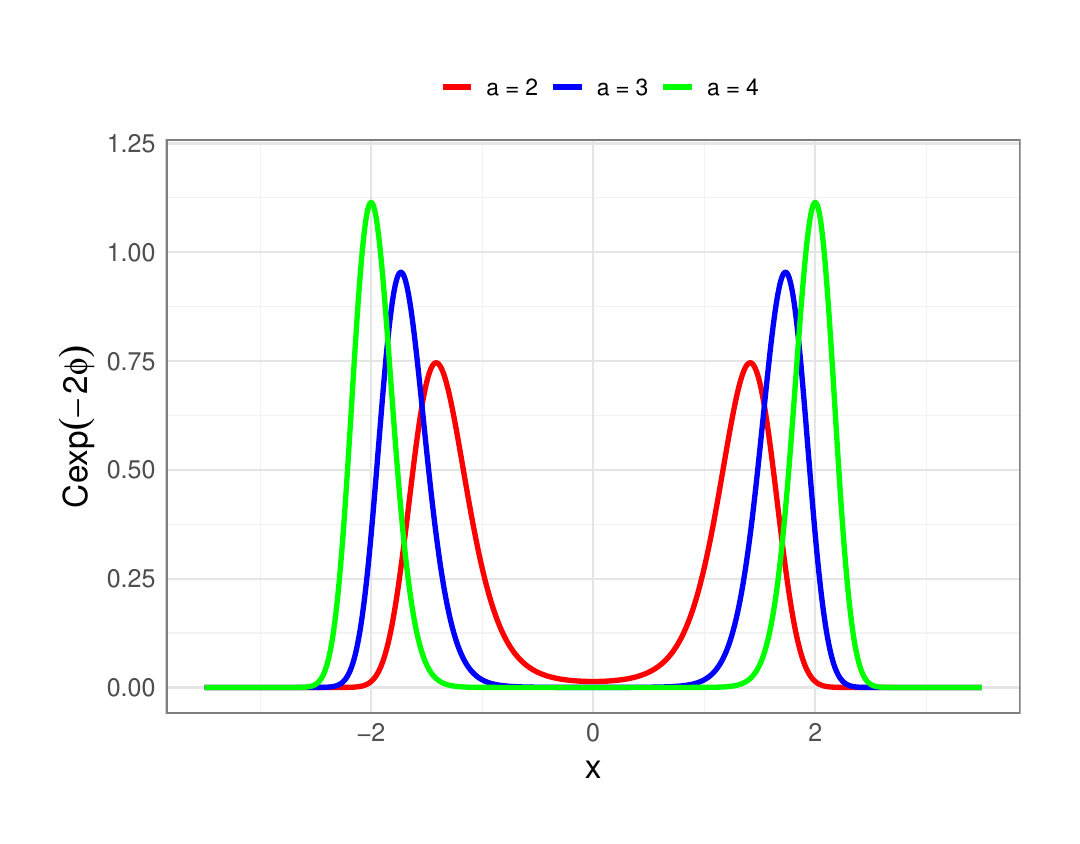}
\caption{$\phi = {\frac{1}2} x^4 - a x^2$, $a= 2, 3, 4$.  Left panel:  $L^1$ normalized $\psi (x)$; Right panel: $L^2$ normalized $\psi(x)$.  We indicate vertical scale differences between the panels.}
\end{center}
\end{figure}

The Fokker-Planck equation and its stationary regime, in the vein analogous to Section I of the present paper, has been analyzed in Ref. \cite{okopinska}.  For the drift potential of the double-well form, the corresponding effective potential ${\cal{V}}(x)$ arises in the sextic form. If adopted to our notation ($D\equiv \nu =1/2$), we have:
\be
\phi(x)= bx^4 - ax^2  \Longrightarrow
{\cal{V}}(x)= [ (8b^2) x^6 - (8ab) x^4 + (2a^2-6b) x^2 + a],
\ee
which for our choice of $\phi(x)= {\frac{1}2}(x^4 - 2a^2 x^2)$, i. e. $b=1/2$ and $a \rightarrow  a^2$, takes the form
\be
{\cal{V}}(x)=  2 x^6 - 4a^2 x^4  + (2a^4 -3) x^2 + a^2.
\ee
By arguments of Section I,  $\psi (x)\sim \exp[- \phi(x)]$  is the eigenfunction of  $H= -(1/2)\Delta + {\cal{V}}(x)$ assigned to the bottom eigenvalue zero.

In the simplified notation of Eq. (60), the  $L(\mathbb{R})$ normalization factor for  $\psi(x)= A \exp[- \phi (x)]$   comes from the integral $\int_R \psi(x) dx =1$. Accordingly:
\be
 \int _R \exp[-(bx^4 - ax^2)]dx = \left({\frac{a}{b}}\right)^{1/2}  {\frac{\pi }{2\sqrt{2}}}
 \exp(a^2/8b)  \left[B_{-1/4}\left({\frac{a^2}{8b}}\right)  + B_{+1/4}\left({\frac{a^2}{8b}}\right)\right]   = {\frac{1}A}   ,
\ee
where $B_{\gamma }(\xi)$ is the modified Bessel function of the first  kind.\\

Let us choose  $b=1/2$. Then, the values of the  normalization constant $A$  read: $a=2; A^{-1}= 14,7436$, $a=3, A^{-1} =138,083$, $a=4, A^{-1} = 3841,28$.\\

\begin{figure}[h]
\begin{center}
\centering
\includegraphics [width=0.45\columnwidth, valign=c] {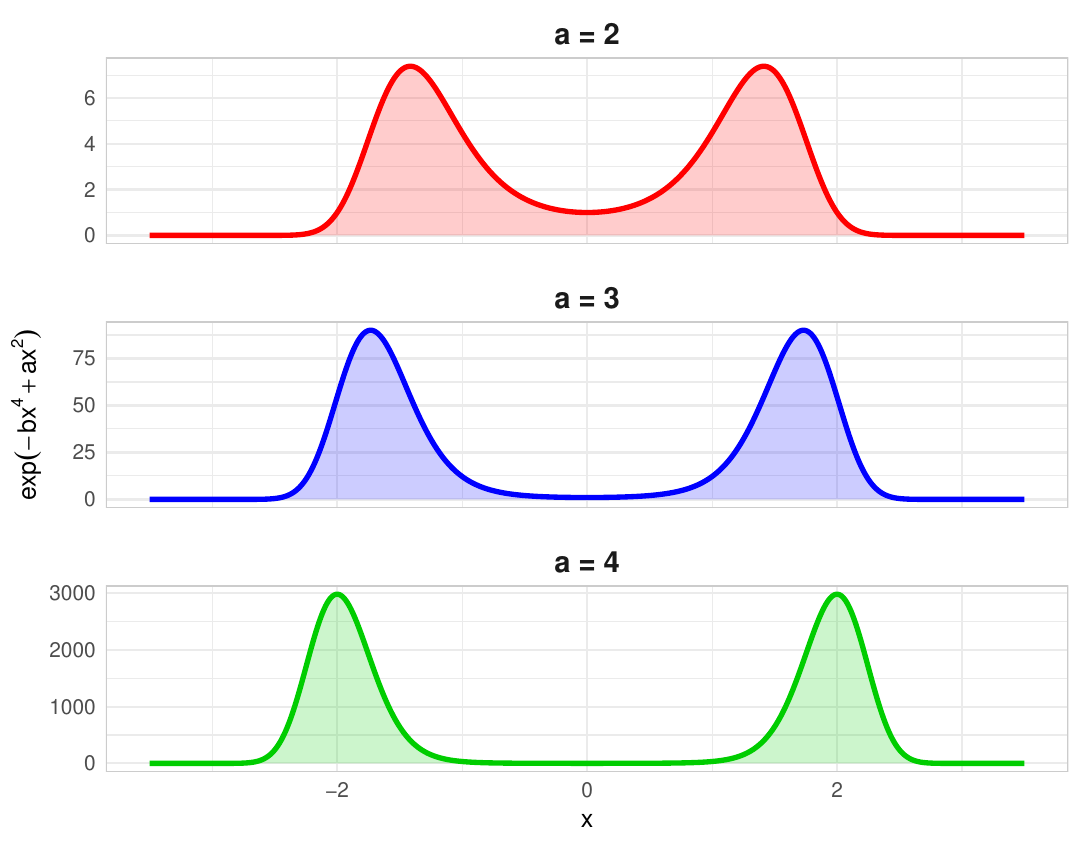}
\caption{The "bare" $\psi (x) = \exp[- \phi(x)]$, $\phi (x)= {\frac{1}2} x^4 -ax^2$, $a=2, 3, 4 $ devoid of  any normalisation factor.  Dramatic vertical  scale differences should be noted. Normalizations depicted in Fig. 16,  strongly push down  the maxima  of $\psi (x)$, while   moving  the minimum ($\psi (0)$  to a vicinity of $0$.}
\end{center}
\end{figure}

To deduce  $K(x)= \psi(0) \psi(x)$ we need the $L^2(R)$ normalized $\psi(x) $, and thence the $L(R)$ normalized $\psi^2(x) \rightarrow \rho_*(x)$.
This comes from the integral
\be
\int_R \exp [- 2\phi]  dx = \int_R \exp[(2a)x^2 - (2b)x^4) = {\frac{1}C}.
\ee
Then $\rho_*(x) = C \exp [- 2\phi] $. The corresponding $\rho ^{1/2}_*(x) = C^{1/2} \exp[-\phi (x)]$.

In analogy with our previous considerations,  cf. Section III.A,  we can introduce (with a tacit assumption $k(x,t) \rightarrow K(x)= \psi_1(0) \psi_1(x)$, where $\psi_1(x)\sim \exp [-\phi(x)]$:
\be
K(x)= \rho_*^{1/2}(0) \rho_*^{1/2}(x) = C \exp[-\phi(x)]   \Longrightarrow   K= \int_R K(x) dx = C/A
\ee
which ultimately allows to recover the  $L^1(R)$-normalized pdf $\rho_*^{norm}(x)$ from $\exp[- \phi (x)]$  alone
\be
\rho_*^{norm}(x) = {\frac{K(x)}K} = A \exp[- \phi (x)].
\ee
We realize that  $\rho _*^{norm}$ actually is the $L^1(\mathbb{R})$-normalized version of  $\exp[-\phi (x)]$, compare e.g. the left panel of Fig. 16.

\subsection{ ${\cal{V}}(x)={\frac{1}2}(x^4 - 2a^2 x^2 - \epsilon_1)$; Path-wise description of taming,  via the killing/branching perturbations of the free Brownian motion.}

\subsubsection{$k(x,t)$ in the unimodal  regime.}

Let us begin a discussion from the pure killing case of the Feynman-Kac discussion. Let us consider the  nonnegative  double-well potential  ${\cal{V}}(x) = (1/2)(x^2-1)^2$, see e.g. Fig. 14.    The obvious outcome of our simulations (we depict the number of alive trajectories at each recorded instant of time) is  the  continual killing of trajectories (as yet in existence):

\begin{figure}[h]
\begin{center}
\centering
\includegraphics [width=0.5\columnwidth, valign=c] {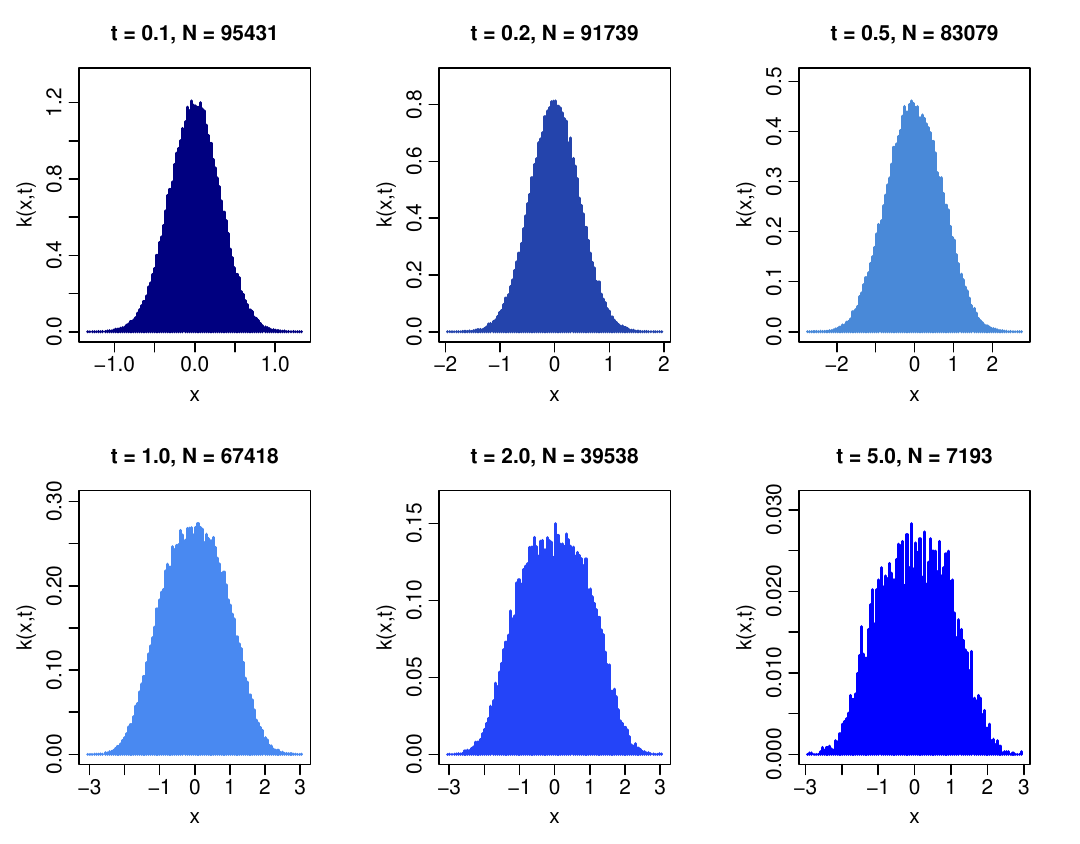}
\includegraphics [width=0.35\columnwidth, valign=c] {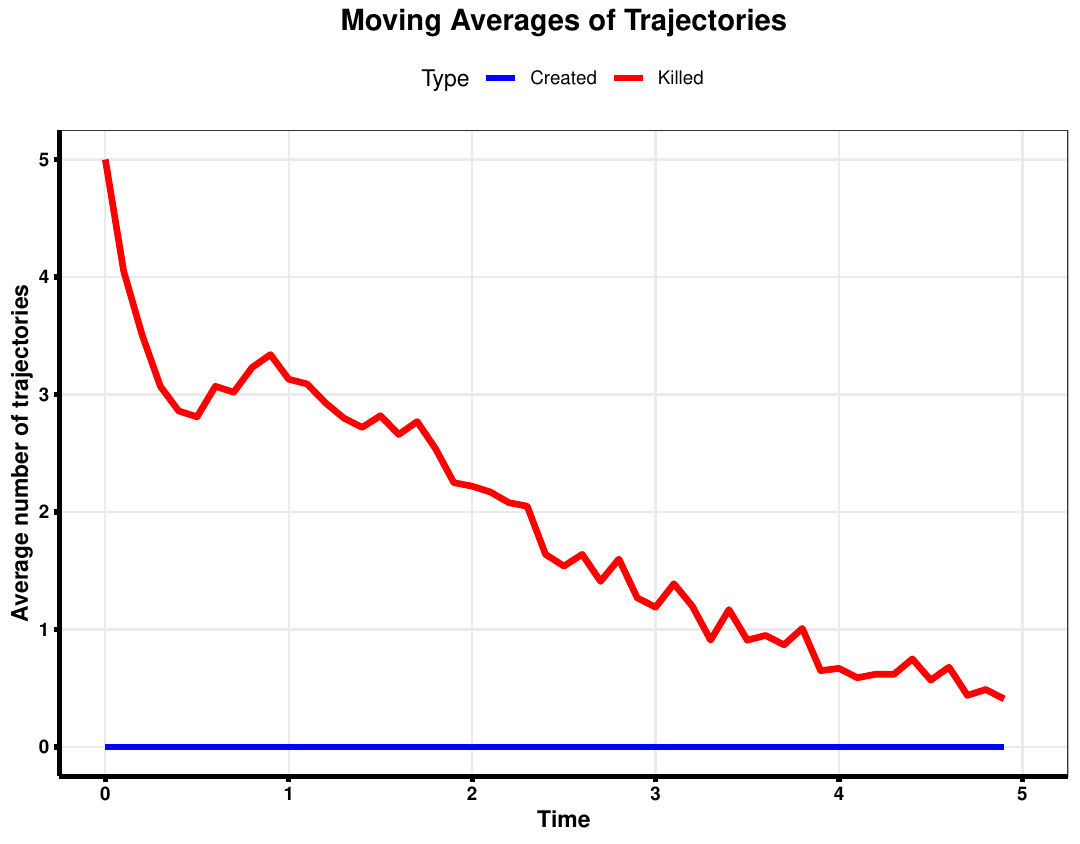}
\caption{$a=1$, ${\cal{V}}(x) \equiv V(x)= (1/2)(x^2-1)^2$, the spectrum of $H$ is positive-definite. The initial number of consecutively released from $x=0$ at time $t=0$ trajectories was $N(0)=10^5$. Left panel:  We have clearly visualized  the decay of $k(x,t)\rightarrow \exp(-t\epsilon_1/2)\cdot\psi_1(0)\psi_1(x)$, by depicted records $N(t)$ of still alive trajectories at times $0.1, 0.2, 0.5, 1.0, 2.0, 5.0$.  Note changes of vertical scales, from figure to figure; Right panel: Pure killing in terms of moving averages.}
 \end{center}
\end{figure}

On the basis of our previous considerations, we anticipate the  presence of the  undoubtful taming behavior  of the Feynman-Kac diffusion,  by passing to  potentials "with subtraction". These secure the bottom eigenvalue $0$ for $H= -(1/2)\Delta + {\cal{V}}(x)$, and by construction have  disjoint branching (negativity) and killing (positivity) subdomains in $\mathbb{R}$.

 \begin{figure}[h]
\begin{center}
\centering
\includegraphics [width=0.45\columnwidth, valign=c] {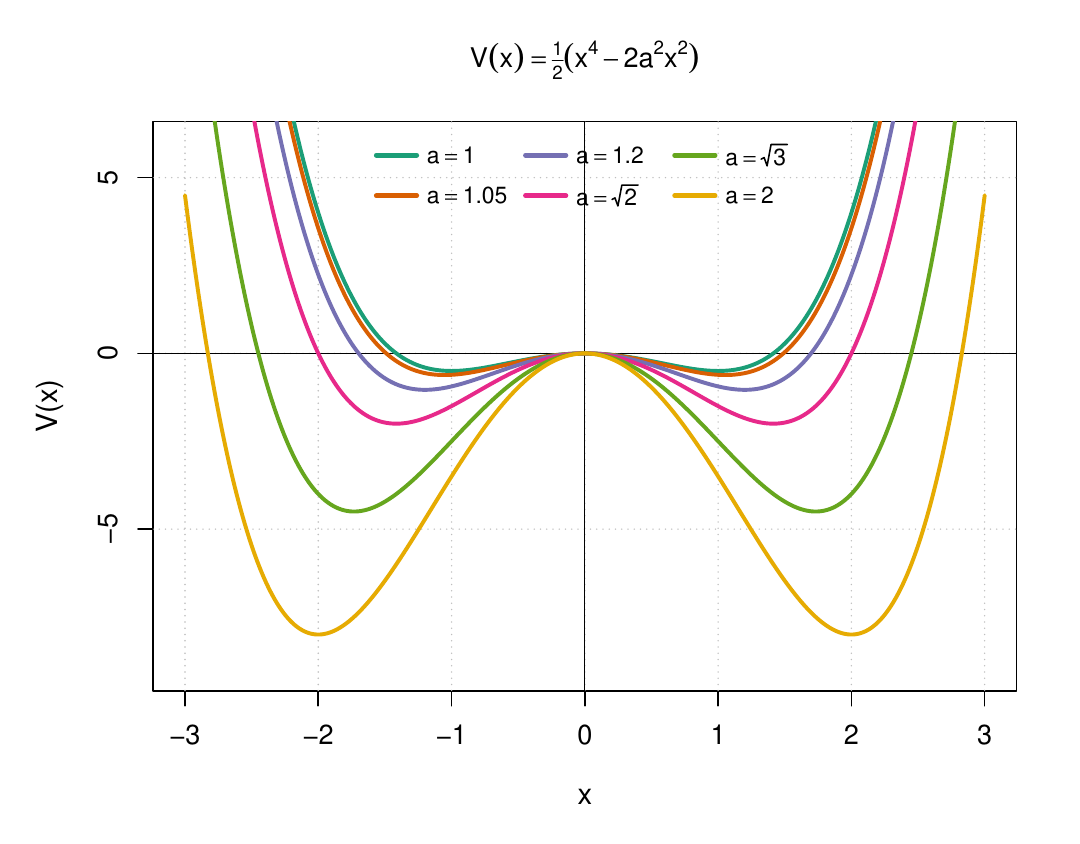}
\includegraphics [width=0.45\columnwidth, valign=c] {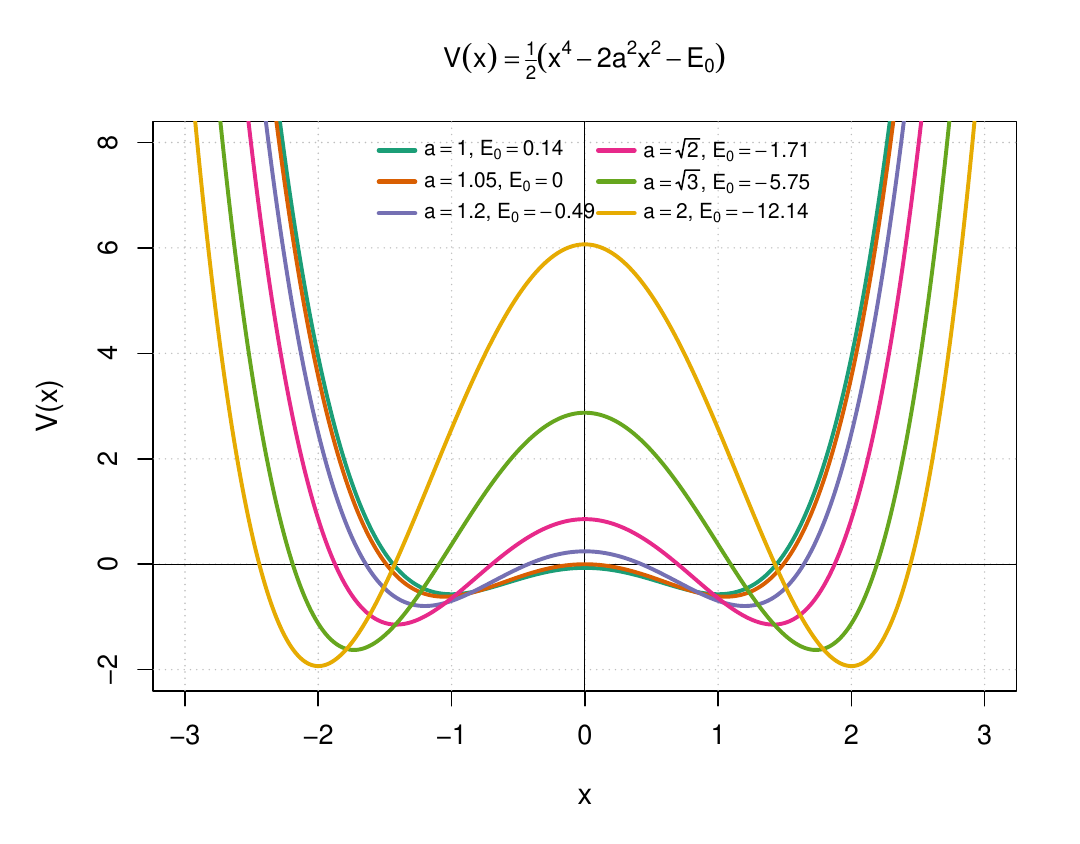}
\caption{The expanded list of potentials   $V(x)= {\frac{1}2}(x^4-2a^2 x^2)$, together  with the  bottom eigenvalues $E_0$ of $H_0= - \Delta + (x^4-2a^2x^2)$ (originally evaluated for the case of $\nu =1$, three more in each panel), complementing those reproduced in Fig. 14. Potentials "with subtraction"   actually  refer to  $H=- (1/2)\Delta + {\cal{V}}(x)$, with   ${\cal{V}}(x)= V(x) - \epsilon_1$, where $\epsilon_1= E_0/2$.  In the figure, we keep the notation $E_0$ instead of $2\epsilon_1= E_0$.}
\end{center}
\end{figure}

The quartic Feynman-Kac potential does not arise in straightforward way in the standard  Fokker-Planck formalism. In Ref. \cite{zaba}, in  Section 5 (see e.g. Figs (11) and (13)), we have demonstrated that the Langevin-driven Brownian motion (actually its invariant pdf, the drift and the ground state function $\rho _*^{1/2}(x)$) can be reconstructed  via the spectral analysis of  $H=-\Delta + [V(x) - \epsilon_1]$, where  $V(x)= x^4- 2a^2 x^2$, with $a = 1, 1.2, 2$  and $a_{critical} \sim 1.053 4677$, (here $D=1$).  The bottom eigenvalues of $H_0 = -\Delta + V(x)$ have been computed, together with corresponding eigenfunctions. The  eigenvalue $\epsilon_1$ has  been  found to take negative values  for $a=1.2$ and $a =2$ ($\nu =1$ in that analysis).  The   reconstruction procedure has been  accomplished  by means of  the computer assistance.

 \begin{figure}[h]
\begin{center}
\centering
\includegraphics [width=0.5\columnwidth, valign=c] {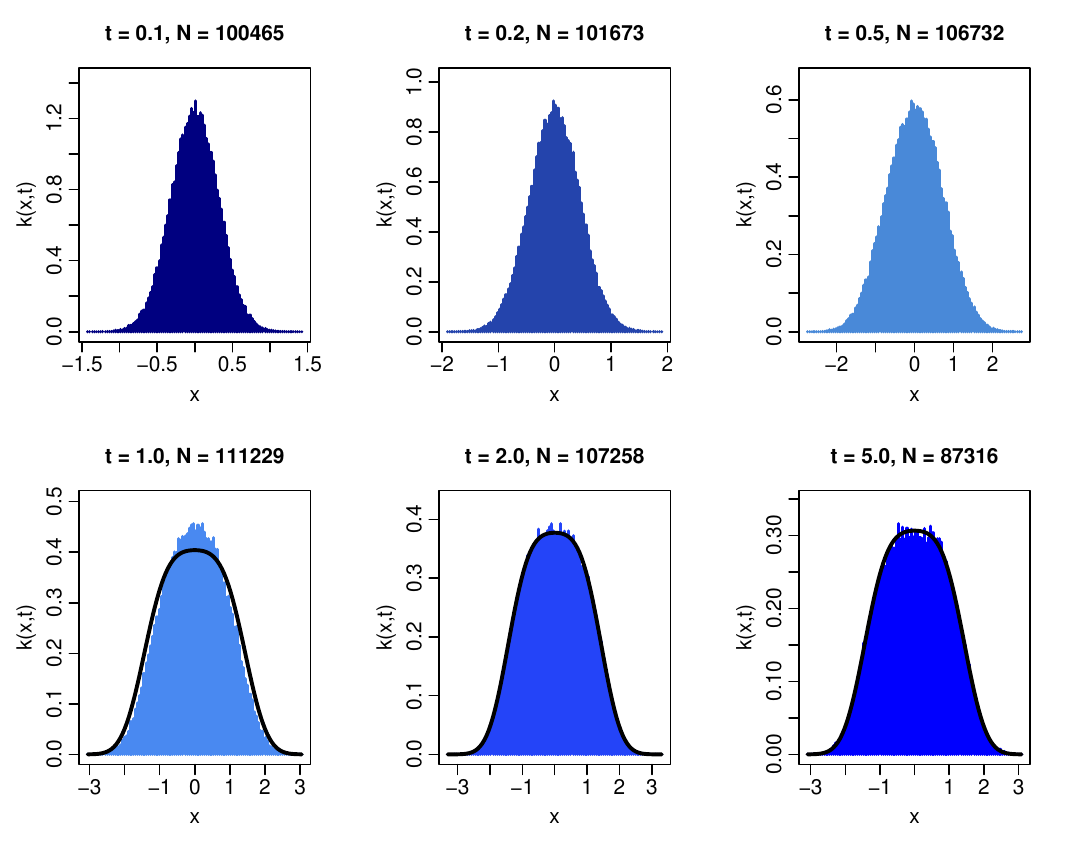}
\includegraphics [width=0.36\columnwidth, valign=c] {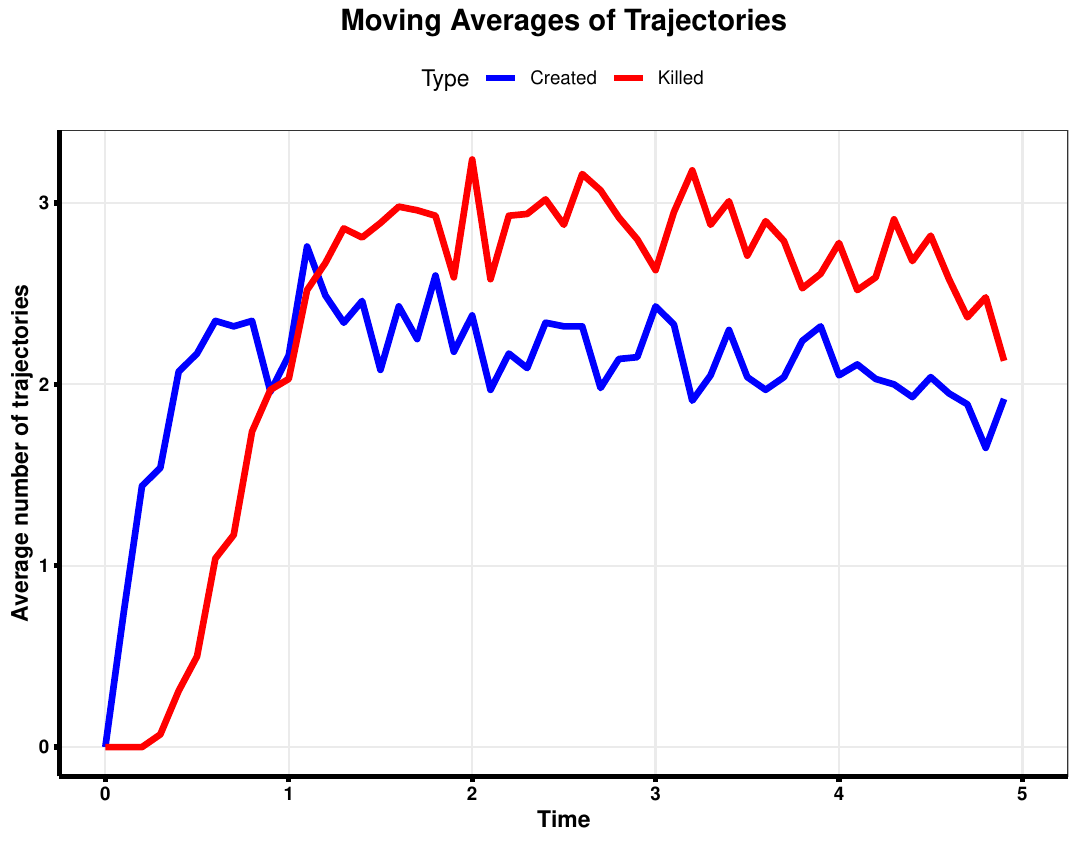}
 \caption{The "canonical" version of the quartic potential  (61) for  $a=1$: ${\cal{V}} \equiv V(x) = (1/2)[(x^2 -1)^2 -1]= (1/2)(x^4 - 2 x^2)$. The bottom eigenvalue of $H$  (computed via Strang splitting method) reads  $\epsilon _1 = E_0/2;  E_0=0.137786$. Left panel: Slow decay of $k(x,t) \rightarrow  \exp(-\epsilon_1 t/2)  \psi_1(0) \psi_1(x)$.  Right panel: The asymptotic  decay of $k(x,t)$ in terms of running averages.  The $-1/2$  shift  ($1/2$ is  not  the bottom eigenvalue of $H_0$)  of the potential of Fig. 14  to the form depicted in the left panel of Fig 15 was insufficient to achieve the relaxation regime.}
\end{center}
\end{figure}

 \begin{figure}[h]
\begin{center}
\centering
\includegraphics [width=0.5\columnwidth, valign=c] {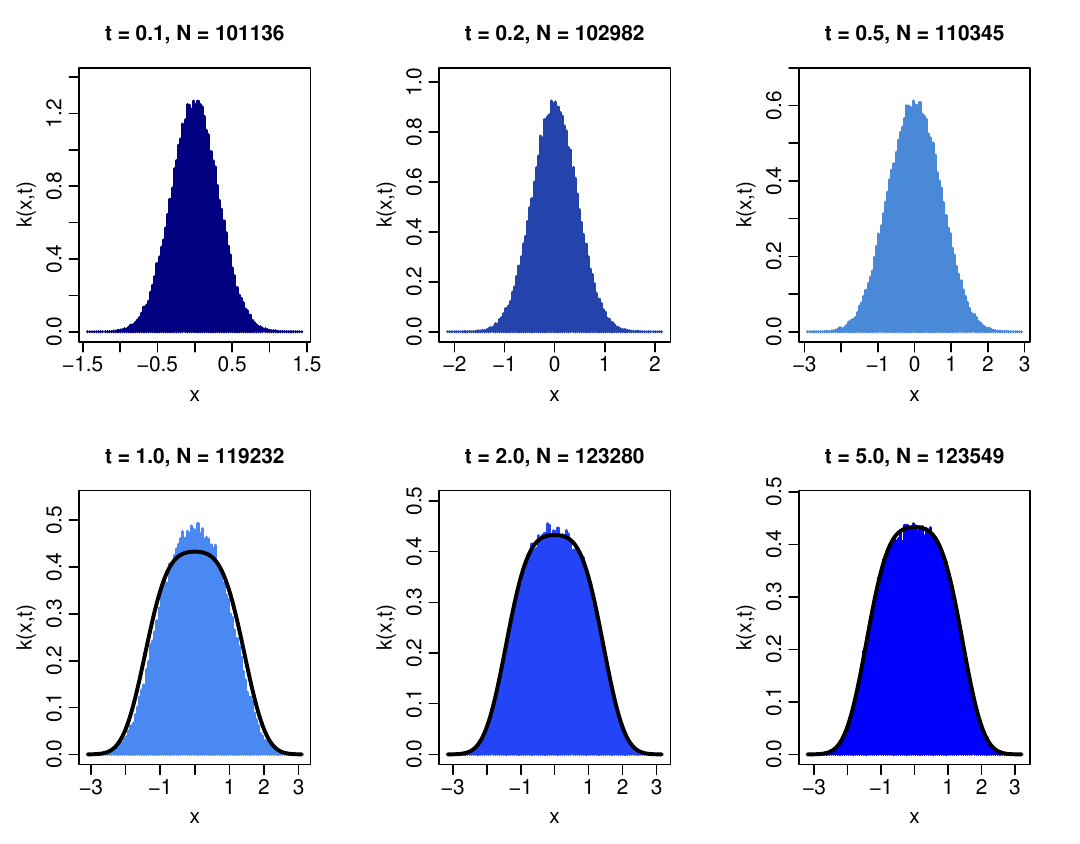}
\includegraphics [width=0.35\columnwidth, valign=c] {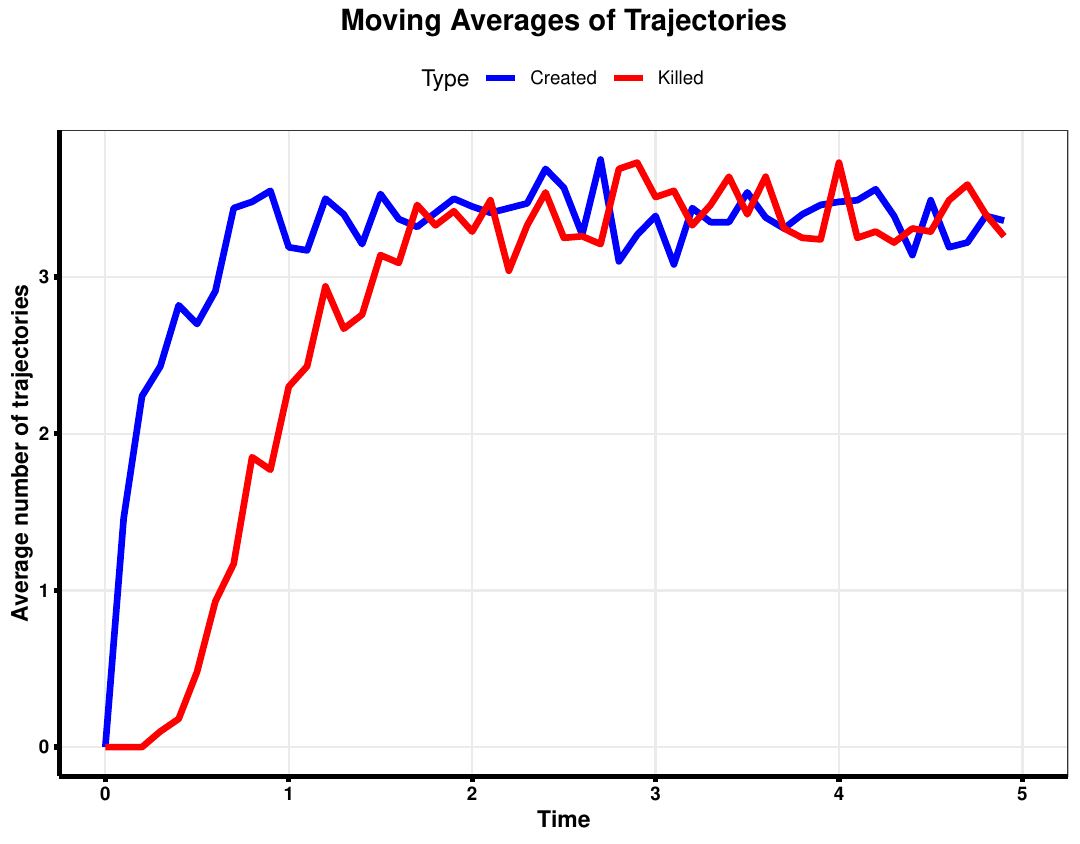}
 \caption{${\cal{V}}(x)$ "with subtraction", for $a=1$ and $E_0 = 0.137786$: ${\cal{V}}(x)= (1/2)(x^4 - 2 x^2  - E_0)$.  Left panel:  Equilibration  of $k(x,t) \rightarrow K(x)=\psi_1(0) \psi_1(x)$, with the stabilization of the number$N(t)$ of alive trajectories at $N \sim 123 500$.  Right panel: The stabilization of  $k(x,t)$ in terms of running averages.}
\end{center}
\end{figure}

We point out that the renormalized operator $H= H_0 - (1/2)\epsilon _1$,  with  $H_0= -(1/2) \Delta + V(x)$,  in our   further  considerations,   contains the effective potential "with subtraction"    ${\cal{V}}(x)= V(x) - (1/2)\epsilon_1$, irrespective of whether the bottom eigenvalue of $H_0$ is positive or not.

From now on, we take the quartic potential $V(x)={\frac{1}2}(x^4 - 2a^2 x^2)$ as  an a priori candidate  to define the   legitimate  Feynman-Kac entry ${\cal{V}}(x) = V(x)- (1/2) \epsilon_1$, i.e. the "potential with subtraction".  This  motivates somewhat   closer  analysis of the path-wise killing/branching scenario, following the lines of Sections II and III. We consider double-well potential examples, listed (together with bottom eigenvalues $\epsilon_1$  of $H$ in Fig. 19.

 \subsubsection{$k(x,t)$ in the bimodal regime.}

   \begin{figure}[h]
\begin{center}
\centering
\includegraphics [width=0.5\columnwidth, valign=c] {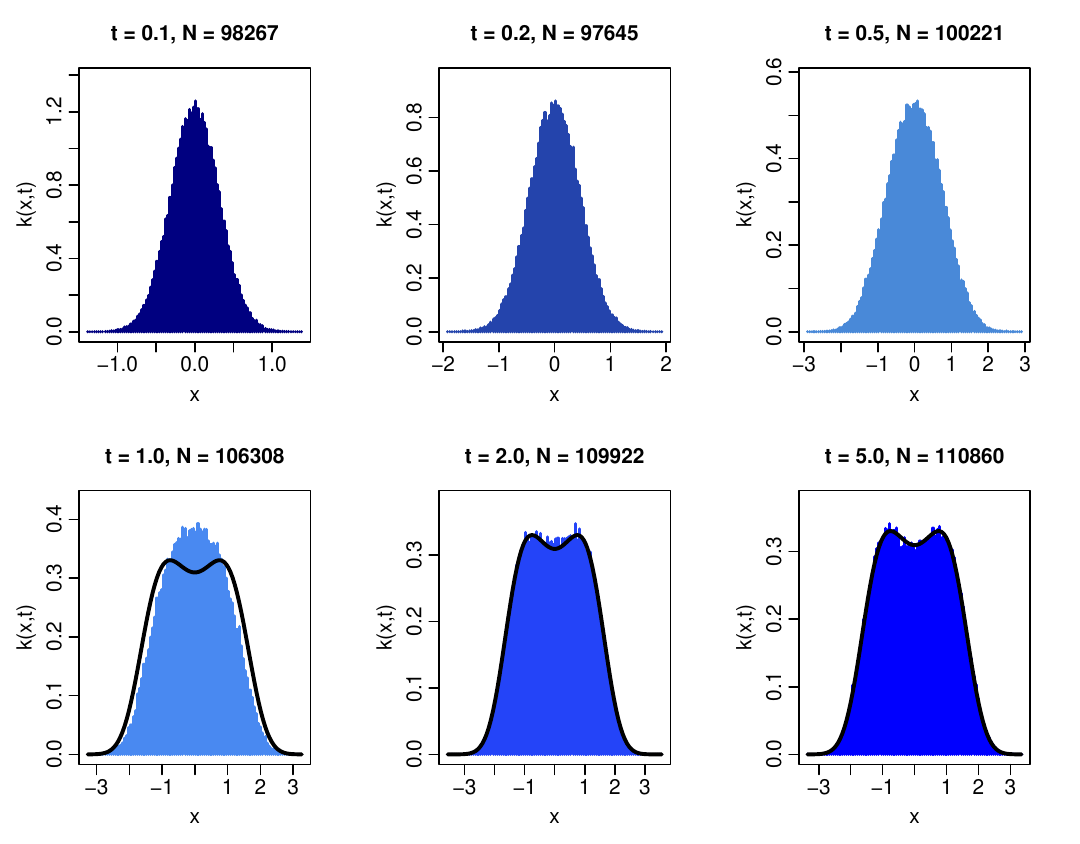}
\includegraphics [width=0.35\columnwidth, valign=c] {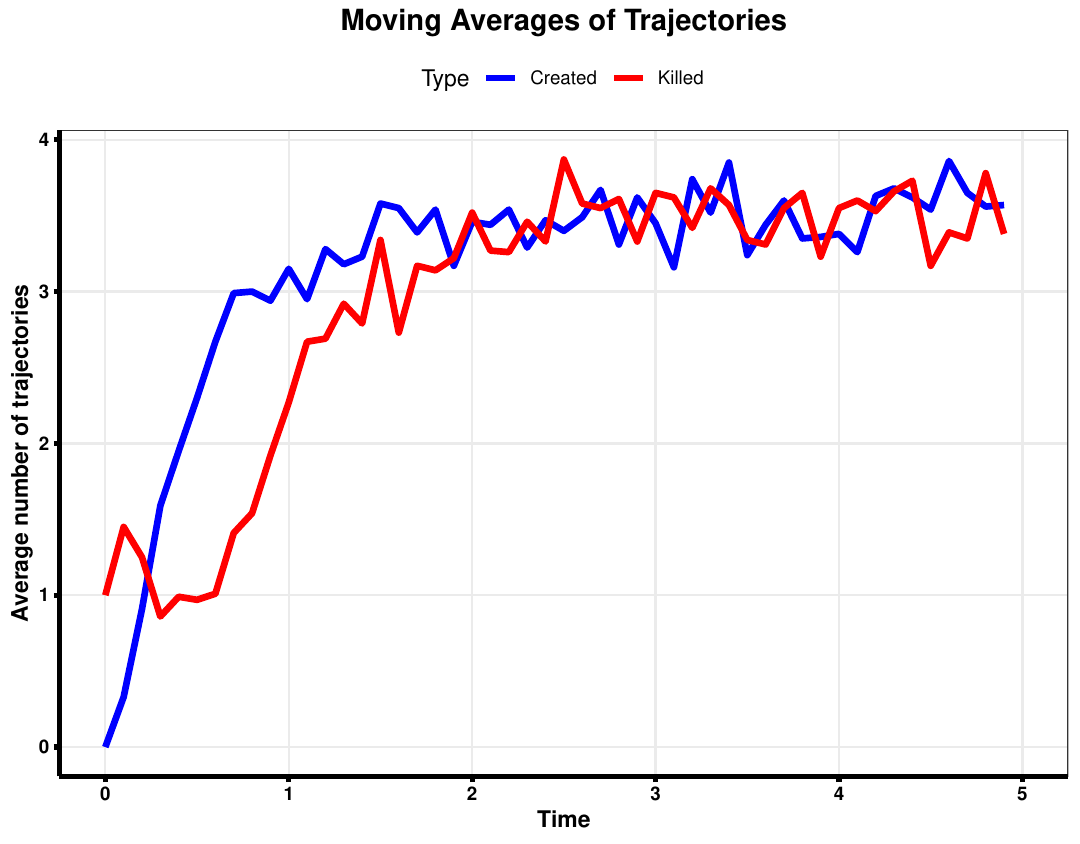}
\caption{Bimodality signatures at equilibrium for $a=1.2$ and $E_0 =  -  0.489604$:  ${\cal{V}}(x) = (1/2)(x^4 - 2.88 x^2  - E_0)$.  Left panel:  Equilibration  of $k(x,t) \rightarrow K(x)= \psi_1(0) \psi_1(x)$, with the stabilization of the number $N(t)$  at $N \sim 110 000$.  Right panel: The killing-branching interplay  in terms of running averages.}
\end{center}
\end{figure}

Unimodality of $k(x,t)$   can be considered as a signature of the existence of a  positive
bottom eigenvalue for the reference potential  $V(x)=(1/2)x^2(x^2-2a^2)$  in $H_0= -(1/2)\Delta + V(x)$.     For the transitional  value $a_{critical}= 1.0534677$ (separating topologically different unimodal and bimodal shape regimes for eigenfunctions),  established in  \cite{turbiner2}, see also \cite{zaba} (Fig. 12), the bottom eigenvalue of $H_0$  equals zero. Hence no "subtraction" is necessary to achieve the Feynman-Kac equilibration. This we have tested numerically, and qualitatively the results do not significantly differ (except for a faster approach to stabilization in the asymptotic number of alive trajectories), from the previously considered $a=1$ case.

To elucidate the emergence of the bimodal regime,  we shall refer to the $a=1.2$ case of Fig. 19. In this case the bottom eigenvalue of $H_0$ reads  $\epsilon_1 =- 0,489604$. The operator
$H= H_0 + (1/2) \epsilon_1$ has the eigenvalue zero, and  the Feynman-Kac equilibrium regime for $k(x,t)$ shows signatures of bimodality.

  \begin{figure}[h]
\begin{center}
\centering
\includegraphics [width=0.5\columnwidth, valign=c] {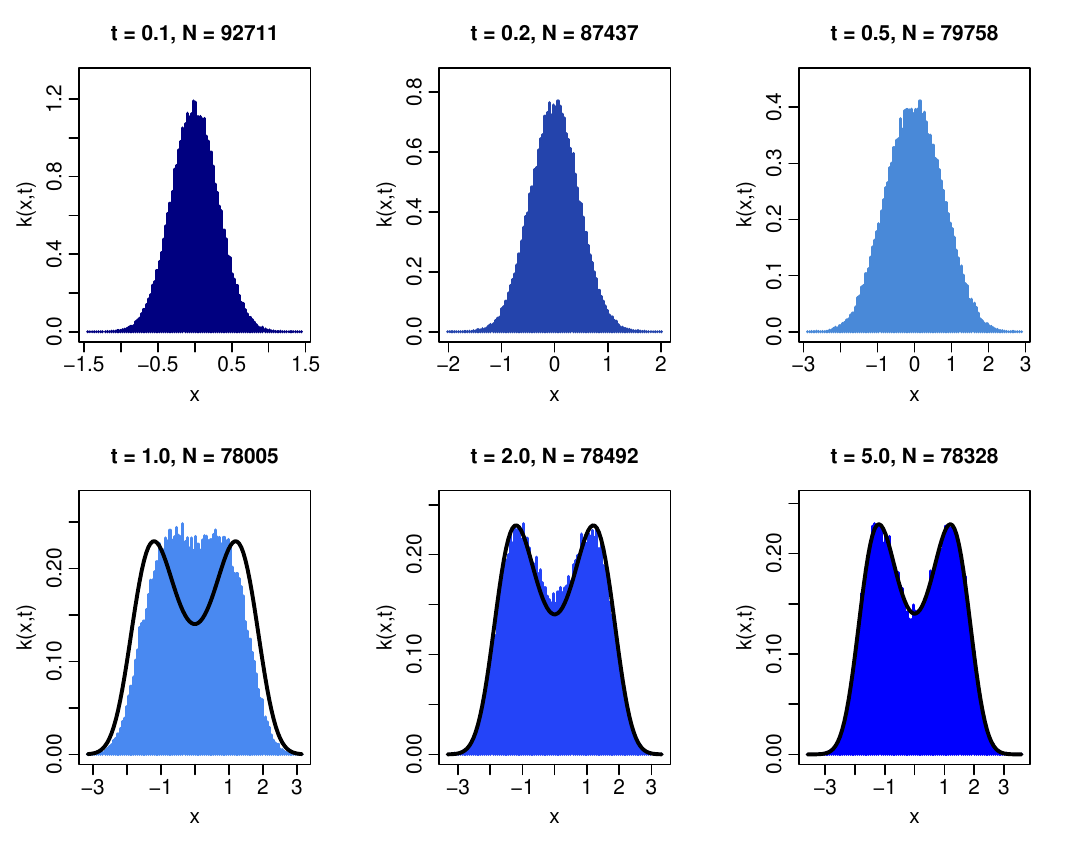}
\includegraphics [width=0.35\columnwidth, valign=c] {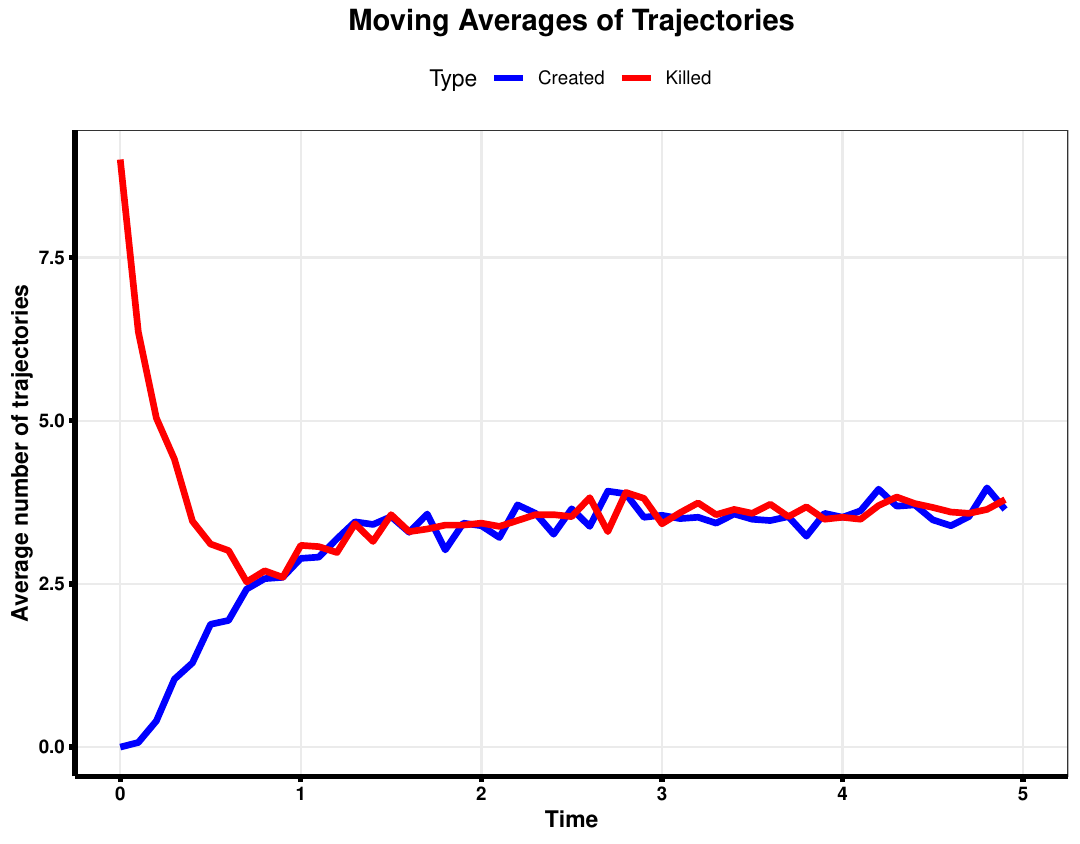}
\caption{Bimodality signatures at equilibrium for $a=\sqrt{2}$ and $E_0 =  - 1.710351$:  ${\cal{V}}(x) = (1/2)(x^4 - 4 x^2  - E_0)$.  Left panel:  Equilibration  of $k(x,t) \rightarrow K(x)= \psi_1(0) \psi_1(x)$, with the stabilization of the number  $N(t)$ of alive trajectories at $N \sim 78 500$.  Right panel: The killing-branching interplay  in terms of running averages.}
\end{center}
\end{figure}

  \begin{figure}[h]
\begin{center}
\centering
\includegraphics [width=0.5\columnwidth, valign=c] {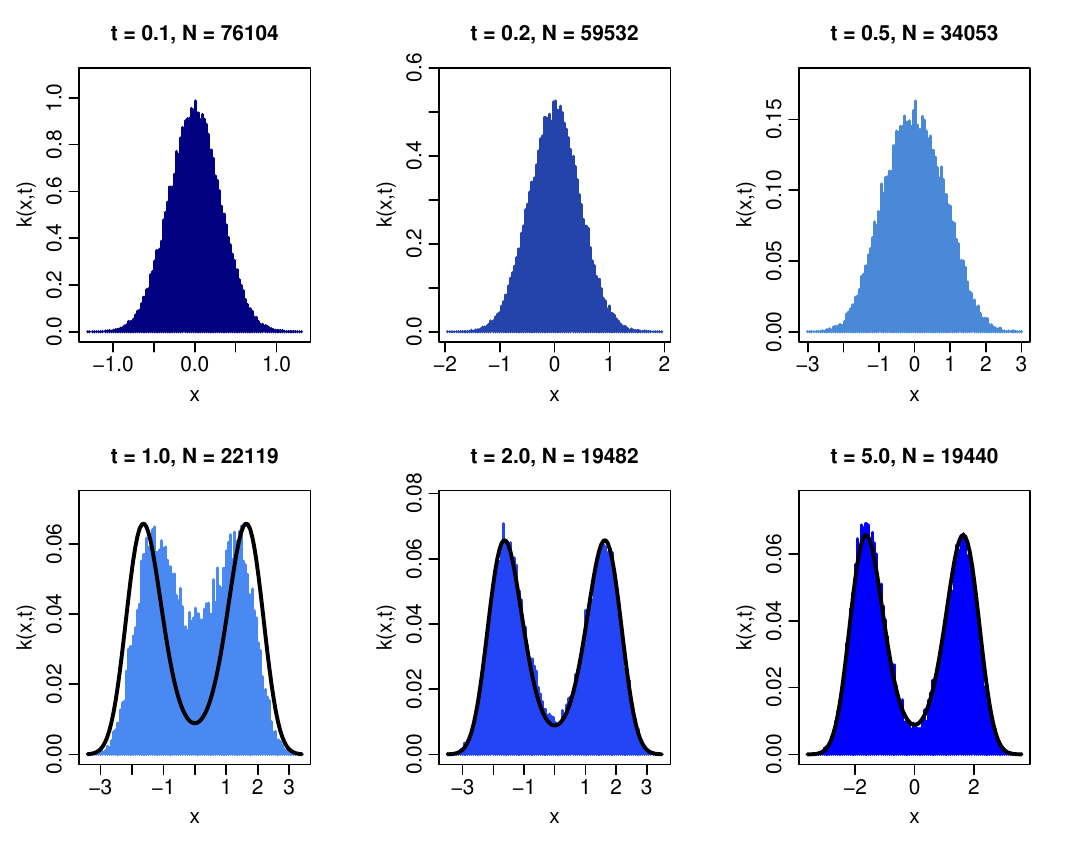}
\includegraphics [width=0.35\columnwidth, valign=c] {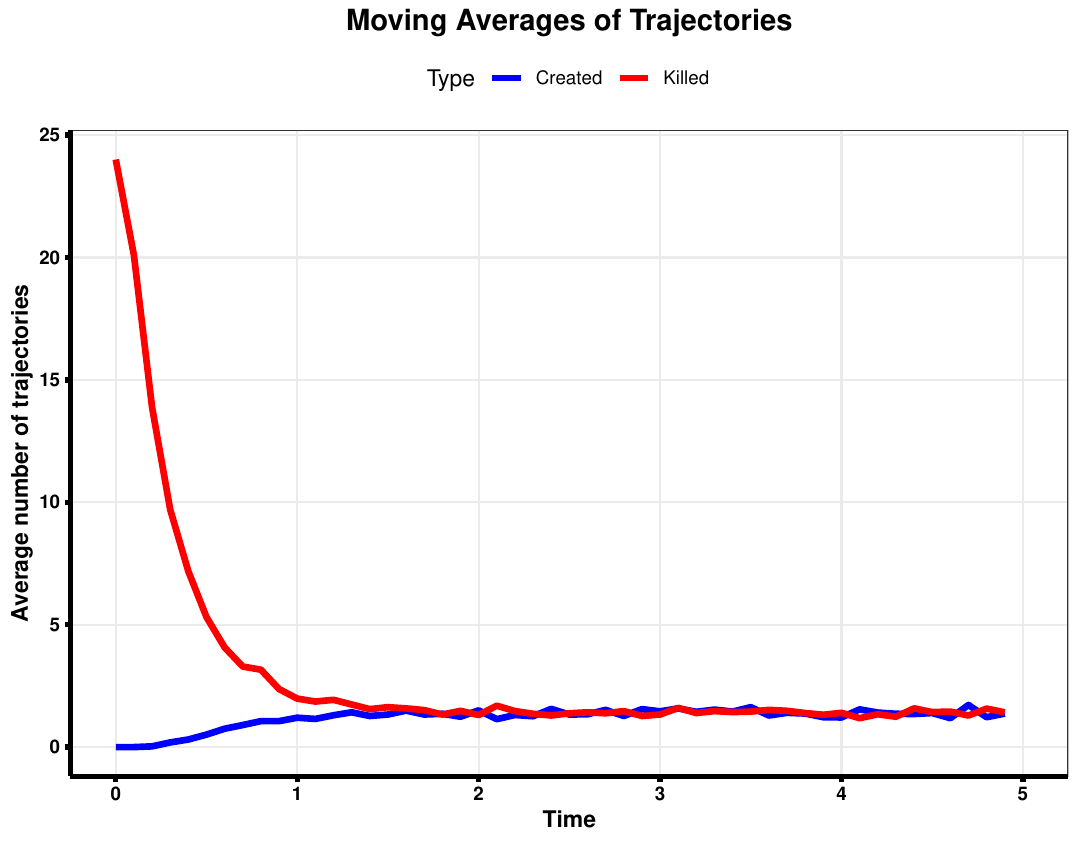}
\caption{Bimodality signatures at equilibrium for $a=\sqrt{3}$ and $E_0 =  - 5.748191 $:  ${\cal{V}}(x) = (1/2)(x^4 - 6 x^2  - E_0)$.  Left panel:  Equilibration  of $k(x,t) \rightarrow K(x)= \psi_1(0) \psi_1(x)$, with the stabilization of the number $N(t)$ at $N \sim 19 500$.  Right panel: The killing-branching interplay  in terms of running averages.}
\end{center}
\end{figure}

 \begin{figure}[h]
\begin{center}
\centering
\includegraphics [width=0.5\columnwidth, valign=c] {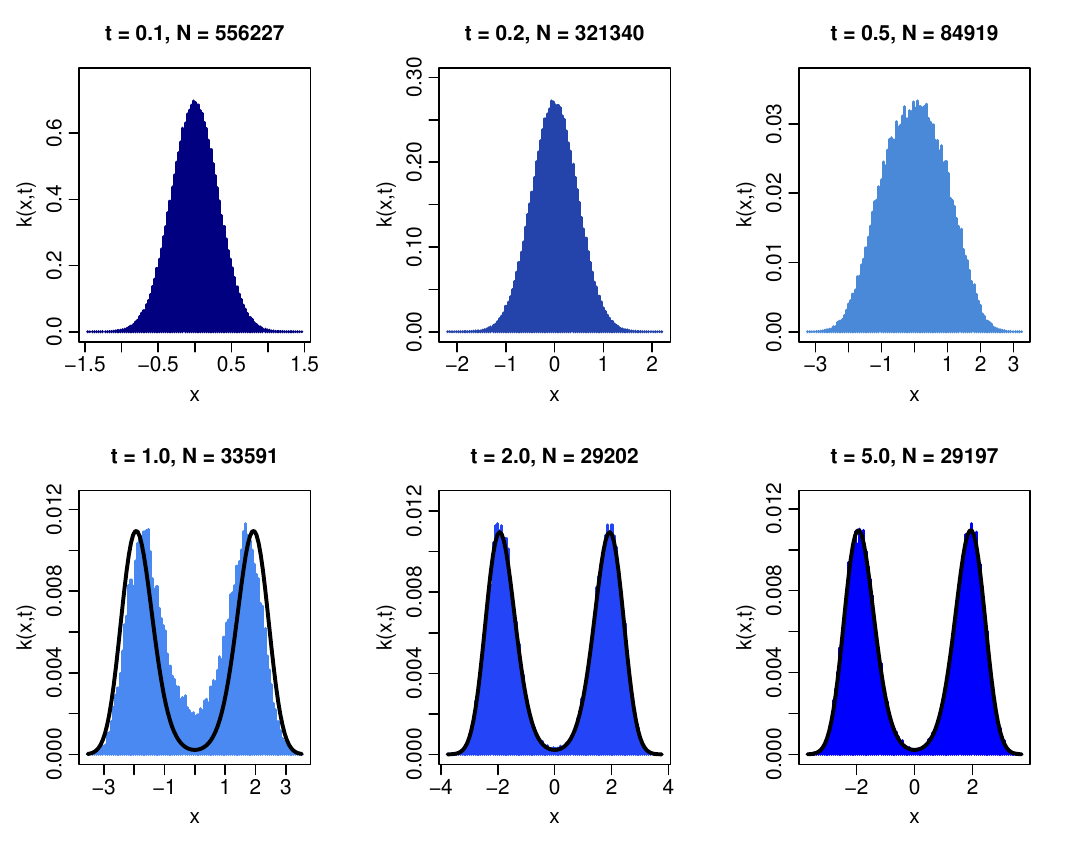}
\includegraphics [width=0.35\columnwidth, valign=c] {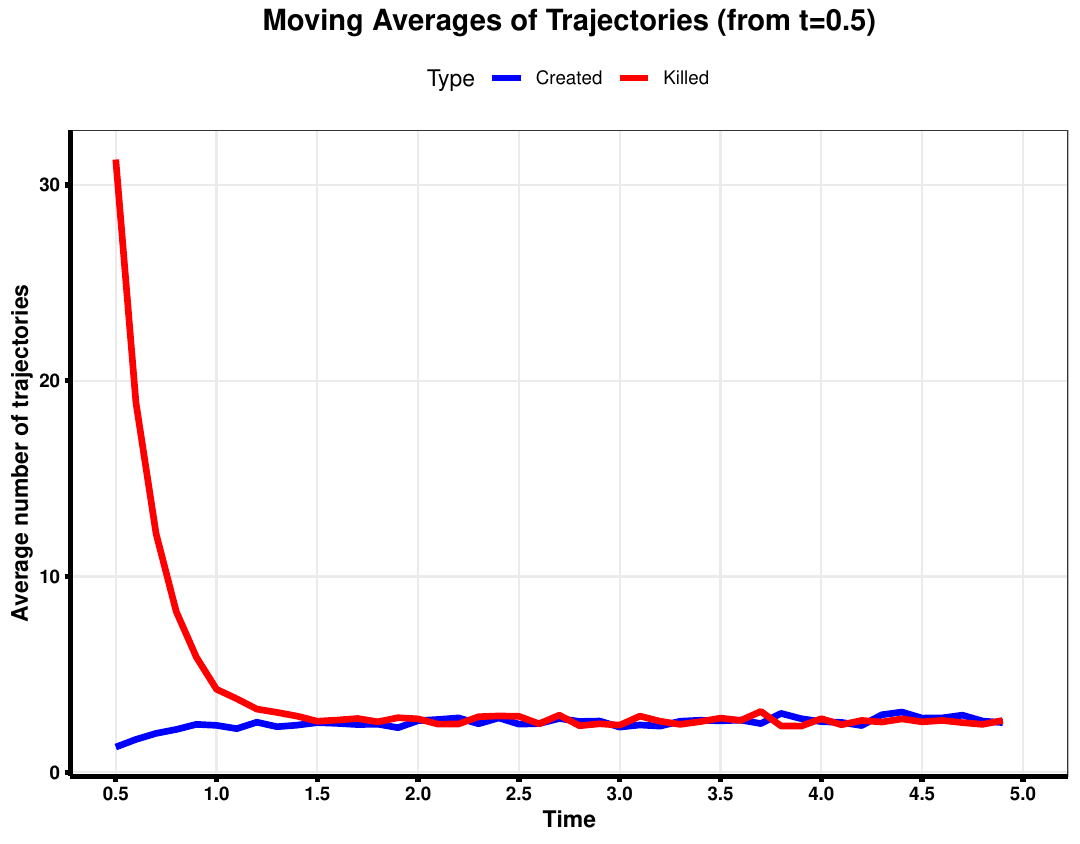}
\caption{The case of $a=2$ and $E_1= -12.1363$ :  ${\cal{V}}(x) = (1/2)(x^4 - 8 x^2  - E_0)$.   Killing effects are initially so strong that  alive trajectories do not form a reliable statistics, while starting from $N(0)=10^5$ (actually, at time $t=5$  we were left with about $3000$ alive trajectories).  To obtain a reliable asymptotic  statistics, we take  $N(0)= 10^6$.  Left panel:  Equilibration  of $k(x,t) \rightarrow K(x)= \psi_1(0) \psi_1(x)$, with the stabilization of the number $N(t)$ of alive trajectories at $N \sim 29 000$.  Right panel: The killing-branching interplay  in terms of running averages,  for clarity  depicted from time $t=0.5$.}
\end{center}
\end{figure}

\subsection{Euclidean connotations: Where have  the instantons gone ?}

\subsubsection{Splitting of bottom energy levels (energy gap) in the  double-well. Deeply non-perturbative  regime.}

In the present section  of the paper  we refer to Hamiltonian operators $H_0= - {\frac{1}2}
\Delta + V(x)$,  with   $\nu = 1/2$ and  the  double-well potential of the form $V(x)= (1/2)(x^4 - 2a^2x^2)$. To pass to $H= {\frac{1}2}[-
\Delta + (x^4 - 2a^2x^2 - E_0)]= -{\frac{1}2}\Delta   + {\cal{V}}(x) $, with $V(x) - \epsilon_1 = {\cal{V}}(x)$, we must know the bottom eigenvalue of $H_0$.   This can be accomplished by means of the Strang splitting method, invoked in the present paper before \cite{zaba,jmp14}.

Before, we have adopted the Strang algorithm to spectral solutions of the  superharmonic  Hamiltonians with $\nu =1$.  Since our $H_0= {\frac{1}2}[- \Delta + (x^4 - 2a^2x^2)]$ differs form the  $\nu = 1$ case merely by an overall multiplication by $1/2$, we realize that the spectral data $E_k$ for $\nu =1$  can be rewritten as the spectral data $\epsilon _k = E_k/2$ for $\nu = 1/2$.  Resulting  bottom eigenvalues were reported in Figs. 15 and 19.

Since it is of interest to know the bottom energy gap in the double well problems, we have employed the Strang method to evaluate first excite (odd) state eigenvalue for each considered case.
We present the computation  outcomes  for the case of $\nu =1$. It suffices to divide them by $2$ to pass to the case of $\nu =1/2$.  \\

\noindent
$a=1; \, E_0=0.137786, \, E_1= 1.713028; \,  \Delta E = 1.575242$,\\
$a=1.2: \, E_0=-0.489604, \, E_1= 0.551566;  \,  \Delta E = 1.04117$,\\
$a=\sqrt{2}, \, E_0= - 1.710351, \, E_1= - 1.247923; \, \Delta E = 0.462428$,\\
$a= \sqrt{3}, \, E_0= -5.748191, \, E_1= - 5.706793; \, \Delta E = 0.041398$,\\
$a=2, \,  E_0=- 12.13630, \, E_1= -12.13481; \, \Delta E =  0.00149$.\\

The bottom levels splitting drops down  surprisingly fast, with the growing  impact  of the negative quadratic term in $V(x)$.

\subsubsection{Instanton as a misnomer versus the lure of instantons.}

It is nowadays a widely accepted routine  to employ  the "Euclideanization" of otherwise intractable (mostly) quantum models.  The  double-well  spectral problem, specifically an issue of the bottom levels splitting, has been addressed  by means of the so-called instanton calculus, which belongs to the standard    Euclidean path integral   inventory. Its various technical aspects are  covered in detail in numerous research papers and  monographs, c.f.  a sample \cite{grosche}-\cite{instanton}. Compare e.g. also \cite{feynman,barton,gar1}.

However, in the present paper, not only the term "euclideanization", or   an explicit setting  of the  Euclidean classical Lagrangian  against its "normal" (e.g. non-Euclidean) version,  and  as well the habitual   phrase "Euler-Lagrange equations in the Euclidean form", see eg. Sections  I.C  and II.B, appear to be a   misnomer.  In this connection, we refer to section V of \cite{gar1} entitled "the illusion of imaginary time".

It is true that a celebrated   text-book  Wick rotation,  represented by a Euclidean map $\exp(-itH_{quant}) \rightarrow  \exp (-tH_{Eucl})$,   executes the transformation  of the   "real time" quantum model into the  corresponding model "in Euclidean time" (with the semigroup dynamics replacing  the unitary one).

Such fairly crude  reasoning, except for deceiving   resemblances on the formal level, does not apply to our discussion  in the present paper,  which is kept in the entirety on the level of stochastic processes, with a  manifestly {\it real}  time clock.    Actually, at no point any Euclidean mapping has been involved and no mappings between entirely distinct (Euclidean vs non-Euclidean)  models  of "anything"  are involved.

One should not be deceived by  the  routine Euclidean lore, when in Section II.B  we explicitly solve the "Euclidean  equation of motion", and  for clarity of exposition  in Figs 1, 14, and  15  we present model curves together  with their "Euclidean  (e.g. inverted)  partners".

 We point out that our "Euclidean trajectory input"     (in terms of Euclidean classical paths)  to the action $S = S(y,0,x,t) = \int_0^t [{\frac{1}2}(\dot{x}^2 + x^2)] d\tau $, c.f. Eqs (20-25)  has involved classical solutions of the standard Euler-Lagrange equations (19), (20), which were introduced  as the direct consequence  of the  formulas (10)-(12). There, the weighted  Feynman-Kac kernel  $k(y,s,x,t)$ has been  associated  with the transition probability density  $p(y.s.x.t)$ of the diffusion process.  The Lagrangian formulation of the path integral   (11), has been the crucial step in the whole analysis of Section I.C.

 The time label throughout the paper remains  exclusively  in $\mathbb{R}_+$, and never refers to any time symmetric (like e.g. $t\in [-T,+T]$  or $t\in \mathbb{R}$)  evolution. Moreover, we are interested in  finite time scenarios beginning from $t_0\geq 0$.

  This  underlies the usage  of $k(x,t)$  where the evolution   refers to $t\in\mathbb{R}_+$,   and we are interested in the asymptotic  $(k(x,t) \rightarrow K(x)$. Even beyond the quadratic  (harmonic)  case, we can try to figure out the contribution  of classical (Eq. (20)) paths to  action $S(x,t)$ in  the  formulas (23-27).
  In particular, Eqs. (24) and  (27) tells us that the   time evolution of $k(x,t) \sim \exp[- S(x,t)]$, with $S(0,0,x,t)=S(x,t) = {\frac{x^2}2} \coth t$  is determined  by that of  the  solution of  (Euclidean-looking)  Eq. (20), see e.g. Fig. (26).

  \begin{figure}[h]
\begin{center}
\centering
\includegraphics [width=0.45\columnwidth, valign=c] {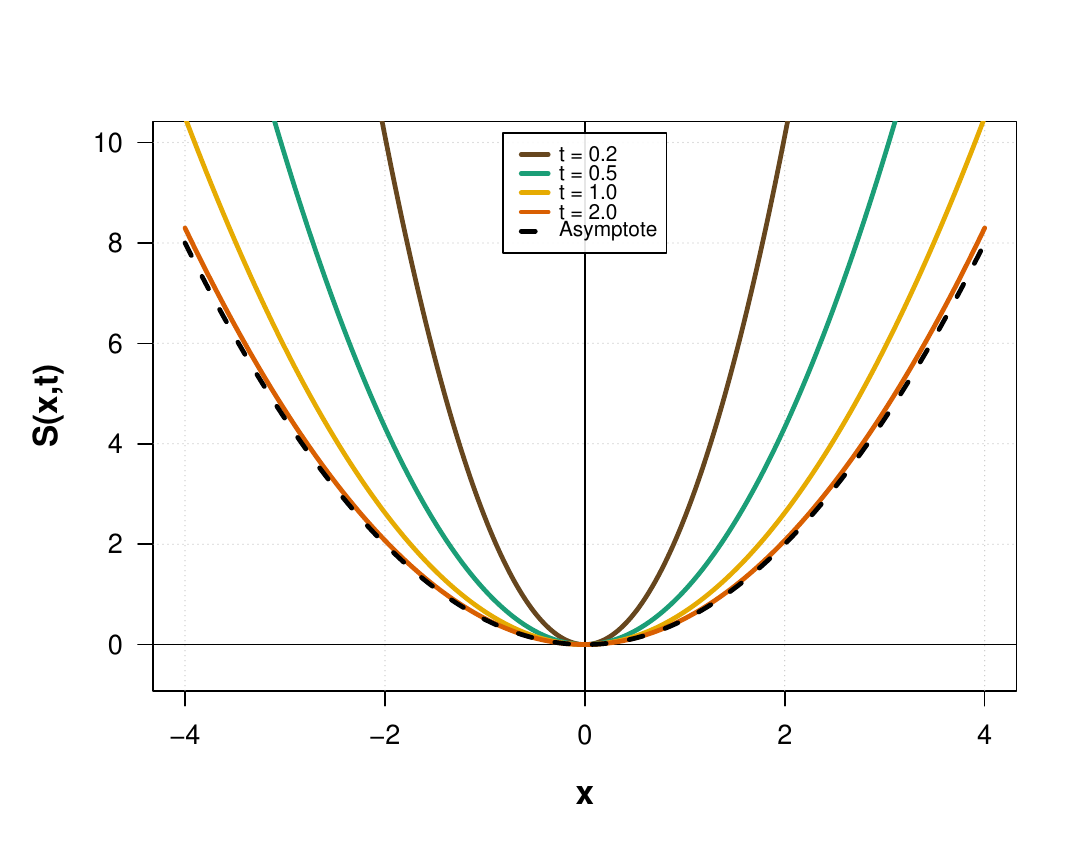}
\caption{A signature  of relaxation: Time evolution of $S(x,t)= \int_0^t {\frac{1}2} [\dot{x}^2 + (x^2 - 1)]d\tau $ in the harmonic case,  cf. (24-28), where  $k(x,t) \sim \exp[- S(x,t)] \rightarrow K(x)\Rightarrow  K= \int_R K(x) dx$. The action  $S(x,t)$ is evaluated in terms of a minimizing classical solution of Eq. (20),  see e.g. (20-25).}
\end{center}
\end{figure}

The phrase  "imaginary time equations of motion" is a semantic artifice telling that local conservation laws of standard diffusion processes (see   our Technical Comment in Section I,  Refs. \cite{klauder,olk,monthus,zaba1} and \cite{physa,qpot,recoil,gar1,burgers}) operate with   force terms of the form $+\nabla {\cal{V}}$ whose sign is  opposite to  customary (of Newtonian origin) force inputs  recovered  through $b(x)= - \nabla \phi $. Compare e.g.  the second Newton law in the form of  Eq. (20).

In fact, the equation  $\ddot{x}= {\frac{\partial {\cal{V}}}{\partial {x}}}$ often  happens to be called  "the Euclideanized equation of motion", only if per force set against the standard  (Newtonian by origin)   $\ddot{x}= -  {\frac{\partial {\cal{V}}}{\partial {x}}}$, both laws involving the very same potential  function ${\cal{V}}(x)$.

In passing, let us mention that throughout the paper, the spectral properties (search for lowest eigenvalues and eigenfunctions) are addressed  exclusively for Hamiltonians of the form $H= {\frac{1}2}[- \Delta + (x^4 - 2a^2x^2)]$  where we encounter the standard  double-well potential. We  have never spectrally addressed  the  inverted problem, like e.g.  the "double-barrier" potential (or the inverted oscillator in the harmonic case, \cite{barton}).

The "inverted double-wells" have  appeared only in conjunction with the implicit path-wise description,  and in particular  with   classical solutions of Eq. (20),   which minimize the action functional $S(x,t)= \int_0^t  {\frac{1}2} [\dot{x}^2 + {\cal{V}}(x)]d\tau $   in the    Lagrangian  path integral for the integral  kernel   $k(y,0,x,t)$, cf. Section I.C.  It is the   path-wise  dynamics, which "looks" Euclidean, see e.g. \cite{klauder}, and Eq. (20) again.

\subsubsection{Some explicit solutions of the Euler-Lagrange equations (20) with the double-well potential   ${\cal{V}}(x)$.}

Although no explicit   imaginary time   transformation  has been  ever  involved in our  discussion, it is worthwhile to examine some    solutions of the Euler-Lagrange equations  (20), amenable to  Euclidean  associations (that according to the current instanton lore, albeit we use an undoubtedly   {\it real} time label).  To this end we shall  employ  double-well potentials ${\cal{V}}(x)$  of Sections IV.A and B.\\

Let us first consider the potential ${\cal{V}}(x)\equiv V(x) = \lambda (x^2-a^2)^2$, depicted for $\lambda = 1/2$ in Figs 14  and 15.  Cf. Fig. 18 for the visualization of the  induced (decaying) semigroup dynamics in case of   $a=1$.  We recall that a subtraction of  half the lowest eigenvalue $E_0/2$    is a must to  achieve  the relaxation regime.\\

{\bf Basic instanton.}\\

\noindent
In the Newton-type second law  (20),  $x(t)$ is a dynamical variable. Therefore, multiplying  from both sides  of Eq. (20) by $\dot{x}$, we  recover  (the customary Newtonian mass parameter $m$ has been scaled out)
\be
{\frac{1}2} {\frac{d}{dt}} (\dot{x}^2) = {\frac{d}{dt}} V(x(t)) \Longrightarrow  {\frac{1}2} \dot{x}^2 = V(x(t)) + c, \, \,  c\geq  0.
\ee

Let us assume that  $c=0$,  when   the total energy  vanishes, ${\cal{E}} = {\cal{T}} - {\cal{V}} ={\frac{1}2} \dot{x}^2 - V(x) = 0$.  With the explicit form of $V(x) = \lambda (x^2-a^2)^2$, we get
\be
\dot{x}  = \pm \sqrt{2\lambda }  (x^2-a^2) \Longrightarrow  x(t)= \pm a \tanh [a \sqrt{2\lambda} \, \, (t-t_0).
\ee
This   form of the solution  $x(t)$  ensures that we can associate  $x(0) =0$ with $t_0=0$,  and secures an asymptotic property  $x(t) \rightarrow \pm a $ as $ t\rightarrow \infty $. The obtained $x(t)$ is known as a {\it basic instanton} solution of the  ("imaginary time")  Newton equation (20).

We have in hands  $\dot{x}= \pm  a^2 \sqrt{2\lambda } \cosh^{- 2} (a \sqrt{2\lambda } \, t)$.  Since,  in the present case , the Lagrangian reads ${\cal{L}}= {\cal{T}}  + {\cal{V}} = 2{\cal{T}} = 2{\cal{V}} =  2\lambda a^4 \cosh^{-4} (a\sqrt{2\lambda }\, \, t)$, the classical path contribution to the action  readily follows:
\be
S(x,t) = \int_0^t  {\cal{L}}(\tau )  d\tau ={\frac{2a^3\sqrt{2\lambda }}3} \tanh(a\sqrt{2\lambda }\, t) \left[ 1+ {\frac{1}2} \cosh^{-2} (a\sqrt{2\lambda } \, t)\right]= {\frac{1}3}  x(t) \left[2a^2\sqrt{2\lambda }   + \dot{x}(t) \right]  .
\ee
For large $t$ (alternatively for large $a$), $\cosh^{-1} (a\sqrt{2\lambda}\,t)$  approaches zero, while   $\tanh (a\sqrt{2\lambda } \, t)$ approaches $1$, both exponentially. Therefore, in any of those regimes we would   have
\be
S(x,t) \sim {\frac{2a^3\sqrt{2\lambda }}3}  \Rightarrow  \exp \left[- {\frac{2a^3\sqrt{2\lambda }}3}\right],
\ee
as a valid contribution of a classical solution of (20) to the path integral.

We point out, that a standard instanton calculus involves typically the integration $\int_{-T}^{+T}$ and eventually  $\int_{-\infty }^{+\infty } $,  instead of  our $\int_0^t$.  Therefore, the  single  instanton outcome would be twice larger, e.g. $S(x,t) \rightarrow 2 S(x,t)$, leading to $\exp[- 2 S(x,t)]$, which in the $t \to \infty $ limit would imply $\exp[- (4/3)a^2 \sqrt{2\lambda}]$,  reminiscent of the WKB (semiclassical) calculations  of the bottom energy  levels splitting in the standard double-well  quantum model.  \\

{\bf Periodic instantons   and the vacuum bounce.}\\

\noindent
Let us integrate (67) again, but presuming that in integration constant is negative and nonzero. Accordingly, we consider, \cite{kirsten,instanton}:
\be
 {\frac{1}2} \dot{x}^2 = V(x(t)) - c, \, \,   0\leq c \leq V
\ee
We decompose $V(x)$ to the form previously utilized throughout Section IV. Namely, we consider
\be
V(x) = \lambda (x^2- a^2)^2 = \lambda (x^4 -2a^2x^2) +V_0,
\ee
where $V_0= \lambda a^4$, and ${\frac{\partial V}{\partial x}}= 4\lambda x(x^2 - a^2)$.

Substituting (72) in (71), we have
\be
\dot{x}^2 = (-2c + 2a^4\lambda )-4\lambda a^2 x^2 + 2\lambda x^4.
\ee
This identity, by means of clever substitutions, can be recast in the form of the nonlinear  m
elliptic equation:
\be
\left({\frac{dy}{dx}}\right)^2= k^2A^2 - (1+m^2)k^2y^2 + {\frac{k^2m^2}{A^2}}y^4 = {\frac{k^2}{A^2}} (A^2-y^2)(A^2-m^2y^2),
\ee
provided we set:
\be
k= \sqrt{\frac{4\lambda a^2}{1+m^2}}, \,   \, \, \,  A^2= {\frac{k^2 m^2}{2\lambda }}, \, \, \, \,
c= b^2 V_0, \, \, \, \,   b = {\frac{1-m^2}{1+m^2}}.
\ee
Eq. (74) admits a family of periodic solutions,  which we recast as solutions  of Eq. (73):
\be
y = A\, sn[k(x-x_0),m]  \Longrightarrow  x(t) = \pm {\sqrt{\frac{2m^2a^2}{1+m^2}}} \, sn \left[ {\sqrt{\frac{4a^2\lambda}{1+m^2}}}(t-t_0),m)\right],
\ee
where we can safely take $t_0=0$. Here $sn (k(x-x_0),m)$ is the Jacobi elliptic sine function with the modulus $0\leq  m\leq 1$,  $x_0$ is arbitrary, and may take the value $0$.

The solution (76) of the equation (20) is  called the {\it basic periodic instanton}.  The period of $sn(x,m)$ is $4{\cal{K}}(m)$ where ${\cal{K}}(m) $  is the complete elliptic integral of the first kind
\be
{\cal{K}}(m) = \int_0^{\pi/2} {\frac{1}{\sqrt{1-m^2 sin^2 \phi}}} d\phi.
\ee
In view of (76), the period of the basic periodic instanton is
 $ T=4 {\cal{K}}(m) [(1+m^2)/(4a^2\lambda)]^{1/2}$.

If $m=1$, then   $b=0$ and thence $c=0$, the solution (76) degenerates into the basic instanton  $x(t)= \pm \tanh [a\sqrt{2\lambda}(t-t_0)]$.

There are more solutions  of Eq. (20)  in the reach.
Leaving aside their  usefulness issue, within the tenets of the instanton calculus, let us mention another example of the {\it periodic instanton},  \cite{kirsten}. Assuming $x(t)= x(t+ T)$, where $T$ stands for the period, one can deduce (that is somewhat intricate in view of the complicated reparametrization):
\be
x(t)= {\frac{\beta (k)}{\sqrt{2\lambda}}}\, dn[\beta (k)(t+t_0),\gamma ].
\ee
The parameter  $0\leq k\leq 1$ follows from $k^2= (1-u)/(1+u)$, where $u^2= c/a^4\lambda $.
Other entries follow:    $\gamma = 2\sqrt{k}/(1+k)$, and  $\beta  =  a(1+k) \sqrt{2\lambda /(1+k^2)}$.  We point out that the  Jacobi  amplitude function $dn$ can be retrieved from $sn$ according to:   $dn(x,\gamma )= \sqrt{1-\gamma ^2 sn^2(x,\gamma)}$.

The major observation is that the Jacobian elliptic function $dn[\beta(k)t,\gamma ]$ has period  $\beta(k) T = 2 n {\cal{K}}(\gamma )$, $n=1,2...$.

Another important observation is that as $c\rightarrow 0$, together with $k\rightarrow 1$, the periodic solution (78) degenerates to the so-called {\it  vacuum bounce}:
\be
x(t)= a\sqrt{2}\,  sech [2a\sqrt{\lambda }(t+t_0)].
\ee
The visualization of solutions  (78) and (79) can be found in Ref. \cite{kirsten}, see also \cite{instanton}.

\section{Outlook.}

In the present endeavour we have largely extended previous observations, \cite{zaba1}-\cite{stef}, concerning the relaxation of  drifted diffusion processes,  where the  principal  dynamical entry of the transition probability density of the diffusion  has been an  integral kernel  of the  Feynman-Kac semigroup operator.  We are guided by intuitions reaching that far as to  Refs. \cite{klauder,ito,faris},   where killed diffusions were  associated  merely  with  nonnegative-definite  Feynman-Kac potentials. It is obvious that  the killing mechanism alone is   incompatible with the  considered  diffusion  relaxation scenario.

Since F-K potentials, understood as continuous bounded from below functions, a priori admit bounded  negativity subdomains in $\mathbb{R}$, we have  addressed the general  issue of the compensating mechanism for killing.  That in part borrows some impetus from the notion of "potentials with subtraction" (e.g. "shifted potentials"),  \cite{faris,vilela}.

 Our proposal is to take seriously not only the killing of random paths but also their  branching, here realised as cloning, effectively realised as  a bifurcation of a random path into two independent branches, (cf. \cite{huillet} for an analogous idea  for the Brownian motion  in the interval with absorbing ends).   We stress that  random  killing events  might happen exclusively in positivity domains  of the F-K potential, while branching in its  negativity domains only,  provided  the reference free  Brownian trajectory visits these mutually   disjoint spatial areas in the  course of evolution. Clearly,  sample paths visiting the positivity domains have lower chances to survive up to the prescribed terminal time $t$.   On the other hand, sample paths visiting the negativity domains have higher chances to survive, since their bifurcations open competing travel options and the number of non-killing  (survival) options definitely proliferates.

 To proceed with  the  consistent  path-wise picture of tamed Feynman-Kac diffusions we find necessary to   abandon an explicit "particle motion" paradigm.  We do not consider hereby the killing of real physical particles, or the birth of new ones in  branching events  (which might be associated with the  creation of mass in random motion, \cite{helms}-\cite{kesten}),  but  concentrate on   the path-wise analysis, understood as switching-off (killing) or opening new  options (through branching) for the admissible  uninterrupted  path  between the  point $y$ left at  $t=0$,  and the terminal point  $x$ to be reached at time $t$.  Concerning the  "opening new   options" through trajectory bifurcation, we cite  a phrase from  conclusions of Ref. \cite{pre24}:\\

\noindent
"The trajectory picture we have described in the present paper, effectively reduces each branching event to the trajectory bifurcation at a random time instant. This, to some
extent, may be interpreted in terms of the metaphor  ("the garden of forking paths", {\color{blue} cf. \cite{borges,frohlich}}),
concerning an uncontrollable multitude of ways allowing to reach a predefined destiny (here a terminal point $x$  at time $t$), from a predefined beginning (starting point $y$ at $t = 0$), along a continuous path, with branching versus killing events happening randomly on the way. A continuity property of the ultimate (uninterrupted) path, is nonetheless preserved and the terminal point of the trajectory can be always reached by
meticulously avoiding path segments with dead ends ('pruned branches')".\\

 One may as well rephrase the term "path" as the term "history", and interpret the Feynman-Kac kernel as a  "sum over histories", where killing represents  a blockade of prohibited path segments, while branching (trajectory bifurcations) opens new, potentially admissible, safe   routes to the ultimate destination.

 In the present paper, we have not provided a precise  analytic {\it derivation of the equivalence} between the renormalized semigroup  and the killing/branching random dynamics. Rather, we heavily rely on the computer-assisted outcomes, {\it  to justify the existence} of a strong link between these two dynamical formalisms.

 Our  computer-assisted path-wise analysis of the "taming F-K potentials" workings,  placed special emphasis on superharmonic and double-well potentials. These were considered   in two complementary roles: of the  conservative   drift-inducing one, and of the Feynman-Kac potential proper.  The major  reference  concepts were  introduced  analytically in the linear drift case.  The level of technical difficulties met  in nonlinear   models, somewhat narrowed our discussion to the asymptotic regime. Nonetheless,   simulations have reproduced a correct   path-wise   dynamical behavior,  as depicted in Figs. (9), (13), (18), (20)-(25).

 The "potentials with subtraction" \cite{faris} have naturally  appeared on the way, and the consistency of the proposed killing-branching  taming  mechanism   appears to be confirmed in a setting much broader than that of Ref. \cite{pre24}.\\

 In somewhat naive lore, one may interpret drifted diffusions  of Section I as an admissible stochastic realization of properly   weighted  (via the Doob-like conditioning, cf. (10) and (13)) tamed Feynman-Kac diffusions  with killing and branching. The reverse route (albeit bypassing  standard growth  restrictions that normally  guarantee a uniqueness and non-explosiveness of the process, \cite{klauder}) could be followed as well: given the Feynman-Kac potential, one may in principle deduce the drifted diffusion realization (10) of the Fokker-Planck dynamics (1).\\

The stochastic processes in question, together with  the   closely related (a priori admissible)   classical  solutions of the Newtonian law of  motion (cf. Eq. (20)) often are interpreted as a  the  "Euclideanization" of the standard Minkowski space classical/quantum  mechanics. We have  paid  attention to the fact that the Newton-type  equation (20), albeit with the  sign-reversed potential, has nothing to do with any Wick rotation of the time label. It is an inevitable consequence of  the considered  stochastic processes, and  their   dynamics  indeed  has a "Euclidean look"  from the outset,  as noticed long ago in Ref. \cite{klauder}.  This  observation  was even promoted to the status  of the "Brownian recoil principle", \cite{qpot,recoil}  in the hydrodynamical reformulation of  random motions.

  Throughout the paper, all arguments refer to the {\it real time} dynamics with $t\geq 0$.   Under the code-name of instantons, we actually  encounter more or less specialized   solutions of the  equation   (20).   The bottom energy splitting in the double-well case, has been established numerically (by means of the Strang splitting method) in the deeply non-perturbative regime. \\

\end{document}